\PassOptionsToPackage{super,numbers,sort&compress}{natbib} 
\documentclass[pdflatex,sn-nature]{sn-jnl}

\usepackage{graphicx}%
\usepackage{multirow}%
\usepackage{amsmath,amssymb,amsfonts}%
\usepackage{amsthm,mathtools,tensor}%
\usepackage{mathrsfs}%
\usepackage[title]{appendix}%
\usepackage{xcolor}%
\usepackage{textcomp}%
\usepackage{manyfoot}%
\usepackage{booktabs}%
\usepackage{enumitem}
\usepackage{algorithmicx}%
\usepackage{algpseudocode}%
\usepackage{listings}%

\theoremstyle{thmstyleone}%
\theoremstyle{thmstyletwo}%

\theoremstyle{thmstylethree}%

\unnumbered

\usepackage{amsmath, amssymb, amsfonts} 
\usepackage{placeins}                   
\usepackage{graphicx}                   
\usepackage{xcolor}                     
\usepackage{hyperref}                   
\hypersetup{
    colorlinks=true, linkcolor=blue, citecolor=blue, urlcolor=blue }
\usepackage{cleveref}                   
\usepackage{bibunits}                   

\usepackage{physics}                     
\usepackage{quantikz}
\usepackage{braket}
\usepackage{tikz}
\usetikzlibrary{quantikz2}
\usepackage{adjustbox}

\usepackage[version=4]{mhchem}                     
\usepackage{siunitx}                    

\usepackage{array, booktabs}            

\definecolor{natpumpkin}{HTML}{E69F00}
\definecolor{nataqua}{HTML}{56B4E9}
\definecolor{natgreen}{HTML}{009E73}
\definecolor{natyellow}{HTML}{F0E442}
\definecolor{natnavy}{HTML}{0072B2}
\definecolor{natorange}{HTML}{D55E00}
\definecolor{natpink}{HTML}{CC79A7}
\definecolor{natblack}{HTML}{000000}

\begin{document}
\begin{bibunit}[sn-nature]

\title[Measurement-efficient energy estimator]{Measurement-efficient energy estimator for quantum chemistry of photodynamic materials}
\author[1]{Connor Lenihan}
\equalcont{These authors contributed equally to this work.}
\author[1]{Kripa Panchagnula}
\equalcont{These authors contributed equally to this work.}
\author[1]{Luca Erhart}
\equalcont{These authors contributed equally to this work.}
\author[2]{Jem Guhit}
\equalcont{These authors contributed equally to this work.}
\author[1]{Carlo A. Gaggioli}
\equalcont{These authors contributed equally to this work.}
\author[1]{Gabriel Greene-Diniz}
\equalcont{These authors contributed equally to this work.}
\author[1]{Gabriel Laude}
\author[1]{Megan Cerys Rowe}
\author[2]{Fabian Finger}
\author[2]{Pranav Kalidindi}
\author[2]{Enrico Rinaldi}
\author[3]{Mikko Muuronen}
\author[4]{Chandrika Mulakala}
\author[1]{Thomas M. Soini}
\author[2]{Elvira Shishenina}
\author[1]{David Muñoz Ramo}
\author*[1]{David Zsolt Manrique}
\email{{\small david.zsolt.manrique@quantinuum.com}}
\affil[1]{\orgname{Quantinuum}, \orgaddress{\street{Terrington House, 13--15 Hills Road}, \city{Cambridge}, \postcode{CB2 1NL}, \country{UK}}}
\affil[2]{\orgname{Quantinuum}, \orgaddress{\street{Partnership House, Carlisle Place}, \city{London}, \postcode{SW1P 1BX}, \country{UK}}}
\affil[3]{\orgname{Johnson \& Johnson}, \orgaddress{\city{Beerse}, \country{Belgium}}}
\affil[4]{\orgname{Johnson \& Johnson}, \orgaddress{\city{South San Francisco}, \state{CA}, \country{USA}}}

\abstract{
Accurate modelling of chemical reaction rates requires precise correlated energy differences.
Quantum computing offers a potential route to their determination, but extracting energies from quantum-prepared states can demand prohibitive sampling.
We demonstrate measurement-efficient correlated energy estimation using an electronically challenging H$_2$DBB--Ru model derived from a Ru-functionalized hafnium metal-organic framework relevant to radiodynamic and photodynamic therapies.
The mixed estimator combines a quantum-prepared target with sparse multiconfiguration references and recovers ground-state corrections using magnitude-only compute-uncompute measurements.
Analytical variance estimates predict 1.6 mHa precision with a few hundred shots, about 100 times fewer than the single-determinant formulation and three orders of magnitude fewer than optimized Hamiltonian-term sampling.
With approximate targets, it reaches comparable accuracy using simpler state preparation than direct expectation-value estimation.
Emulator and trapped-ion experiments on Quantinuum Helios validate the approach up to (10,10), including a 50-qubit circuit combining AI-assisted state preparation with error detection, providing a measurement-efficient route to correlated energy differences for challenging chemical problems. }

\maketitle

\section{Introduction}
Photodynamic therapy (PDT) and radiodynamic therapy (RDT) are emerging as spatially controlled and minimally invasive approaches to cancer treatment through the sensitized production of reactive oxygen species.\cite{maharjan2022singlet,li2016photosensitized,cui2024singlet}
Nanoscale metal--organic frameworks (MOFs) have attracted particular interest for these applications because their modular structures enable systematic tuning of photosensitizers, radiation-responsive components and the surrounding chemical environment.\cite{furukawa2013chemistry,ni2018nanoscale,lu2018lowdose}
Improved control over where and how these materials generate reactive oxygen species could ultimately support efforts to increase therapeutic activity while limiting unintended effects on healthy tissue.
The therapeutic activity of these materials is governed by a complex interplay of electronically excited and non-adiabatic processes, making mechanistic understanding of their excited-state reactivity a prerequisite for the rational design of next-generation PDT and RDT materials.

Obtaining a predictive description of these excited-state processes remains challenging.
Density-functional approximations (DFAs) and single-reference correlated wavefunction methods often provide practical approaches for modelling excited-state reactivity, but systems containing near-degenerate electronic states may exhibit substantial multireference character, challenging the reliability of such methods.\cite{cohen2008insights,cohen2012challenges}
In transition-metal systems, these limitations may lead not only to quantitative errors in relative energies\cite{wardzala2026oxo} but also to qualitatively different descriptions of the underlying electronic structure, including changes in state ordering and in the balance between metal-to-ligand charge-transfer (MLCT) and metal-centred (MC) states.\cite{soriano2021importance}
Multiconfigurational wavefunction methods provide a systematic treatment of such effects, but their computational cost increases rapidly with active-space size.\cite{gaggioli2019beyond}
Quantum computing offers a potential route to extending correlated electronic structure calculations beyond classically accessible active spaces (ASs), provided that the resulting energies can be resolved with sufficient precision to distinguish chemically relevant states.

Realizing this potential involves several challenges, including preparing an accurate correlated state, implementing coherent evolution where required by the algorithm, and achieving sufficient statistical precision in the final readout.\cite{mcardle2020quantum}
Quantum phase estimation extracts eigenvalues with favourable sampling complexity, but requires non-negligible initial-state overlap and long, coherent controlled time evolution, placing its principal advantages in fault-tolerant regimes.\cite{abrams1999quantum,babbush2018encoding}
Adiabatic preparation,\cite{kremenetski2021adiabatic} quantum imaginary-time evolution,\cite{motta20} dissipative preparation,\cite{li2025dissipative} and quantum eigenstate filtering\cite{lin2020eigenstatefiltering} instead provide routes to correlated target states without prescribing a high-precision energy readout.
Accompanying the state preparation, energy calculation is conventionally carried out from repeated measurements of the Hamiltonian.\cite{peruzzo2014variational,mcclean2014locality}
Grouping compatible terms and optimizing shot allocation reduce this cost, but the sampling required for chemically relevant precision can remain prohibitive as the active space grows.\cite{verteletskyi20,gonthier2022measurements}
Thus, preparing a correlated state does not by itself provide a chemically precise energy: measurement remains a distinct bottleneck.

The mixed estimator (ME) provides an alternative energy-readout strategy when a target eigenstate is available on the quantum processor and a suitable reference state can be obtained classically.
Lenihan et al. formulated a quantum mixed estimator using a single-determinant reference and classical-shadow reconstruction.\cite{lenihan2025excitationamplitudesamplinglow,Huang2020}
Building on this work, we combine sparse multiconfiguration references with various direct-overlap measurements, including a purpose-built protocol, to reduce sampling and runtime costs.
A notable property of the direct-overlap ME formulation is that, for an exact ground-state target, the fixed sign of the energy correction allows it to be recovered from magnitude-only measurements, avoiding more expensive phase reconstruction.
We show that these improvements over the original application of the mixed estimator reduce the shot requirement by orders of magnitude.
The resulting workflow evaluates electronic energies from quantum-prepared states without directly sampling individual Hamiltonian terms or requiring the coherent time evolution used in phase estimation.

\begin{figure*}
    \centering
    \resizebox{\textwidth}{!}{%
        \input{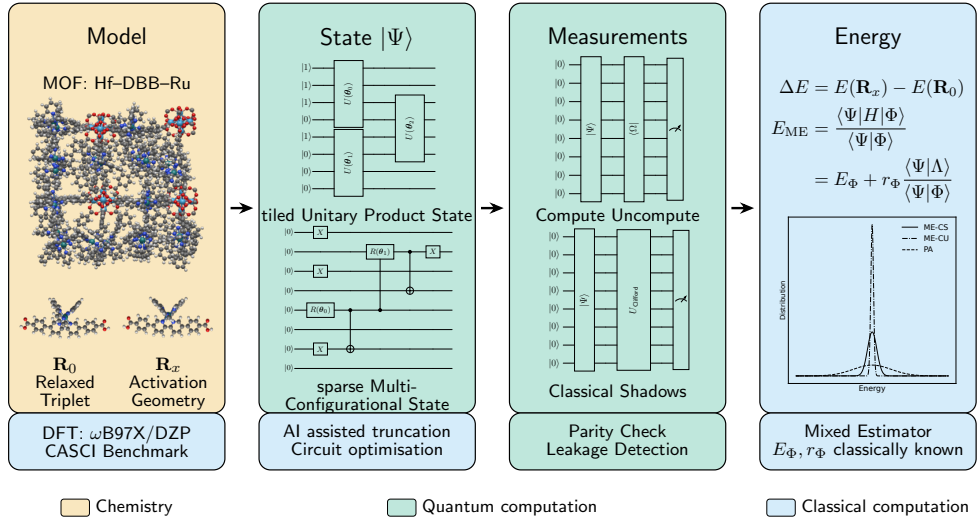}
    }
    \caption{End-to-end workflow from chemical modelling to precise energy differences, illustrated here for an H$_2$DBB--Ru molecular model extracted from the Hf--DBB--Ru framework.
    Classical calculations define the relevant geometries and AS Hamiltonians.
    A quantum state-preparation method provides a target eigenstate or approximation, here we use a variant of tUPS: pp-tUPS optimized by statevector VQE; and sparse multiconfiguration states.
    Direct measurements of overlaps with classically constructed sparse reference and residual states, particularly compute--uncompute, reduce sampling cost and variance relative to Pauli averaging.
    The mixed estimator combines these overlaps with classically known reference quantities to obtain electronic energies and their chemically relevant differences.}
    \label{fig:workflow}
\end{figure*}

We demonstrate the ME workflow using singlet-oxygen generation in a H$_2$DBB--Ru molecular model extracted from the Ru-functionalized Hf--DBB--Ru framework (Fig.~\ref{fig:workflow}).\cite{ni2018nanoscale}
For this system, different DFAs predict markedly different electronic-state energetics, leading to singlet-oxygen generation rates that span more than five orders of magnitude as discussed below.
These differences arise not only from quantitative changes in the relative energetics, but also from qualitatively different descriptions of the underlying electronic structure, including changes in state ordering and in the balance between metal-to-ligand charge-transfer (MLCT) and metal-centred (MC) states.
This makes H$_2$DBB--Ru a chemically relevant benchmark exhibiting challenging excited-state electronic structure, providing a realistic setting in which to develop and evaluate quantum methodologies for resolving energy differences that challenge conventional electronic structure approaches.

We test the readout using exact and systematically approximate sparse multiconfiguration states (sMCS) and perfect-paired tiled unitary product states (pp-tUPS) across (6,6), (8,8), and (10,10) active spaces.
Finite-shot simulations and experiments on Quantinuum Helios-1E and Helios-1, including leakage and spin-parity error detection, then evaluate its sampling requirements and behaviour under device noise.

The resulting workflow is transferable to photochemical energetics relevant to rate calculations in other RDT and PDT photosensitizers, while also addressing the broader challenge of estimating correlation energies from quantum-prepared states with sufficient precision to resolve chemically relevant differences.

\section{Results}
\label{sec:results}

\subsection{Mixed estimator with sparse multiconfiguration references}
The mixed estimator is an established non-variational energy functional used in Quantum Monte Carlo calculations.\cite{molecular_AFQMC,developments_in_afqmc,mixed_estimator_VMC_PMC,Huggins2022_unbiasing}
Lenihan et al. recently demonstrated its evaluation on a quantum computer using a single-determinant reference and classical-shadow reconstruction.\cite{lenihan2025excitationamplitudesamplinglow}
Here we extend this framework to sparse \cite{MultiConfigSparse} multiconfiguration references and compare several direct and shadow-based overlap protocols.

Let $\ket{\Psi}$ be the normalized target state and let $\ket{\Phi}=\sum_{x\in S_{\Phi}}a_x\ket{x}$ be a normalized sparse reference with $\braket{\Psi|\Phi}\neq0$, where $S_{\Phi}$ is the set of determinants retained in the sparse expansion.
The mixed estimator is
\begin{equation}
E_{\text{ME}} = \frac{\mel{\Psi}{H}{\Phi}}{\braket{\Psi|\Phi}},
\end{equation}
which gives the exact eigenvalue when $\ket{\Psi}$ is an eigenstate of $H$.
To separate the reference contribution from its orthogonal correction, define the projector $Q_{\Phi}=I-\ket{\Phi}\bra{\Phi}$ and write
\begin{equation}
H\ket{\Phi}=E_{\Phi}\ket{\Phi}+r_{\Phi}\ket{\Lambda},
\end{equation}
where $E_{\Phi}=\mel{\Phi}{H}{\Phi}$, $r_{\Phi}=\norm{Q_{\Phi}H\ket{\Phi}}$, and $\ket{\Lambda}=Q_{\Phi}H\ket{\Phi}/r_{\Phi}$ is the normalized residual state, with $\braket{\Phi|\Lambda}=0$.
The mixed estimator can therefore be written as
\begin{equation}
E_{\text{ME}} = \frac{\mel{\Psi}{H}{\Phi}}{\braket{\Psi|\Phi}} = E_{\Phi} + r_{\Phi} \frac{\braket{\Psi|\Lambda}}{\braket{\Psi|\Phi}},
\label{eq:perp-estimator-final}
\end{equation}
where the quantum processor needs to prepare $\ket{\Psi}$, and the measurement protocols estimate the overlaps $h=\braket{\Psi|\Lambda}$ and $s=\braket{\Psi|\Phi}$.
For a sparse reference, $E_{\Phi}$, $r_{\Phi}$, and the determinant expansion of $\ket{\Lambda}$ are obtained classically from the action of $H$ on $\ket{\Phi}$.

Remarkably, $h$ and $s$ are complex in general, but their relative phase cancels in Eq.~\eqref{eq:perp-estimator-final} because the energy correction $r_{\Phi}h/s$ must be real.
The only remaining ambiguity in reconstructing the correction from the overlap magnitudes is therefore its sign.
If the target state is the lowest-energy eigenstate in the same symmetry sector as $\ket{\Phi}$, the variational principle fixes this sign
\begin{equation}
E_{\mathrm{ME}}-E_{\Phi}
=r_{\Phi}\frac{h}{s}
\leq0,
\label{eq:sign_of_ratio}
\end{equation}
therefore magnitude-only overlap measurements are sufficient.
There are various direct and shadow-based protocols that can access either the complex overlaps or their magnitudes.
Choosing among them is not a secondary implementation detail.
The reference determines the scale of the correction and the cost of preparing its residual, whose worst-case support scales as $\mathcal{O}(mn^4)$ for an $m$-determinant reference over $n$ spin orbitals, the prepared target controls the systematic error, and the nonlinear readout determines how finite-shot fluctuations combine in the final energy.
Together, these choices can change the resources required to reach a fixed energy accuracy by orders of magnitude.
The following sections trace this chain from correlated AS benchmarks and reference construction, through approximate target-state preparation, to the sampling efficiency and hardware performance of the competing overlap protocols.

\subsection{AS calculations define correlated benchmarks}

In this section, we discuss the classical calculations using correlated AS methods which generate benchmark energies for the quantum computing workflow.
We use an H$_2$DBB--Ru molecular model extracted from the Hf--DBB--Ru framework as a tractable model of the Ru-centred electronic structure problem considered here, as it reproduces the reported absorption and phosphorescence characteristics of the full MOF.\cite{ni2018nanoscale}
To this aim, we performed calculations using the Semistochastic Heat-Bath Configuration Interaction (SHCI) method with self-consistent orbital optimization, denoted SHCISCF.\cite{sharma2017semistochastic}
This selected-CI method optimizes the orbitals together with approximate CI coefficients, providing a scalable strategy to generate natural-orbital basis for subsequent quantum calculations.
We carried out SHCISCF calculations for two geometries denoted $\mathbf{R}_0$ and $\mathbf{R}_x$, corresponding to the relaxed triplet-state geometry and the geometry associated with singlet-oxygen activation, respectively.
Although the complete kinetic model also depends on electronic couplings and reorganization energies, we use $\Delta E_{\mathbf{R}_x-\mathbf{R}_0}=E_{\mathbf{R}_x}-E_{\mathbf{R}_0}$ as the final benchmark quantity because it constitutes the largest contribution to the activation energy, which enters the Marcus rate exponentially.\cite{bai2017divide}
We identified these geometries using density-functional approximations (DFAs) following the protocol described in Ref. \cite{bai2017divide} and in Section~\ref{Classical rate modeling}.

We performed triplet calculations employing (6,6), (8,8), and (10,10) active spaces, optimizing the orbitals for each geometry and active space.
These orbitals were then used to generate the corresponding quantum Hamiltonians and used as an orbital basis for CASCI (Complete AS Configuration-Interaction) benchmarks, providing exact electronic energies within each selected active space.
To calculate the energies, the lowest triplet-sector was used throughout, with $(N_\alpha,N_\beta)=(4,2)$, $(5,3)$, and $(6,4)$ targeting the high spin component in the three respective active spaces.
For the (10,10) active space (AS; orbitals shown in Supplementary Fig.~\ref{SI-fig:T1_and_Rx_orbs_1010}), the two singly occupied orbitals of $\mathbf{R}_0$ are located on a ruthenium $d$ orbital and a ligand $\pi^*$ orbital, consistent with $^3$MLCT character.
At $\mathbf{R}_x$, both are predominantly ruthenium $d$ orbitals, one with $t_{2g}$ and the other with $e_g$ character, consistent with the $^3$MC region of the triplet potential energy surface.
The remaining active orbitals are principally ligand $\pi$ and $\pi^*$ orbitals.

For the (6,6) calculations, the SHCISCF energies lie approximately 40~mHa below the Hartree--Fock values at both geometries.
The dominant determinant has a squared coefficient of approximately $0.93$, while the natural-orbital occupations depart from two and zero by about $0.07$.
These values indicate modest but non-negligible multireference contributions.\cite{wardzala2024organic}
Supplementary Table~\ref{SI-tab:shciscf_data} reports the energies, occupation numbers, and leading configurations of the wavefunction.

CASCI calculations using the geometry-specific SHCISCF orbitals give $\Delta E_{\mathbf{R}_x-\mathbf{R}_0}\simeq-6.3$~mHa in both the (6,6) and (8,8) active spaces, followed by $+0.1$~mHa in the (10,10) space (while this energy is between +35 and +47 mHa when second order $N$-electron valence perturbation theory (NEVPT2) correction is added, see Supplementary Table~\ref{SI-tab:casci_data}).
The change at (10,10) exposes sensitivity to the correlated orbital space and shows that these calculations are not yet converged with AS size.
They therefore provide controlled benchmarks for quantum energy estimation rather than converged physical energies.

Using these exact benchmarks as targets, we first examine the effect of reference quality by varying only $\ket{\Phi}$ for the two (6,6) AS Hamiltonians corresponding to the $\mathbf{R}_x$ and $\mathbf{R}_0$ geometries.
For each size $m$, the $m$ largest-amplitude determinants of the exact state define a selected subspace, and $\ket{\Phi_m}$ is the lowest eigenstate of $H$ within that subspace.
In a practical calculation, the support could instead be obtained from a selected-CI calculation or by sampling a prepared target state in the determinant basis, as in quantum-selected configuration interaction (QSCI) or sample-based quantum diagonalization (SQD), making the sampled subspace directly available as a ME reference.\cite{kanno2026qsci,robledomoreno2025sqd}

\begin{figure*}[t]
\centering
    \includegraphics[width=\textwidth]{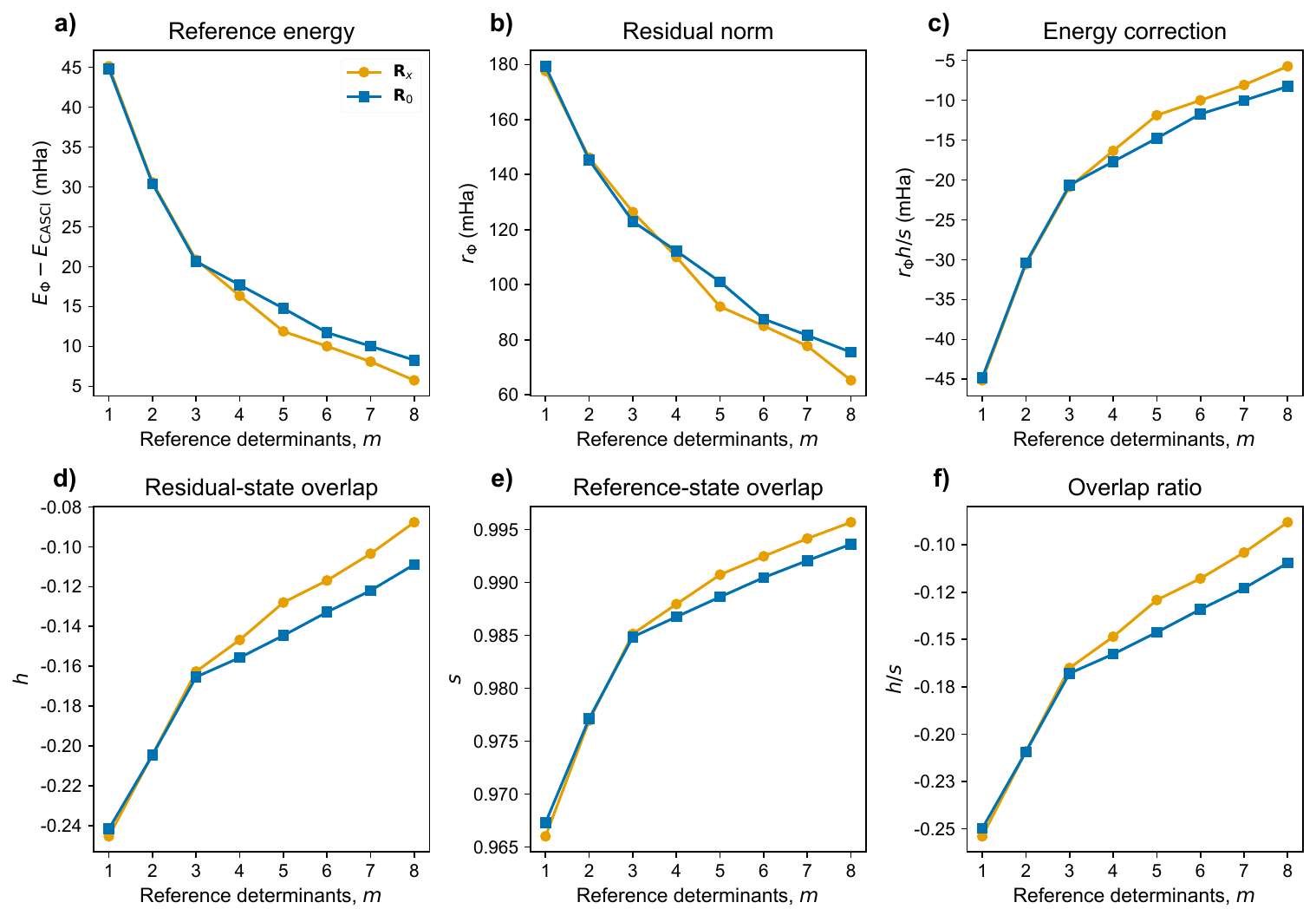}
\caption{Exact ME quantities for the triplet-spin ground states of the two (6,6) AS Hamiltonians corresponding to the $\mathbf{R}_x$ and $\mathbf{R}_0$ geometries.
For each $m$, the reference is obtained by diagonalizing the Hamiltonian in the subspace of the $m$ largest-amplitude CASCI determinants, from which the corresponding normalized residual state is constructed.
\textbf{a}, Reference-energy error $E_\Phi-E_{\mathrm{CASCI}}$.
\textbf{b}, Residual norm $r_\Phi=\norm{Q_\Phi H\ket{\Phi}}$.
\textbf{c}, Energy correction $r_\Phi h/s$.
\textbf{d}, Residual-state overlap $h=\braket{\Psi|\Lambda}$.
\textbf{e}, Reference-state overlap $s=\braket{\Psi|\Phi}$.
\textbf{f}, Ratio of the two overlaps.}
\label{fig:exact_mixed_estimator}
\end{figure*}

Figure~\ref{fig:exact_mixed_estimator} shows that increasing $m$ improves the reference in two complementary ways: $r_{\Phi}$ decreases, and $|h|$ decreases while $s$ increases, reducing $|h|/s$.
Both factors in the magnitude of the quantum correction, $|E_{\mathrm{ME}}-E_{\Phi}|=r_{\Phi}|h|/s$, therefore decrease with $m$.
These trends drive the reduction in shot requirements and other quantum resources, as the quantum measurable part is smaller.\cite{lenihan2025excitationamplitudesamplinglow}
Increasing $m$ is not free, however, because it also increases the preparation cost of $\ket{\Phi}$ and $\ket{\Lambda}$ in the direct overlap methods, producing a trade-off between circuit resources and measurement cost.

\FloatBarrier

\subsection{Mixed estimator with approximate target states}

In practice, the prepared target state may not be the exact CASCI eigenstate.
We therefore test the mixed estimator with two systematically improvable target-state preparations.
The perfect-paired tiled unitary product state (pp-tUPS) is a shallow, symmetry-preserving ansatz controlled by its number of layers.\cite{burton2024tups} Its parameters may be found variationally by leveraging the variational quantum eigensolver (VQE).\cite{peruzzo2014variational}
The sparse multiconfiguration state (sMCS) retains and renormalizes the $l$ largest-amplitude determinants of the CASCI state.\cite{MultiConfigSparse,MultiConfigSparse_ggd}

\begin{figure}[!htbp]
    \centering
    \includegraphics[width=0.95\linewidth]{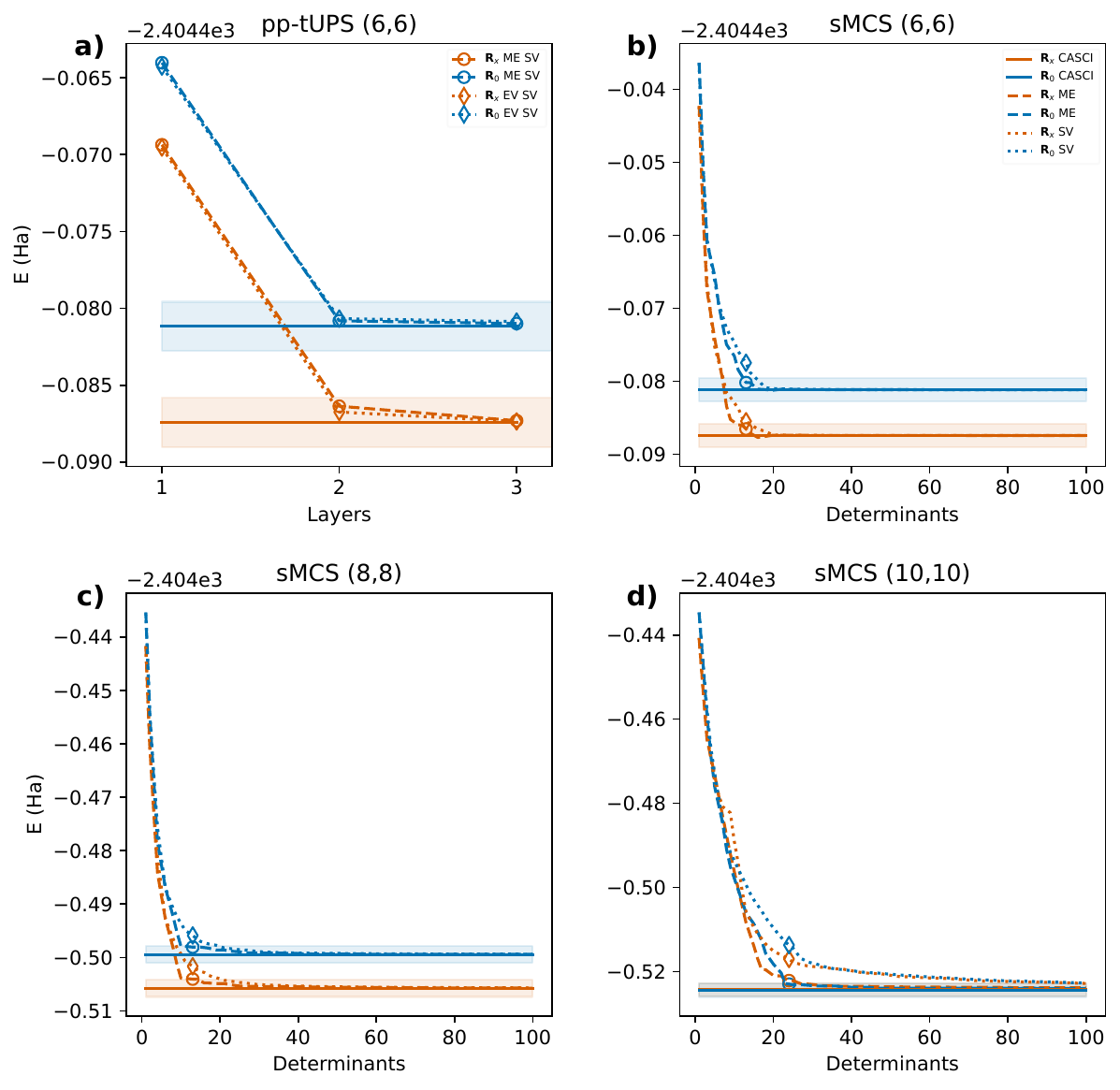}
    \caption{Statevector energy convergence of mixed estimator and expectation value for the approximate target states: \textbf{a)} pp-tUPS targets in the (6,6) active space.
    \textbf{b)}--\textbf{d)}, sMCS targets in the (6,6), (8,8), and (10,10) active spaces.
    Dashed and dotted curves correspond to ME and direct EV values, with solid lines marking the CASCI energy, and the shaded region marking $E_{\text{CASCI}}\pm1.6$ mHa.
    Orange and blue correspond to the $\mathbf{R}_x$ and $\mathbf{R}_0$ geometries.
    Scatter markers for sMCS indicate the truncations used for emulator and hardware runs as defined by Eq.~\eqref{eq:residual_truncation_condition}.}
    \label{fig:tups_mc_mixed_estimator_convergence}
\end{figure}

Figure~\ref{fig:tups_mc_mixed_estimator_convergence} compares the direct expectation value and the mixed estimator using the same sequence of approximate target states for all three active spaces.
For pp-tUPS, two layers place both estimators within the $\pm1.6$ mHa interval for both geometries.
For sMCS, the mixed estimator generally approaches the CASCI energies with fewer retained determinants than the direct expectation value, with the clearest separation in the larger ASs.
Because the state is not exact in this test and the mixed estimator is non-variational, it is not guaranteed to converge monotonically or to be closer at every truncation.
The result, however, demonstrates that the mixed estimator can reduce the target-state preparation required for a specified energy accuracy.
For subsequent discussions, we restrict ourselves to the (6,6) and (10,10) ASs, as the former is small enough to test a variety of methods with the larger pushing the boundaries of other methods.
More (8,8) data is presented in the SI.

\begin{figure}[!htbp]
    \centering
    \includegraphics[width=0.95\linewidth]{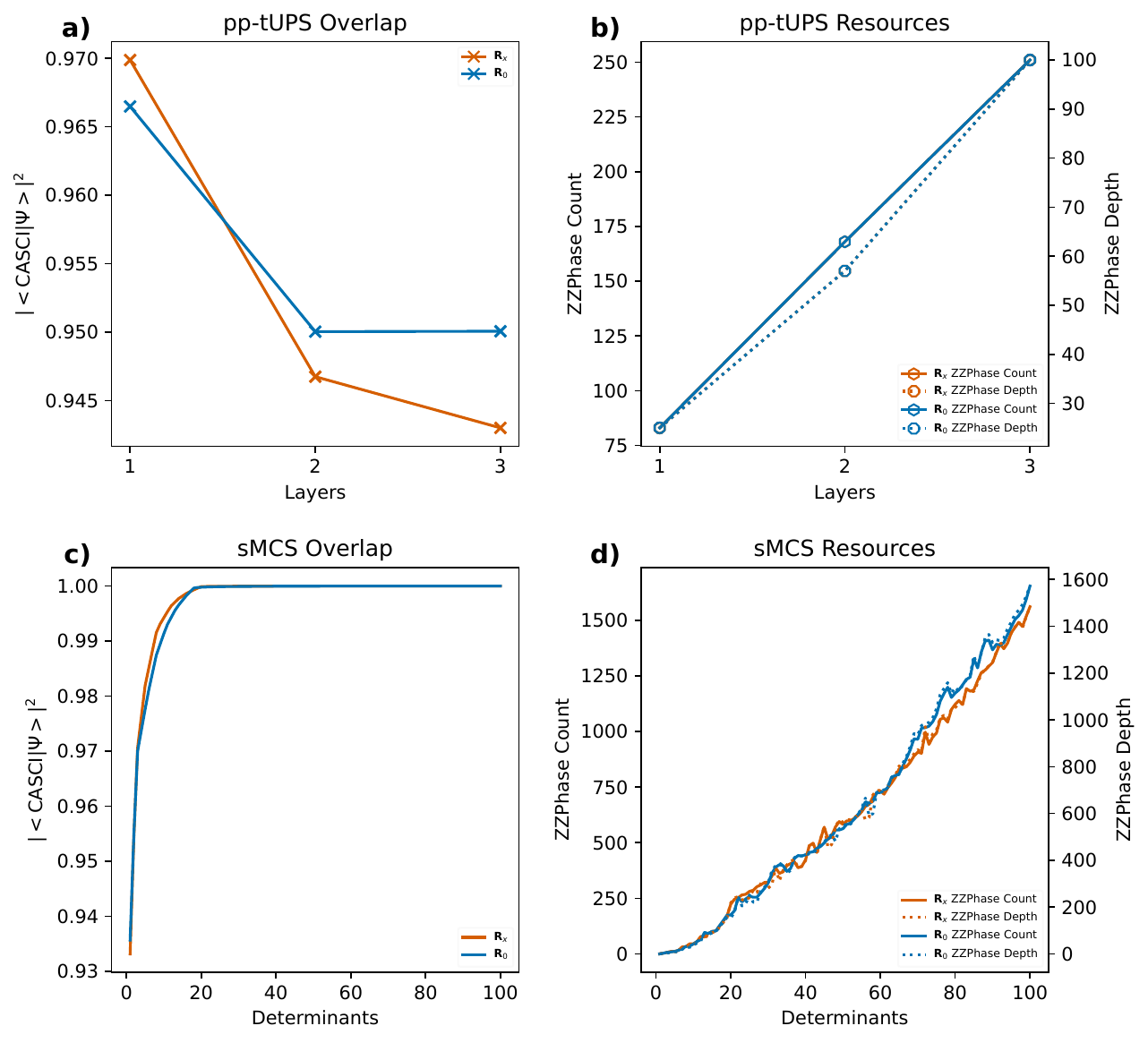}
    \caption{State-preparation quality and circuit resources for the (6,6) active space.
    \textbf{a)}, \textbf{b)} indicate the overlap of the CASCI and pp-tUPS state, alongside the two-qubit pp-tUPS resources. By definition of the ansatz (\cref{fig:tups_parity_circuits}), resources for both geometries are identical.
    \textbf{c)}, \textbf{d)} are analogous for the sMCS states.
    Orange and blue correspond to the $\mathbf{R}_x$ and $\mathbf{R}_0$ geometries. pp-tUPS resources for both states are identical; a feature of the ansatz with two-qubit gate count as hexagons and depth as octagons. sMCS gate count is shown as a solid line, with dotted corresponding to depth.}
    \label{fig:tups_mc_cas66_stateprep}
\end{figure}

Figure~\ref{fig:tups_mc_cas66_stateprep} quantifies the associated target-state overlap and two-qubit gate (ZZPhase gate) resources for the (6,6) active space.
For a fixed active space, the pp-tUPS resources depend only on the number of layers, with two layers requiring 168 two-qubit gates at a two-qubit depth of 57.
The sMCS direct expectation values reach chemical accuracy with 14 and 16 determinants for $\mathbf{R}_x$ and $\mathbf{R}_0$, respectively, corresponding to 92 and 103 two-qubit gates and depths of 82 and 95.
Thus, the gate counts are lower for sMCS, while pp-tUPS is shallower in this case.
Across the tested sMCS cases, reaching the $1.6$ mHa criterion with the mixed estimator reduces the target-state number of determinants to 11 and 13 respectively lowering the two-qubit gate counts to 56 and 96 with depths of 47 and 91 respectively.
As expected, the sMCS overlap with the CASCI state increases monotonically with the number of retained determinants.
The pp-tUPS overlap is not monotonic with layer number even though its energy improves, which may be rationalised as the early layers better filtering out the high energy excited states without fully optimising the coefficients of the lower energy determinants.
Due to the ansatz universality,\cite{burton2024tups} we expect this overlap to increase as more layers are introduced.

For the (10,10) AS, direct expectation-value estimation requires 96 determinants for $\mathbf{R}_x$ and more than 100 for $\mathbf{R}_0$ to reach the target accuracy (Fig.~\ref{fig:tups_mc_mixed_estimator_convergence}).
The corresponding circuits require more than 1600 two-qubit gates and depths of similar magnitude (Supplementary Fig.~\ref{fig:mc_cas881010_stateprep}).
This is where we most markedly see the gain of the mixed estimator as chemical accuracy is achieved for 27 and 23 determinants even in the largest AS, corresponding to 224 and 159 two-qubit gates, with depths of 211 and 134 respectively.

The comparisons above vary only the target-state preparation, thereby isolating its approximation error.
Direct overlap measurement additionally requires circuits for $\ket{\Phi}$ and $\ket{\Lambda}$.
Because $\ket{\Phi}$ comes from diagonalization in a selected determinant subspace and $\ket{\Lambda}$ follows classically from $Q_{\Phi} H\ket{\Phi}$, both are known determinant expansions well suited to sMCS state preparation.
Their circuits add gates and depth beyond the target-state circuit, depending on the measurement protocol, as discussed later.

The residual state can also be truncated before circuit synthesis.
We denote the three state-preparation sizes by $(m,k,l)$, where $m$ is the number of determinants in the reference $\ket{\Phi_m}$, $k$ is the number retained in the residual $\ket{\Lambda_{m,k}}$, and $l$ is the pp-tUPS layer count or the number of determinants in the sMCS target $\ket{\Psi_l}$.
For fixed $m$ and $l$, a conservative condition for residual truncation to contribute less than a target energy accuracy $\epsilon$ is
\begin{equation}
\frac{r_{\Phi_m}}{|s_{m,l}|}
\norm{\ket{\Lambda_{m,k}}-\ket{\Lambda_m}}
<\epsilon,
\label{eq:residual_truncation_condition}
\end{equation}
where $s_{m,l}=\braket{\Psi_l|\Phi_m}$, $\ket{\Lambda_m}$ and $\ket{\Lambda_{m,k}}$ are the full and truncated residual states, respectively, and the state-difference norm can be calculated classically.
This condition is conservative because it assumes the worst-case alignment of the discarded determinants with the target state.
Approximate information about the target state can allow more selective discarding.
The required reference and residual sizes are coupled and vary nontrivially, leaving scope for further joint optimization of the state sizes and determinant support.
The truncation scans in Supplementary Fig.~\ref{fig:mixed_estimator_truncation} show that smaller residual states can be used in practice, but also that the feasible state-size landscape becomes more constrained as the active space grows.

The smaller active spaces were already tractable using simple truncation, while the (10,10) implementation provided greater scope for resource reduction.
We used AI-assisted truncation and circuit optimization as an enabling step for this largest calculation, as described in Methods, with the resource metric $2G(\Psi_l)+G(\Lambda_{1,k})$, where $G(\alpha)$ is the number of ZZPhase gates in the compiled preparation circuit for $\ket{\alpha}$.
The search was constrained by energy accuracy and spin symmetry.
The optimization left the $\mathbf{R}_x$ construction unchanged at $(m,k,l)=(1,21,24)$ and selected a smaller $(1,21,22)$ construction for $\mathbf{R}_0$, with the target support formed from the union of the reference- and residual-state supports.
Across both geometries, the support selection and circuit optimization reduced the combined ZZPhase-gate count from 857 to 755, approximately $12\%$, and reduced the $\mathbf{R}_0$ two-qubit depth from 351 to 262.
Further details of the $(m,k,l)$ landscape, selected supports, and resource accounting are given in Methods and Supplementary Section~\ref{SI-sec:mixed_estimator_truncation}.
The resulting state pair was used in the emulator and hardware tests below.

\subsection{Direct overlap measurements reduce sampling cost}\label{sec:ME_variance}

For the symmetry-sector ground states considered here, the known sign of the correction reduces the mixed estimator to two squared overlaps.
Let $U_{\alpha}\ket{0}^{\otimes n}=\ket{\alpha}$ prepare an $n$-qubit state $\ket{\alpha}$.
For $\ket{\Omega}\in\{\ket{\Phi},\ket{\Lambda}\}$, compute--uncompute (CU), also known as the vacuum test, prepares $\ket{\Psi}$, applies $U_{\Omega}^{\dagger}$, and measures the system register.\cite{havlicek2019supervised}
Its all-zero probability is
\begin{equation}
p_{\Omega}=P(0^n)=\abs{\braket{\Psi|\Omega}}^2.
\end{equation}
Separate circuits with $\Omega=\Phi$ and $\Omega=\Lambda$ estimate $p_{\Phi}=|s|^2$ and $p_{\Lambda}=|h|^2$, respectively.
CU uses one system register, no ancilla, and no controlled target preparation.
Using the sign determined in Eq.~\eqref{eq:sign_of_ratio} gives
\begin{equation}
E_{\mathrm{ME}}
=E_{\Phi}-r_{\Phi}\sqrt{\frac{p_{\Lambda}}{p_{\Phi}}}.
\label{eq:me_absolute_overlap}
\end{equation}

\begin{figure}[!htbp]
\centering
\resizebox{\columnwidth}{!}{%
\begin{quantikz}[
    column sep=0.34cm,
    row sep=0.22cm
]
\lstick{$\ket{0}_{q_1}$}
    & \gate[4]{U_{\Psi}^{\phantom{\dagger}}}
      \gategroup[
        wires=5,
        steps=4,
        style={
            draw,
            dotted,
            thick,
            rounded corners=1pt,
            inner xsep=4pt,
            inner ysep=3pt
        },
        background,
        label style={
            label position=below,
            anchor=north,
            yshift=-2mm
        }
      ]{$\text{while } f=1$}
    & \gate[4]{U_{\Phi}^{\dagger}}    & \octrl{4}    & \qw    &[0.65cm]      \gate[4]{U_{\Phi}^{\phantom{\dagger}}}    & \gate[4]{U_{\Lambda}^{\dagger}}    & \meter{} \\
\lstick{$\ket{0}_{q_2}$}    & \qw    & \qw    & \octrl{3}    & \qw    & \qw    & \qw    & \meter{} \\
\lstick{$\vdots$}    & \qw    & \qw    & \octrl{2}    & \qw    & \qw    & \qw    & \meter{} \\
\lstick{$\ket{0}_{q_n}$}
    & \qw    & \qw    & \octrl{1}    & \qw    & \qw    & \qw    & \meter{} \\
\setwiretype{n}
    & \setwiretype{n}
    & \lstick{$\ket{0}_{f_\Phi}$} &
    \targ{}\setwiretype{q}
    & \meter[
        style={
            label={[]above:{$f$}},
            label={[xshift=3mm]right:{$f=0$}}
        }
      ]{}
\end{quantikz}
}

\caption{Flagged compute--uncompute circuit.
The target-state preparation is followed by a reference-state flag and uncomputation against an orthogonal residual state.
The flag event estimates $|s|^2$, while the disjoint all-zero unflagged event estimates $|h|^2$.}
\label{fig:four_qubit_fcu_circuit}
\end{figure}
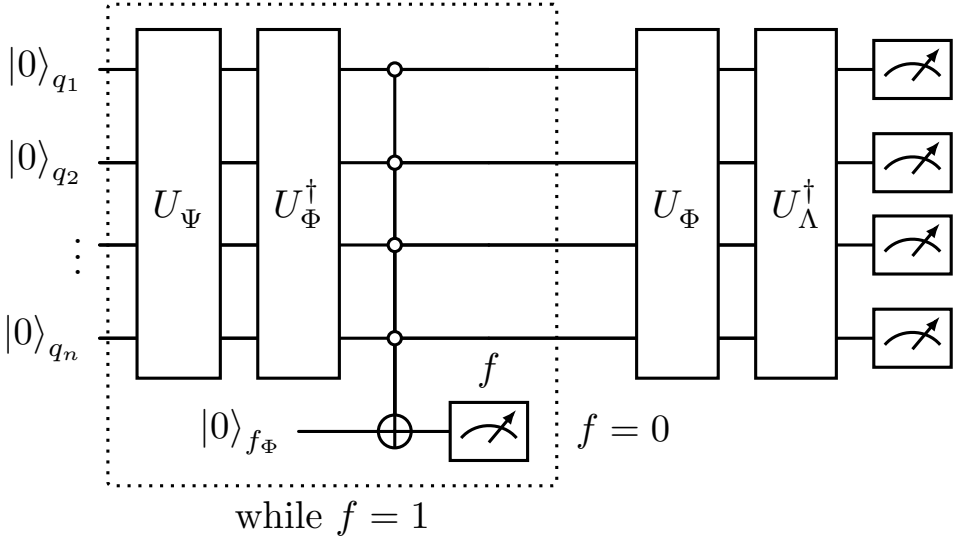

The same reference--residual orthogonality also permits a joint-circuit variant of the direct measurement.
In the flagged compute--uncompute (FCU) protocol introduced here, the circuit in Fig.~\ref{fig:four_qubit_fcu_circuit} first flags the component of $\ket{\Psi}$ along $\ket{\Phi}$ and then uncomputes the remaining branch against $\ket{\Lambda}$.
Each shot returns one of three mutually exclusive outcomes (reference, residual, or other) so a common circuit outcome estimates both $p_{\Phi}$ and $p_{\Lambda}$.
Accounting for finite covariance, FCU has no greater shot variance than optimally allocated separate CU measurements.
FCU requires one flag qubit, any work qubits needed for the zero-state test, and greater circuit depth.
The simple estimator also requires the prepared reference and residual states to remain orthogonal, details in the Supplementary Information.

We compare CU and FCU with signed- and squared-overlap alternatives.
The real Hadamard test (RHT) and real interferometric-swap test (RIST) directly estimate $\operatorname{Re}\braket{\Psi|\Omega}$, a complementary setting can recover the imaginary part when required.\cite{ekert2002direct,mitarai2019methodology}
The standard swap test (SST) and destructive swap test (DST) instead estimate $|\braket{\Psi|\Omega}|^2$ with two system registers.
SST uses an ancilla and a controlled-SWAP network, whereas DST replaces these operations with pairwise Bell-basis measurements.\cite{buhrman2001fingerprinting,ekert2002direct,garciaescartin2013swap,cincio2018overlap}

Classical shadows (CS) provide reusable amplitude information rather than a dedicated overlap measurement.
For sparse $\ket{\Phi}=\sum_xa_x\ket{x}$ and $\ket{\Lambda}=\sum_xb_x\ket{x}$, estimates of the real target amplitudes give $s=\sum_xa_x\braket{x|\Psi}$ and $h=\sum_xb_x\braket{x|\Psi}$.
We prepare $\ket{\tau}=(\ket{0}^{\otimes n}+\ket{\Psi})/\sqrt{2}$, apply a sampled global Clifford, and measure in the computational basis.\cite{Huang2020,lenihan2025excitationamplitudesamplinglow}
The same shadow record can be contracted with many references or observables.
Low-depth Clifford ensembles can reduce the circuit overhead, although the covariance used here assumes global Clifford measurements.\cite{low_depth_clifford}

As a direct-energy baseline, Pauli averaging (PA) writes the mapped Hamiltonian as
\begin{equation}
H=\sum_{\ell}\alpha_{\ell}P_{\ell}=\sum_gH_g,
\label{eq:qubit_mapped_pauli_hamiltonian}
\end{equation}
and measures compatible terms in a common basis.\cite{mcclean2014locality,peruzzo2014variational,verteletskyi20}
The benchmark uses the Jordan--Wigner mapping, a deterministic greedy qubit-wise-commuting grouping, and optimized allocation across the groups.
Although $\ket{\Psi}$ is an eigenstate of $H$, it is generally not an eigenstate of each independently measured $H_g$, so the grouped measurements retain finite variance.\cite{gonthier2022measurements}

\begin{figure}
\centering
\includegraphics[width=0.9\linewidth]{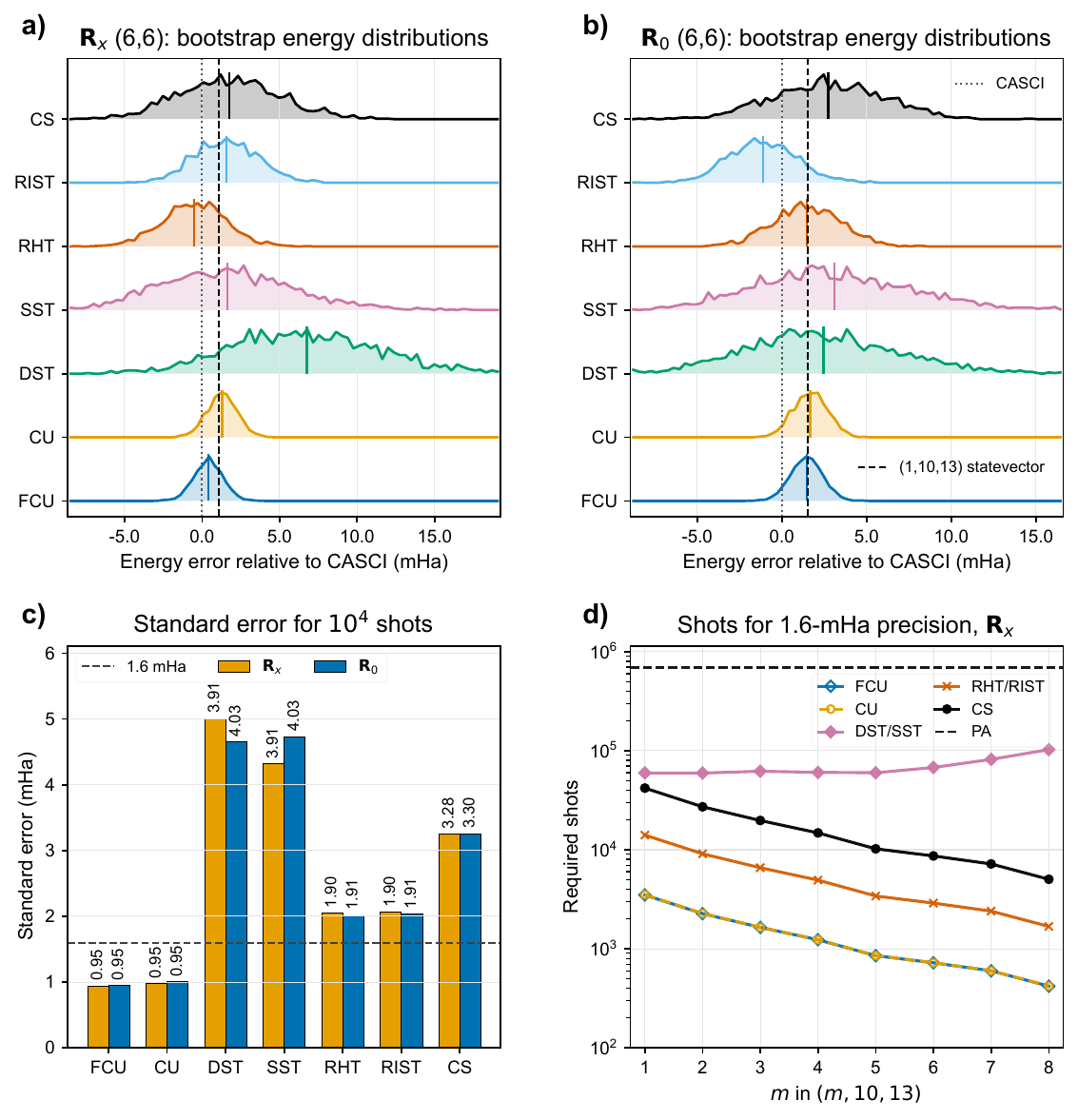}
\caption{
Noiseless ME simulator error analysis for the two (6,6) AS Hamiltonians corresponding to the $\mathbf{R}_x$ and $\mathbf{R}_0$ geometries, using $(m,k,l)=(1,10,13)$.
(a,b) Bootstrap energy distributions for the $\mathbf{R}_x$ and $\mathbf{R}_0$ Hamiltonians, shown as errors relative to $E_{\mathrm{CASCI}}$.
The dotted vertical line marks CASCI, and the dashed vertical line marks the exact $(1,10,13)$ statevector ME value.
(c) Bootstrap standard errors for a $10\,000$-shot effective experiment, using $8000$ residual-state and $2000$ reference-state shots for the separate-circuit protocols.
Numbers above the bars give the analytical standard errors with optimal allocation of the same total shot count.
(d) Idealized shot count required to reach $1.6$ mHa precision for $\mathbf{R}_x$ as a function of $m$ in $(m,10,13)$.
The dashed horizontal line shows the Pauli-averaging reference shot count.}
\label{fig:result_protocol_energy_histograms}
\end{figure}

For real nonzero $h$ and $s$, exact orthogonality gives $h^2+s^2\leq1$.
In the high-quality-reference regime $s\to1$ and $\abs{h}\ll\abs{s}$, the leading large-shot requirements for a target standard error $\epsilon$ are
\begin{equation}
N_{\rm FCU}\le N_{\rm CU}
\approx\frac{1}{4}\frac{r_{\Phi}^2}{\epsilon^2},
\end{equation}
\begin{equation}
N_{\rm RHT}=N_{\rm RIST}
\approx\frac{r_{\Phi}^2}{\epsilon^2},
\end{equation}
\begin{equation}
N_{\rm CS}
\approx 3\frac{r_{\Phi}^2}{\epsilon^2},
\end{equation}
and
\begin{equation}
N_{\rm DST}=N_{\rm SST}
\approx\frac{1}{4}\frac{r_{\Phi}^2}{\epsilon^2} \frac{1}{h^2}.
\end{equation}
Thus RHT and RIST require approximately four times as many shots as CU/FCU, while CS requires approximately 12 times as many.
SST and DST carry an additional $1/h^2$ factor and therefore become statistically unfavourable as the residual overlap decreases.
Magnitude-only protocols require a known correction sign, and the CS expression assumes global Clifford measurements.
Full covariance expressions, optimal allocation rules, and the non-orthogonal FCU case are given in Methods and Supplementary Section~\ref{SI-sec:si_mixed_estimator_details}.
The bootstrap calculations in Fig.~\ref{fig:result_protocol_energy_histograms}a--c for a fixed total of $10\,000$ shots are in good agreement with the expressions.

We test these predictions at fixed state preparation.
Figure~\ref{fig:result_protocol_energy_histograms} uses $(m,k,l)=(1,10,13)$ and a total budget of $10\,000$ shots to estimate the energy of each of the two (6,6) AS Hamiltonians corresponding to the $\mathbf{R}_x$ and $\mathbf{R}_0$ geometries.
The distributions are obtained from the corresponding per-shot protocol statistics using the sampling and bootstrap procedure described in Methods.
At equal sample count, CU and FCU give the narrowest distributions, with standard deviations of $0.95$--$1.01$ mHa.
RHT and RIST give values close to 2 mHa, while SST and DST give approximately $4.2$--$4.4$ mHa.
The centers of the distributions should be compared with the statevector ME values, whereas their widths quantify the finite-shot uncertainty.
CU was therefore selected for the reference scans, AS calculations, emulator runs, and hardware measurements below because it combines low sampling variance with an $n$-qubit, ancilla-free circuit.

Figure~\ref{fig:result_protocol_energy_histograms}d shows that reference improvement and overlap-protocol efficiency compound.
Increasing the reference from one to eight determinants reduces the CS requirement from approximately $40\,000$ to $5000$ shots.
Using CU with the eight-determinant reference reduces the requirement further to approximately $400$ shots, about 100 times fewer than single-determinant CS and approximately $1700$ times fewer than the grouped PA baseline.

Shot efficiency alone does not determine runtime because the protocols have different state-preparation, control, and ancillary-qubit requirements.
We selected CU for the experimental benchmarks because the state-projection and flagging operations in FCU add circuit depth on current hardware.
FCU becomes attractive when this overhead is offset by mid-circuit early termination, particularly when residual-state preparation is substantially more expensive than the reference projection.
Combining optimized shot allocation with ideal circuit durations shows that FCU can outperform CU when the flag is measured mid-circuit and flagged shots terminate before reference restoration and residual-state uncomputation.
In the high-quality-reference limit, this early-termination crossover reduces to
\begin{equation}
\tau_{\Phi}+\tau_{\mathrm{flag}}<\tau_{\Lambda},
\end{equation}
where $\tau_{\Phi}$ and $\tau_{\Lambda}$ are the durations of the reference- and residual-state preparation blocks, with the same durations assumed for their inverses, and $\tau_{\mathrm{flag}}$ covers the all-zero flag and mid-circuit measurement.
This condition is plausible for a parallel flag construction, where $\tau_{\mathrm{flag}}\propto\log n$ at the cost of work qubits proportional to $n$, whereas applying a two-body electronic Hamiltonian to an $m$-determinant reference can give the untruncated residual a worst-case support of $\mathcal{O}(mn^4)$ determinants and thereby make residual-state preparation substantially more expensive.
The full runtime expressions and the resource scaling of the other protocols are given in Supplementary Section~\ref{SI-sec:runtime_comparison}.

The appropriate protocol also depends on the target and available hardware resources.
Excited states without a rigorously known sign require a signed protocol, while complex states require both overlap quadratures.
RHT and RIST provide this information at greater control overhead.
SST, DST, and RIST use a second state register, allowing parallel state preparation and additional symmetry checks, fan-out can exchange ancillary qubits for reduced controlled-SWAP depth.
CS is less efficient for a single energy in the present regime.

Other overlap methods include direct fidelity estimation with importance-sampled Pauli measurements, randomized-measurement correlations, collective measurements over several copies, and amplitude-estimation refinements of overlap-probability circuits.\cite{flammia2011direct,elben2020crossplatform,fanizza2020beyond,brassard2002amplitude,grinko2021iterative}
Our numerical comparison is restricted to the protocols implemented here.

\subsection{Mixed estimator energy differences on emulator and hardware}

\begin{figure}
    \centering
    \includegraphics[width=0.95\linewidth]{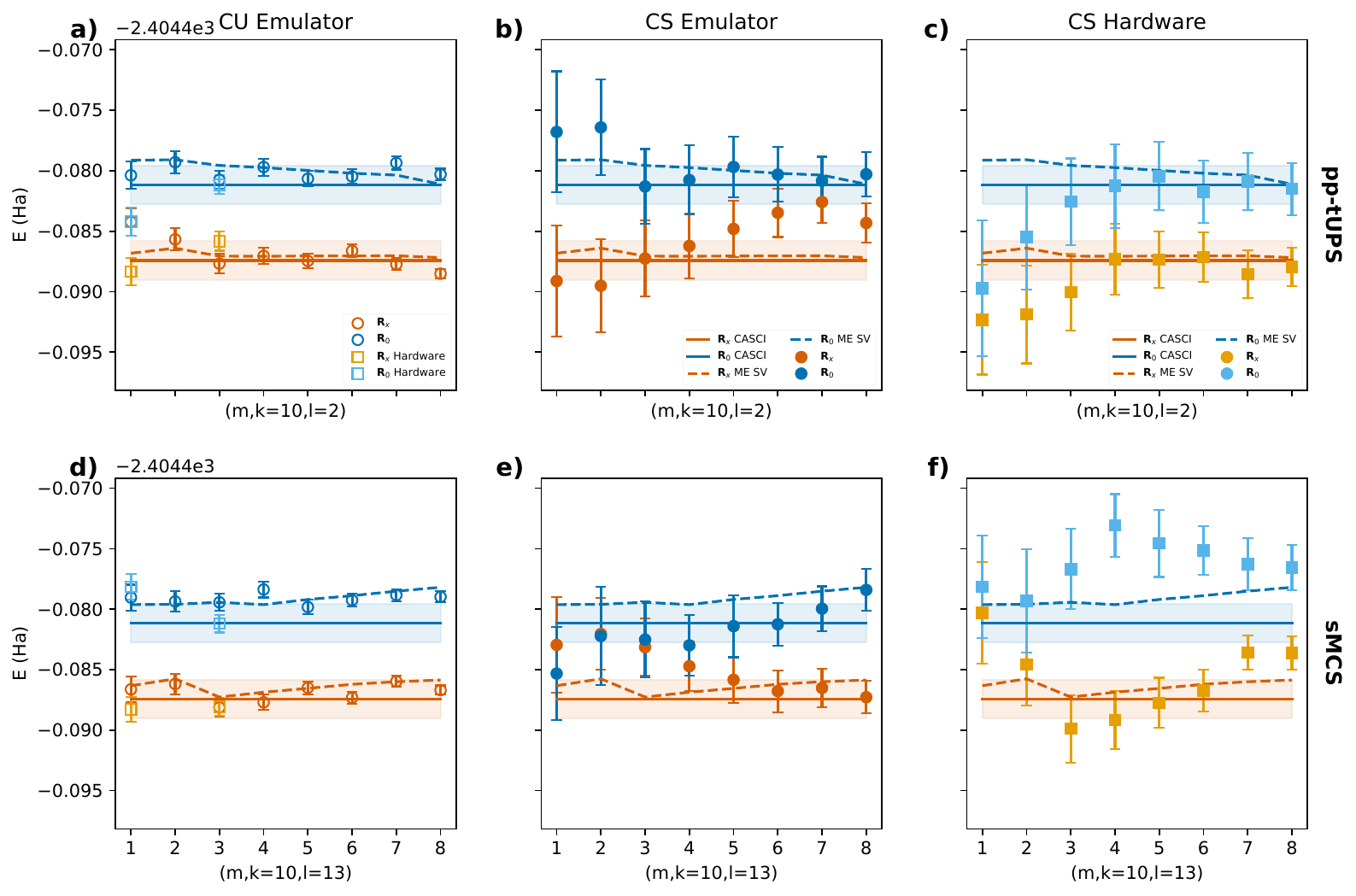}
    \caption{Reference size dependence of ME energies in the (6,6) AS for CS and CU on emulator and hardware.
    Orange and blue correspond to the $\mathbf{R}_x$ and $\mathbf{R}_0$ geometries; solid line showing the CASCI energy, with the shaded region indicating $E_{\text{CASCI}}\pm1.6$ mHa and dashed line indicating the SV value.
    Circles correspond to Helios-1E, squares to Helios-1.
    Empty and filled shapes are CU and CS respectively.
    \textbf{a)}--\textbf{c)} are pp-tUPS using $(m,10,2)$, and \textbf{d)}--\textbf{e)} sMCS using $(m,10,13)$.
    \textbf{a)}, and \textbf{d)} are CU on Helios-1E, with Helios-1 data shown for $m\in\{1,3\}$.
    \textbf{b)}, and \textbf{e)} are CS on Helios-1E with Helios-1 equivalent in \textbf{c)}, and \textbf{f)}.
    CU uses $10\,000$ shots, with CS using a single dataset of $10\,000$ shots reprocessed for each $m$.
    Error bars indicate one standard deviation.
    CS pp-tUPS runs were executed in between January and February 2026, CU in June--August 2026.
    CS sMCS emulator runs in June 2026, with hardware in July--August 2026.}
    \label{fig:tups_mc_cu_cs}
\end{figure}

Having established the sampling advantage of CU for targeted energy estimation, we next test it under realistic execution conditions on the Helios-1E emulator and Helios-1 hardware.
The (6,6) active space provides a controlled comparison of CU and CS across reference sizes and target-state constructions.
Figure~\ref{fig:tups_mc_cu_cs} varies the reference size $m$ at fixed $k=10$.
The pp-tUPS calculations use $(m,10,2)$ and the sMCS calculations use $(m,10,13)$.

CU gives the smallest uncertainties, typically at or below $1$~mHa, compared with approximately $1$--$5$~mHa for CS.
The Helios-1 results confirm this precision advantage for both target-state constructions.
Increasing the CU reference from $m=1$ to $m=3$ reduces the uncertainty in all four hardware energies and generally moves the central values closer to CASCI, most clearly for $\mathbf{R}_0$.
CS uncertainties likewise decrease with $m$ on Helios-1E and Helios-1, and the hardware results reproduce the emulator trend, although the central-value accuracy is not monotonic.

Each CU reference requires new reference- and residual-overlap circuits, so the $m=1$ and $m=3$ hardware points are independent measurements.
One set of classical shadows can instead be contracted with every classically specified reference and residual state, reconstructing the complete $m$ scan without further quantum measurements.
This provides greater flexibility, but CU gives substantially smaller uncertainty at the same submitted-shot count and, as shown above, requires fewer shots for a targeted energy.
CU is therefore preferable for a small number of precise energies, while CS is useful when many references must be explored from a common dataset.
We next consider the (10,10) AS, focusing on the sMCS state pair selected through the AI-assisted workflow, denoted AI-sMCS.
The AS comparison uses $(1,10,13)$ for $(6,6)$ and $(1,21,24)$ for $(10,10)$.
In AI-sMCS, $\mathbf{R}_x$ retains $(1,21,24)$, while for $\mathbf{R}_0$ the AI-assisted optimization further reduces the target support to $(1,21,22)$.

\FloatBarrier
\begin{table*}[!t]
    \caption{Compute--uncompute overlaps for the AI-selected (10,10) sMCS states.
    All Helios results use ternary leakage detection and postselection, with $N_s=2000$ and $N_h=8000$ submitted shots per geometry on hardware and twice these numbers on the emulator.
    Emulator uncertainties are normalized to the hardware allocation.\cite{ransford2026helios}
    CPC and MBPC denote additional coherent and measurement-based parity checks.
    The $p_{\Phi}$ and $p_{\Lambda}$ columns give postselected all-zero probabilities.
    The accepted-shot columns give the retained percentages for the $s$ and $h$ circuits.
    Uncertainties are one standard error, and the uncertainty in $|h/s|$ is propagated from the independent $s$ and $h$ measurements.
    Total qubits include data and ancillary qubits.}
    \label{tab:helios1_tups_mc_me_energies}
    \centering
    \footnotesize
    \setlength{\tabcolsep}{4pt}
    \resizebox{\textwidth}{!}{%
    \begin{tabular}{@{}llcccrrr@{}}
        \toprule
        Backend & Detection & Total qubits & Accepted $s$ (\%) & Accepted $h$ (\%) & $p_{\Phi}=|s|^2$ & $p_{\Lambda}=|h|^2$ & $|h/s|$\\
        \midrule
        \multicolumn{8}{@{}l}{\textit{Geometry} $\mathbf{R}_x$, (10,10) active space}\\
        \cmidrule{1-8}
        Statevector & N/A & 20 & -- & -- & $0.8974\phantom{{}\pm0.0000}$ & $0.0957\phantom{{}\pm0.0000}$ & $0.3266\phantom{{}\pm0.0000}$\\
        Helios-1E & Leakage & 20 & 90.3 & 82.6 & $0.8452\pm0.0085$ & $0.0793\pm0.0033$ & $0.3064\pm0.0066$\\
        Helios-1 & Leakage & 20 & 92.4 & 85.7 & $0.7829\pm0.0096$ & $0.0733\pm0.0031$ & $0.3059\pm0.0068$\\
        Helios-1E & Leakage + CPC & 22 & 86.3 & 78.4 & $0.8653\pm0.0082$ & $0.0857\pm0.0035$ & $0.3147\pm0.0067$\\
        Helios-1 & Leakage + CPC & 22 & 86.5 & 78.8 & $0.8242\pm0.0092$ & $0.0782\pm0.0034$ & $0.3080\pm0.0069$\\
        Helios-1 & Leakage + MBPC & 50 & 85.2 & 78.8 & $0.8256\pm0.0092$ & $0.0773\pm0.0034$ & $0.3059\pm0.0069$\\
        \midrule
        \multicolumn{8}{@{}l}{\textit{Geometry} $\mathbf{R}_0$, (10,10) active space}\\
        \cmidrule{1-8}
        Statevector & N/A & 20 & -- & -- & $0.8878\phantom{{}\pm0.0000}$ & $0.1073\phantom{{}\pm0.0000}$ & $0.3477\phantom{{}\pm0.0000}$\\
        Helios-1E & Leakage & 20 & 92.4 & 84.6 & $0.8398\pm0.0085$ & $0.0960\pm0.0036$ & $0.3382\pm0.0065$\\
        Helios-1 & Leakage & 20 & 94.1 & 87.5 & $0.8236\pm0.0088$ & $0.0808\pm0.0033$ & $0.3133\pm0.0065$\\
        Helios-1E & Leakage + CPC & 22 & 87.8 & 81.2 & $0.8510\pm0.0085$ & $0.0994\pm0.0037$ & $0.3418\pm0.0066$\\
        Helios-1 & Leakage + CPC & 22 & 88.0 & 82.7 & $0.8351\pm0.0088$ & $0.0762\pm0.0033$ & $0.3021\pm0.0067$\\
        Helios-1 & Leakage + MBPC & 50 & 87.0 & 81.3 & $0.8470\pm0.0086$ & $0.0821\pm0.0034$ & $0.3113\pm0.0066$\\
        \bottomrule
    \end{tabular}
    }
\end{table*}

Table~\ref{tab:helios1_tups_mc_me_energies} reports the two measured probabilities for the AI-selected (10,10) sMCS pair.
The energies are then obtained from Eq.~\eqref{eq:me_absolute_overlap} using $E_{\Phi}$ and $r_{\Phi}$ calculated classically.
The quantum correction depends on the ratio $|h/s|=\sqrt{p_{\Lambda}/p_{\Phi}}$.
If noise changes both probabilities by the same fraction, the ratio is unchanged.
Only unequal changes in $p_{\Phi}$ and $p_{\Lambda}$ bias the energy correction.
The effect of approximately depolarizing noise can therefore cancel partially in the ratio when the two circuits have similar effective error rates.\cite{urbanek2021mitigating}
The leakage-only hardware results show this cancellation.
At $\mathbf{R}_x$, $p_{\Phi}$ and $p_{\Lambda}$ are $12.8\%$ and $23.4\%$ below their statevector values, while $|h/s|$ is lower by only $6.3\%$.
At $\mathbf{R}_0$, the corresponding deviations are $7.2\%$, $24.7\%$, and $9.9\%$.
The ratio therefore reduces the overlap bias, but it does not cancel errors that affect the two probabilities differently.

To demonstrate how additional error detection changes the ME energies through the measured overlap probabilities, we applied a coherent parity check (CPC) and a measurement-based parity check (MBPC) on top of leakage postselection, as described in Methods.
Table~\ref{tab:helios1_tups_mc_me_energies} shows that these checks generally move the individual overlap probabilities closer to their statevector values, but the improvement is not uniform across both overlaps and geometries.
On Helios-1E, CPC brings $|h/s|$ closer to the statevector value at both geometries.
On Helios-1, however, CPC slightly improves the ratio at $\mathbf{R}_x$ but worsens it at $\mathbf{R}_0$, while MBPC moves the $\mathbf{R}_0$ ratio back toward its statevector value relative to CPC.
Error-detection performance should therefore be evaluated using the resulting ME energy, rather than either overlap alone.

\begin{figure}[!tbp]
    \centering
    \includegraphics[width=\textwidth]{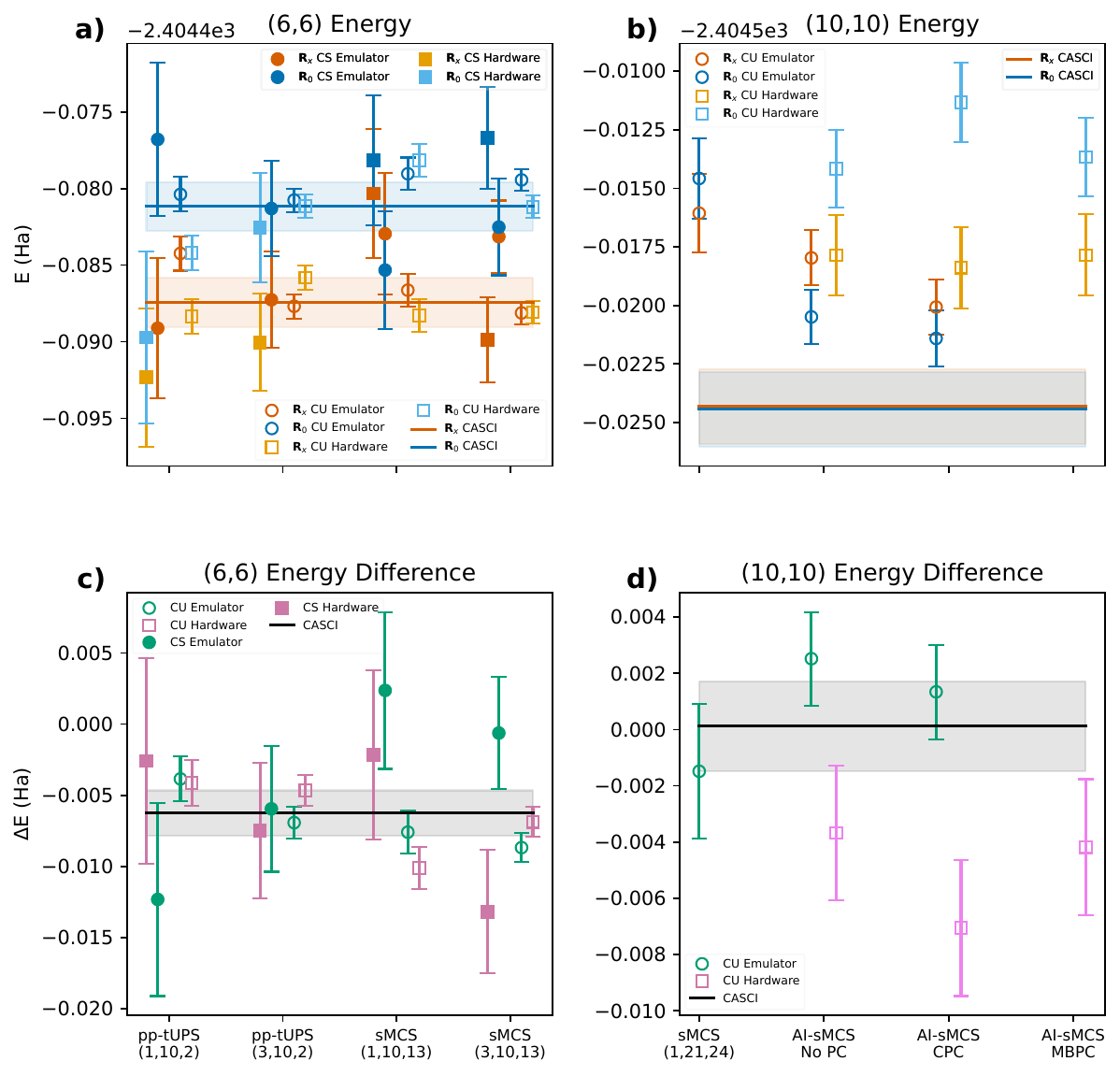}
    \caption{ME energies and energy differences ($\Delta E =E_{\mathbf{R}_x}-E_{\mathbf{R}_0}$) on Helios-1E and Helios-1 for (6,6) and (10,10) ASs.
    For \textbf{a)}, and \textbf{b)}, orange and blue correspond to the $\mathbf{R}_x$ and $\mathbf{R}_0$ geometries; solid line showing the CASCI energy, with the shaded region indicating $E_{\text{CASCI}}\pm1.6$ mHa.
    Light and dark colours correspond to emulator and hardware respectively.
    \textbf{c)}, and \textbf{d)} show the energy difference, with pink and green being Helios-1 and Helios-1E respectively; solid black line being the CASCI value with the shaded region indicating $E_{\text{CASCI}}\pm1.6$ mHa.%
    \textbf{a)} shows the (6,6) AS using pp-tUPS and sMCS for CS (filled) and CU (empty).
    \textbf{b)} shows the (10,10) AS, with a common (1,21,24) sMCS reference used as a starting point for AI state truncation.
    Different error detections on the AI-sMCS state are shown.
    \textbf{c)} and \textbf{d)} show the (6,6) and (10,10) AS energy differences respectively.
    Spread on the $x$-axis for \textbf{a)} and \textbf{c)} is purely for visual aid.
    $10\,000$ shots were used for each state preparation method, and error bars being one standard deviations.
    (6,6) CS pp-tUPS runs were executed in between January and February 2026, CU in June--August 2026.
    CS sMCS emulator runs in June 2026, with hardware in July--August 2026.
    (10,10) runs were executed in August 2026.}
    \label{fig:result_active_space}
\end{figure}

Figure~\ref{fig:result_active_space} converts the measured overlaps into individual energies and the energy difference $\Delta E_{\mathbf{R}_x-\mathbf{R}_0}$ for the (6,6) and (10,10) active spaces.
Where CU and CS are compared in the (6,6) results, CU retains the tighter finite-shot uncertainty.
On Helios-1, the (10,10) CU measurements reached individual-energy standard errors of $1.66$--$1.73$ mHa with $10\,000$ submitted shots per geometry, close to the $1.6$ mHa precision target.
By comparison, the ideal grouped PA baseline requires an estimated $2.5$--$3.8$ million shots per geometry to reach $1.6$ mHa, as detailed in Supplementary Table~\ref{SI-tab:pauli_averaging_shots}.
The quoted standard errors describe finite-shot precision, whereas agreement with CASCI depends on the central values.
Across the emulator and hardware results, the absolute deviations of the individual-energy estimates from CASCI span approximately $0.5$--$13$ mHa, while those of $\Delta E_{\mathbf{R}_x-\mathbf{R}_0}$ span approximately $0.2$--$9$ mHa.
In several matched geometry pairs, errors at the two geometries partially cancel, most clearly for the AI-selected (10,10) result, although this behaviour is not universal.

Together, Table~\ref{tab:helios1_tups_mc_me_energies} and Fig.~\ref{fig:result_active_space} show two conditional cancellations: similar relative errors in $p_{\Phi}$ and $p_{\Lambda}$ can cancel in their ratio, and similar errors in $E_{\mathbf{R}_x}$ and $E_{\mathbf{R}_0}$ can cancel in their difference.
Parity checks did not consistently improve the ME energy difference.
Even when a parity check improved an individual overlap measurement, unequal changes in $p_{\Phi}$ and $p_{\Lambda}$ could worsen the ratio or affect the two geometries differently.
The central error decreased on Helios-1E, but neither CPC nor MBPC improved the central $\Delta E_{\mathbf{R}_x-\mathbf{R}_0}$ on Helios-1 relative to leakage-only postselection.
Error-detection strategies optimized for the overlap ratio, rather than for either circuit separately, may therefore be more effective.
Ultimately, full error correction would provide a systematic route beyond this conditional cancellation.

\section{Discussion}

Combining sparse multiconfiguration references with direct overlap measurements substantially reduces the quantum resources required to estimate correlated energies from quantum-prepared states.
For H$_2$DBB--Ru, we apply this strategy to resolve $\Delta E_{\mathbf{R}_x-\mathbf{R}_0}$, an electronically challenging energy difference that strongly influences the predicted singlet-oxygen generation kinetics within the kinetic model employed in this work.
Better references reduce the required quantum correction, while direct overlap measurements lower its sampling cost.
The methodology we develop is applicable to any eigenstate, but for ground-state targets, the fixed correction sign makes magnitude-only CU sufficient, substantially reducing the sampling cost.
With optimal shot allocation, analytical variance estimates predict approximately 400 CU shots for a multideterminant reference to reach 1.6 mHa precision, about 100 times fewer than single-determinant CS and approximately $1700$ times fewer than the grouped PA baseline used here.
This extends the single-determinant, classical-shadow formulation of Lenihan et al.~\cite{lenihan2025excitationamplitudesamplinglow} to a regime in which the classical reference and the quantum readout can be optimized together.

Benchmarks with approximate pp-tUPS and sMCS targets show that this readout advantage is not restricted to exact eigenstates.
The mixed estimator reaches a given energy accuracy with simpler target-state preparation than direct expectation-value estimation, with the largest gain in the (10,10) active space.
Direct overlap measurements introduce additional preparation costs, but truncation and circuit optimization can reduce this overhead.
State preparation and energy readout should therefore be designed together.

Measuring the required overlaps involves trade-offs between shot count, circuit depth, qubit count, data reuse, and access to the overlap sign.
CU is efficient for a small number of targeted energies, whereas CS can evaluate many references from one dataset.
FCU combines both overlaps in a single circuit and can further reduce sampling and runtime through early termination.
The best choice therefore depends on whether the target is a ground or excited state, total runtime, available qubits, and how the results will be used, rather than on shot count alone.

The Helios-1E and Helios-1 experiments test the resulting readout under device noise.
Errors can partially cancel both in the overlap ratio and between the two geometry energies, making the energy difference more accurate than either absolute energy in some cases.
This cancellation is not guaranteed, and the parity checks tested here did not consistently improve the final energy difference.
Error detection should therefore be designed around the complete estimator, with full error correction providing the systematic long-term route.

The broader significance of mixed estimation is that it does not depend on a particular route to the target state.
An eigenstate or sufficiently accurate approximation produced by adiabatic preparation, quantum imaginary-time evolution, state filtering, other projector methods, or engineered dissipation~\cite{kremenetski2021adiabatic,motta20,lin2020groundstate,verstraete2009dissipation} can be connected to the same energy readout, provided that a good reference remains available classically.
An approximate target introduces systematic bias, so state preparation, reference construction, device noise, AS convergence, and chemical-model convergence remain separate requirements.
Remaining chemical-model uncertainties include finite active spaces and selected-CI convergence thresholds, missing solvation, the DFA-derived pathway and geometries, restricted sampling of molecular structure, and nuclear-dynamical effects.
The present benchmarks isolate energy readout and do not establish a converged reaction rate or quantum advantage.
By separating correlated-state preparation from energy readout, mixed estimation becomes a natural complement to future state-preparation methods and offers a general route for translating their advances into precisely resolved correlated energy differences relevant to challenging chemistries, without reinstating the full sampling burden of Pauli averaging.

\FloatBarrier

\section{Methods}
\subsection{Classical electronic structure calculations}
All DFT calculations were performed with the ADF 2025.1 package.\cite{baerends2025amsterdam}
All geometry optimizations were carried out with the PBE functional \cite{perdew1996generalized} and the TZP (for Ru) and DZP (for the other atoms) basis set with the D3 dispersion correction and Becke--Johnson damping factor.\cite{grimme2010consistent,grimme2011effect}
The scalar relativistic Zero Order Regular Approximation (ZORA) Hamiltonian has been employed \cite{van1996relativistic,van1999geometry}, together with Core Large and NumericalQuality Good.
The carbon atoms of the COOH groups were frozen during the optimizations.
For DFT single point calculations, we employed three functionals, namely $\omega$B97X, \cite{chai2008systematic} B3LYP, \cite{stephens1994ab} and PBE functionals. We used the same computational details as for the geometry optimizations, except for the use of Core None.
SHCISCF calculations were performed with the Dice implementation \cite{smith2017cheap} in the PySCF package \cite{sun2018pyscf,sun2020recent} using the crenbl basis set and the crenbl electron core potential (ECP) for Ru \cite{fernandez1985ab,ermler1991ab} and the 6-31G** basis set for the other atoms.
Three sweep iterations with $\epsilon$=[0.01,0.001,0.0001] were used.
CASCI and NEVPT2 calculations were carried out with PySCF and same basis set as SHCISCF.
All Hamiltonians used for subsequent quantum computing calculations have been generated with the InQuanto software platform.~\cite{inquanto}

\subsection{State preparation with pp-tUPS}
\label{sec:tups}

The mixed estimator requires target-state preparation, for which we used pp-tUPS in the (6,6) active space and sMCS in the larger active spaces.

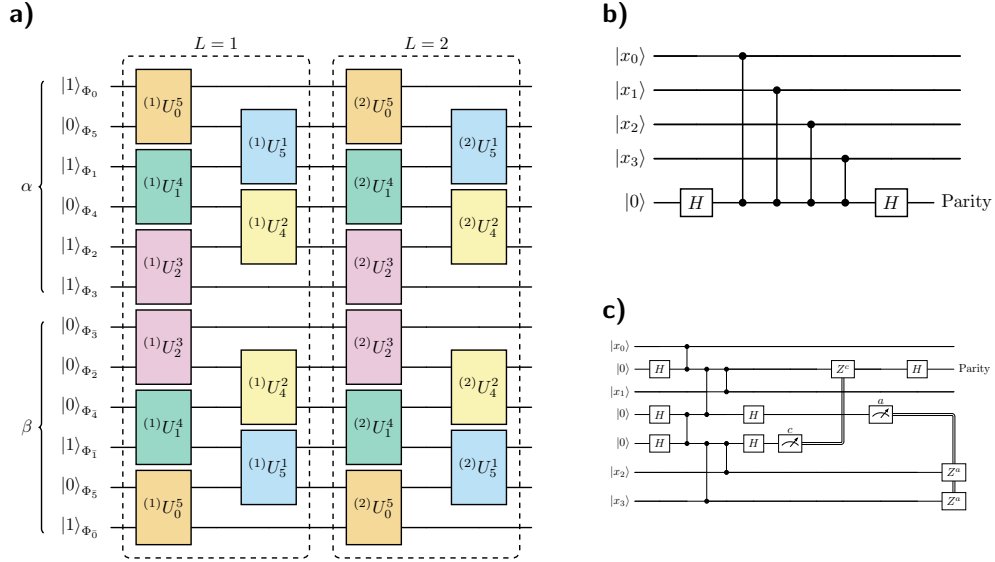
\begin{figure*}[!htbp]
    \centering
    \begin{minipage}[t]{0.54\textwidth}
    \vspace{0pt}
    \noindent\makebox[\linewidth][l]{\textsf{\textbf{a)}}}\par
    \resizebox{\columnwidth}{!}{%
        \begin{quantikz}[row sep={0.8cm,between origins}]
            \lstick[6]{$\alpha$}\midstick{$\ket{1}_{\Phi_0}$}&\gate[2,style={fill=natpumpkin!40}]{^{(1)}U_{0}^{5}}\gategroup[12, steps=3, style={dashed, rounded corners}]{$L=1$}&&&&\gate[2,style={fill=natpumpkin!40}]{^{(2)}U_{0}^{5}}\gategroup[12, steps=3, style={dashed, rounded corners}]{$L=2$}&&&\\
            \midstick{$\ket{0}_{\Phi_5}$}&&&\gate[2,style={fill=nataqua!40}]{^{(1)}U_{5}^{1}}&&&&\gate[2,style={fill=nataqua!40}]{^{(2)}U_{5}^{1}}&\\
            \midstick{$\ket{1}_{\Phi_1}$}&\gate[2,style={fill=natgreen!40}]{^{(1)}U_{1}^{4}}&&&&\gate[2,style={fill=natgreen!40}]{^{(2)}U_{1}^{4}}&&&\\
            \midstick{$\ket{0}_{\Phi_4}$}&&&\gate[2,style={fill=natyellow!40}]{^{(1)}U_{4}^{2}}&&&&\gate[2,style={fill=natyellow!40}]{^{(2)}U_{4}^{2}}&\\
            \midstick{$\ket{1}_{\Phi_2}$}&\gate[2,style={fill=natpink!40}]{^{(1)}U_{2}^{3}}&&&&\gate[2,style={fill=natpink!40}]{^{(2)}U_{2}^{3}}&&&\\
            \midstick{$\ket{1}_{\Phi_3}$}&&&&&&&&\\
            \lstick[6]{$\beta$}\midstick{$\ket{0}_{\Phi_{\bar{3}}}$}&\gate[2,style={fill=natpink!40}]{^{(1)}U_{2}^{3}}&&&&\gate[2,style={fill=natpink!40}]{^{(2)}U_{2}^{3}}&&&\\
            \midstick{$\ket{0}_{\Phi_{\bar{2}}}$}&&&\gate[2,style={fill=natyellow!40}]{^{(1)}U_{4}^{2}}&&&&\gate[2,style={fill=natyellow!40}]{^{(2)}U_{4}^{2}}&\\
            \midstick{$\ket{0}_{\Phi_{\bar{4}}}$}&\gate[2,style={fill=natgreen!40}]{^{(1)}U_{1}^{4}}&&&&\gate[2,style={fill=natgreen!40}]{^{(2)}U_{1}^{4}}&&&\\
            \midstick{$\ket{1}_{\Phi_{\bar{1}}}$}&&&\gate[2,style={fill=nataqua!40}]{^{(1)}U_{5}^{1}}&&&&\gate[2,style={fill=nataqua!40}]{^{(2)}U_{5}^{1}}&\\
            \midstick{$\ket{0}_{\Phi_{\bar{5}}}$}&\gate[2,style={fill=natpumpkin!40}]{^{(1)}U_{0}^{5}}&&&&\gate[2,style={fill=natpumpkin!40}]{^{(2)}U_{0}^{5}}&&&\\
            \midstick{$\ket{1}_{\Phi_{\bar{0}}}$}&&&&&&&&
        \end{quantikz}
    }
    \end{minipage}
    \hfill
    \begin{minipage}[t]{0.40\textwidth}
    \vspace{0pt}
    \noindent\makebox[\linewidth][l]{\textsf{\textbf{b)}}}\par
    \vspace{0.35em}
    \centering
    \resizebox{\linewidth}{!}{%
    \begin{quantikz}
       \lstick{$\ket{x_0}$}  & \qw & \ctrl{4} & \qw & \qw & \qw & \qw & \qw & \qw \\
       \lstick{$\ket{x_1}$}  & \qw & \qw & \ctrl{3} & \qw & \qw & \qw & \qw & \qw \\
       \lstick{$\ket{x_2}$}  & \qw & \qw & \qw & \ctrl{2} & \qw & \qw & \qw & \qw \\
       \lstick{$\ket{x_3}$}  & \qw & \qw & \qw & \qw & \ctrl{1} & \qw & \qw & \qw \\
        \lstick{$\ket{0}$} & \gate{H} & \control{} & \control{} &\control{} & \control{} & \gate{H} & \rstick{\text{Parity}} \\
    \end{quantikz}%
    }

    \vspace{1.75em}

    \noindent\makebox[\linewidth][l]{\textsf{\textbf{c)}}}\par
    \vspace{0.35em}
    \centering
    \resizebox{\linewidth}{!}{%
\begin{quantikz}[
row sep=0.35cm,
column sep=0.45cm
]
\lstick{$\ket{x_0}$}
& \qw
& \ctrl{1}
& \qw
& \qw
& \qw
& \qw
& \qw
& \qw
& \qw
& \qw
& \qw
\\
\lstick{$\ket{0}$}
& \gate{H}
& \control{}
& \ctrl{2}
& \ctrl{1}
& \qw
& \qw
& \qw
& \gate{Z^{c}}\wire[d][3]{c}
& \qw
& \gate{H}
& \rstick{\text{Parity}}
\\
\lstick{$\ket{x_1}$}
& \qw
& \qw
& \qw
& \control{}
& \qw
& \qw
& \qw
& \qw
& \qw
& \qw
& \qw
\\
\lstick{$\ket{0}$}
& \gate{H}
& \ctrl{1}
& \control{}
& \qw
& \gate{H}
& \qw
& \qw
& \qw
& \meter[
    label style={
        yshift=-0.08cm
    }
  ]{a}
& \setwiretype{c}
& \cbend{180}
& \setwiretype{n}
\\
\lstick{$\ket{0}$}
& \gate{H}
& \control{}
& \ctrl{2}
& \ctrl{1}
& \gate{H}
& \meter[
    label style={
        yshift=-0.08cm
    }
  ]
  {c}& \setwiretype{c}
  & \cbend{90}
\\
\lstick{$\ket{x_2}$}
& \qw
& \qw
& \qw
& \control{}
& \qw
& \qw
& \qw
& \qw
& \qw
& \qw
& \gate{Z^{a}}
\\
\lstick{$\ket{x_3}$}
& \qw
& \qw
& \control{}
& \qw
& \qw
& \qw
& \qw
& \qw
& \qw
& \qw
& \gate{Z^{a}}\wire[u][3]{c}
\end{quantikz}
    }
    \end{minipage}
    \caption{Quantum circuits used for state preparation and parity verification.
    \textbf{a)}, Two-layer pp-tUPS circuit, with gate excitations defined in Eq.~\eqref{eq:tups_tile}.
    The gate colours identify the four spin orbitals involved in each spin-blocked excitation.
    Qubits are indexed 0--11 from top to bottom, with the corresponding spin orbitals given as subscripts of the occupation kets.
    Our definition of pp-tUPS does not include orbital optimisation as part of the state preparation.
    Four-data-qubit examples of the parity-check circuits are also shown.
    \textbf{b)}, CPC circuit using 1 ancilla qubit.
    The more general $A>1$ parallelized construction is not shown.
    \textbf{c)}, Constant-depth MBPC circuit.
    The measurement outcomes $a$ and $c$ determine the Z-Pauli corrections applied to the data qubits.}
    \label{fig:tups_parity_circuits}
\end{figure*}

The tiled unitary product state (tUPS) is a chemically motivated variational ansatz defined through repeated applications of a ``tile''\cite{burton2024tups}
\begin{equation}
    {}^{(l)}U_{P}^{Q}(\theta_{PQ,0}^{(l)},\theta_{PQ,1}^{(l)},\theta_{PQ,2}^{(l)})=e^{\theta_{PQ,0}^{(l)}(\kappa_p^q+\kappa_{\bar{p}}^{\bar{q}})}e^{\theta_{PQ,1}^{(l)}\kappa_{p\bar{p}}^{q\bar{q}}}e^{\theta_{PQ,2}^{(l)}(\kappa_p^q+\kappa_{\bar{p}}^{\bar{q}})}.\label{eq:tups_tile}
\end{equation}
$P,Q$ denote spatial orbitals and the anti-Hermitian fermionic excitation generators are\cite{burtonExactElectronicStates2023}
\begin{align}
    \kappa_p^q+\kappa_{\bar{p}}^{\bar{q}} &= (a^{\dag}_qa_p^{\phantom{\dag}}-a^{\dag}_pa_q^{\phantom{\dag}}) + (a^{\dag}_{\bar{q}}a_{\bar{p}}^{\phantom{\dag}}-a^{\dag}_{\bar{p}}a_{\bar{q}}^{\phantom{\dag}}),\label{eq:spin_adapted_single}\\
    \kappa_{p\bar{p}}^{q\bar{q}}&=a^{\dag}_qa^{\dag}_{\bar{q}}a_{\bar{p}}^{\phantom{\dag}}a_p^{\phantom{\dag}} - a^{\dag}_pa^{\dag}_{\bar{p}}a_{\bar{q}}^{\phantom{\dag}}a_q^{\phantom{\dag}},\label{eq:paired_double}
\end{align}
where, $p$ and $\bar{p}$ denote the $\alpha$ and $\beta$ spin orbitals associated with spatial orbital $P$, respectively.
The tiles act on nearest-neighbour spatial orbitals and are arranged in repeated layers containing two offset tilings.\cite{burton2024tups}
The spin-adapted generators in Eqs.~\eqref{eq:spin_adapted_single} and \eqref{eq:paired_double} commute with $\hat{N}$, $\hat{S}_z$, and $\hat{S}^2$.
The ansatz therefore preserves the particle number and spin quantum numbers of the chosen reference.\cite{burton2024tups}
The target at each geometry was the lowest-energy state in the triplet sector, so pp-tUPS was initialized from a spin-pure $M_S=1$ reference.

For a Hartree--Fock reference with energy-ordered orbitals, the first tUPS layer contains redundant occupied--occupied and virtual--virtual excitations.
We therefore used a perfect-paired ordering, successively pairing the lowest- and highest-energy SHCISCF orbitals at each geometry.
We also spin blocked the ansatz to reduce the two-qubit gate count, giving the pp-tUPS circuit shown in Fig.~\ref{fig:tups_parity_circuits}a.
The geometry-specific SHCISCF orbitals were held fixed throughout the optimisation, so our implementation does not include the orbital optimisation used in the original pp-tUPS construction.\cite{burton2024tups}

For the (6,6) active space, we used statevector variational quantum eigensolver (VQE) calculations to optimise the excitation parameters independently at $\mathbf{R}_x$ and $\mathbf{R}_0$.
The AS Hamiltonians and statevector expectation-value objectives were constructed using InQuanto.\cite{inquanto}
Basin-hopping with 100 iterations and a temperature of 0.001 combined local VQE optimization using the the Limited-memory Broyden--Fletcher--Goldfarb--Shanno minimizer with random parameter perturbations accepted or rejected according to a Metropolis criterion.\cite{wales1997basinhopping}
For each layer count, we retained the lowest-energy converged parameter set.
The two-layer solutions were used as the pp-tUPS targets in the finite-shot simulator, Helios-1E emulator, and Helios-1 hardware calculations.
The optimisation used exact statevector energies, while the reported ME energies were reconstructed separately from the specified overlap-measurement outcomes.

The optimised pp-tUPS preparation circuits were compiled with pytket to the Helios native gate set of single-qubit rotations and ZZPhase gates.
The compilation included circuit simplification and rebasing before execution, using InQuanto 5.1.0, pytket 2.8.0 with optimization level 3.
Gate counts and two-qubit depths were evaluated after compilation, with the reported two-qubit resources counting ZZPhase gates and parallel ZZPhase layers, respectively.
The optimised pytket circuits were loaded into guppy~\cite{Koch2024} and the relevant algorithms, that consequently converted to programs for emulator and hardware execution.

\subsection{State preparation with sMCS}
\label{sec:multiconfig_stateprep}

For the larger active spaces, (8,8) and (10,10), statevector VQE calculations are expensive, while the target state $\ket{\Psi}$ is available from classical CASCI calculations.
CI states can be represented as linear combinations of computational-basis states, which correspond to Slater determinants under the Jordan--Wigner mapping.
Such states can be prepared efficiently when the linear combinations are sparse.\cite{MultiConfigSparse}
Here, ``sparse'' corresponds to $|S|\in o(2^n/n)$ for a support $S$ of $n$-qubit basis states.
The sparse synthesis uses $O(|S|n)$ CNOT gates and $O(|S|\log|S|+n)$ single-qubit rotations.\cite{MultiConfigSparse}
This technique has previously been applied to quantum-chemistry use cases\cite{MultiConfigSparse_ggd} and is used here to prepare systematic approximations to $\ket{\Psi}$.

The accuracy of the target-state approximation is controlled by truncating determinants relative to the CASCI state,
\begin{equation}
\ket{\Psi_l}
=
\sum_{x\in S_l}c_x^{(l)}\ket{x},
\label{eqn:psi_l}
\end{equation}
where $S_l\subset S_\Psi$ contains the $l=|S_l|$ largest-amplitude determinants of $\ket{\Psi}$ and the corresponding coefficients $c_x^{(l)}$ are normalised to unity.
We used sMCS preparation\cite{MultiConfigSparse} for several values of $l$ in the (6,6), (8,8), and (10,10) active spaces.
This permits systematic comparison with pp-tUPS approximations to $\ket{\Psi}$ and tests the effect of target-state truncation on the mixed estimator relative to direct expectation-value estimation.

The overlap protocols also require preparation circuits for the sparse reference $\ket{\Phi_m}$ and residual $\ket{\Lambda_{m,k}}$ states.
The reference support $S_m$ contains the $m$ largest-amplitude determinants of the CASCI state, while its coefficients are obtained by diagonalising $H$ within the selected subspace,
\begin{equation}
\ket{\Phi_m}
=
\sum_{x\in S_m}a_x^{(m)}\ket{x}.
\label{eqn:phi_m}
\end{equation}
The corresponding normalised residual is $\ket{\Lambda_m}=Q_{\Phi_m}H\ket{\Phi_m}/r_{\Phi_m}$.
It is subsequently truncated to its $k$ largest-amplitude determinants and renormalised,
\begin{equation}
\ket{\Lambda_{m,k}}
=
\sum_{x\in S_{m,k}}b_x^{(m,k)}\ket{x}.
\label{eqn:chi_k}
\end{equation}
Thus, $(m,k,l)$ gives the numbers of determinants in the reference, residual, and sMCS target states, respectively.
The determinants within the state were ordered by absolute value of the coefficient, truncated by number of determinants as shown in Fig.~\ref{fig:tups_mc_mixed_estimator_convergence}.

All supports were constructed at the required electron number and $M_S=1$.
Determinant truncation therefore preserves particle number and $M_S$, but it need not preserve an exact eigenvalue of $\hat{S}^2$.
For the AI-assisted (10,10) selection, we evaluated $\langle\hat{S}^2\rangle$ and the spin residual for the target and residual states using the criteria described below.

The sparse states were synthesised using InQuanto's \texttt{MultiConfigurationStateSparse} implementation and compiled with pytket to the Helios native gate set.
The same circuit generation, compilation, and optimisation techniques as used for the pp-tUPS were applied to the sMCS states.
The compiled circuits were checked against the intended sparse states by statevector comparison.
ZZPhase-gate counts and two-qubit depths were evaluated after compilation using the same definitions as for pp-tUPS, and similarly, the optimised pytket circuits were loaded into guppy for emulator and hardware execution.

\subsection{AI-assisted sparse-state and circuit optimization}
\label{sec:ai_sparse_optimization}
We used large-language-model agents within a human-supervised iterative search to explore target- and residual-state truncations together with their compiled circuit implementations.
The agents did not minimize a formal numerical loss function.
At each iteration, we supplied the sparse determinant catalog, deterministic ME results, compiled circuit resources, available Helios-1E measurements, and constraints on particle number, spin projection, and total spin.
The agents proposed candidate-construction rules, assisted with search and analysis code, and reviewed preliminary rankings; researchers decided which proposals to implement and advance.

The agent-assisted work used OpenAI Codex v0.139.0 with GPT-5.5 for candidate generation, analysis, and critical review, accessed in June 2026.
No numerical result was taken directly from the model output.
For each shortlisted candidate, deterministic code reconstructed the sparse states, recomputed the ME energies and symmetry diagnostics, synthesized the state-preparation circuits with InQuanto, and compiled them using pytket.
Only candidates whose calculated energies, symmetry diagnostics, and compiled circuit resources had been verified were advanced to emulator evaluation.
For those candidates, the emulator statistics were calculated directly from the Helios-1E measurements.

The reference size was fixed at $m=1$.
For each geometry, $\mathbf{R}_x$ and $\mathbf{R}_0$, the search independently varied the residual-state size $k$, the target-state size $l$, and the determinant supports used to construct $\ket{\Lambda_{1,k}}$ and $\ket{\Psi_l}$.
We considered three target-support constructions: largest-amplitude truncation, spin-connected completion, and residual-support guarding.
Spin-connected completion added determinants related by the relevant spin-exchange operations; these candidates were evaluated but not selected.
In the residual-support-guarded construction, the target state was restricted to determinants appearing in either the reference or the truncated residual state:
\begin{equation}
S_{1,k}^{\mathrm{guard}}
=
S_{\Psi}^{\mathrm{cat}}
\cap
\left(
S_{\Phi_1}\cup S_{\Lambda_{1,k}}
\right),
\label{eq:ai_residual_support_guard}
\end{equation}
where $S_{\Psi}^{\mathrm{cat}}$ is the available CASCI-derived determinant catalog, $S_{\Phi_1}$ is the single-determinant reference support, and $S_{\Lambda_{1,k}}$ is the support of the truncated residual state.
The resulting target-state size is $l=|S_{1,k}^{\mathrm{guard}}|$.
We also tested addback variants in which selected target determinants removed by this restriction were restored.
Each candidate state was renormalized before its overlaps and ME energy were recomputed.
The selected $\mathbf{R}_0$ state used $k=21$ and $l=22$, with no determinant addback.

Every candidate was constructed in the ten-electron, $M_S=1$ sector.
We verified that the total numerical weight outside this sector was below $10^{-8}$.
The soft and hard total-spin thresholds were applied to the target state $\ket{\Psi}$.
For the triplet target state, we preferred candidates satisfying
\begin{equation}
\left|\langle\hat{S}^2\rangle-2\right|
\leq 6\times10^{-3}
\end{equation}
and rejected candidates for which this deviation exceeded $2\times10^{-2}$.
The same expectation-value deviation was calculated for the residual state $\ket{\Lambda}$, but was recorded only as a diagnostic and was not used to exclude candidates. We additionally calculated the spin residuals of the target and residual states,
\begin{equation}
\norm{\left(\hat{S}^2-2\right)\ket{\Psi}}^2,
\qquad
\norm{\left(\hat{S}^2-2\right)\ket{\Lambda}}^2,
\end{equation}
as shown in Fig.~\ref{fig:helios1e_1010_stateprep}e,f.

The initial Pareto search retained candidates with a maximum deterministic absolute energy error of $1.6$~mHa across the two geometries and an absolute error in $\Delta E_{\mathbf{R}_x-\mathbf{R}_0}$ below $0.5$~mHa.
A subsequent circuit-focused search allowed an individual energy error of up to $2.0$~mHa while tightening the energy-difference error to $0.35$~mHa.
These thresholds were used to identify candidates for compilation and emulator evaluation.
Emulator performance was assessed separately from the deterministic criteria.

Shortlisted states were synthesized using
\texttt{MultiConfigurationStateSparse} and compiled to the Helios
native gate set.
We repeatedly applied the \texttt{pytket} optimization pass
\texttt{GreedyPauliSimp}\cite{pytket_greedypaulisimp} until the
two-qubit gate count no longer decreased.
Within the AI-assisted workflow, the agents helped formulate and implement the search over determinant-term ordering and spin-orbital mapping.
We evaluated four determinant-term orderings: decreasing and increasing coefficient magnitude, and ascending and descending determinant bitstring.
The spin-orbital search combined identity, reversal-based, and parity-grouped mappings with previously identified mappings and reproducibly seeded random permutations, yielding menus of up to 200 mappings.
The selected circuits used the identity mapping at $\mathbf{R}_x$ and a non-identity mapping at $\mathbf{R}_0$; the exact $\mathbf{R}_0$ permutation is reported in Supplementary Section~S2.4.
The resulting circuits were checked against their intended sparse states by statevector comparison.
For the reported $(10,10)$ sMCS campaign, circuit compilation and optimization used \texttt{pytket} v2.18.0.
The optimized circuits were imported using \texttt{guppylang} v0.21.16~\cite{Koch2024}, compiled to HUGR v0.16.0, and submitted to the Helios-1E emulator and Helios-1 hardware through QNexus v0.43.2.

For each geometry $g\in\{\mathbf{R}_x,\mathbf{R}_0\}$, we summarised the compiled resources using the aggregate ZZPhase-gate metric
\begin{equation}
G_g
=
2G(\Psi_l)+G(\Lambda_{1,k}),
\label{eq:ai_gate_score}
\end{equation}
where $G(\alpha)$ is the number of ZZPhase gates in the compiled preparation circuit for $\ket{\alpha}$.
The corresponding aggregate two-qubit-depth metric was
\begin{equation}
D_g
=
2D(\Psi_l)+D(\Lambda_{1,k}).
\label{eq:ai_depth_score}
\end{equation}
where $D(\alpha)$ is the compiled two-qubit depth of the preparation circuit for $\ket{\alpha}$.
The target-state preparation contributes twice because it appears in both overlap circuits, while the single-determinant reference requires no ZZPhase gates.
The aggregate depth metric combines the depths of two separate circuits and is not the execution depth of either circuit.

Noiseless state metrics and compiled circuit resources were used to prune the candidate set, but they did not determine the final selection because gate count and depth alone did not reliably predict the noise-induced energy bias.
The remaining candidates were evaluated on Helios-1E using $20\,000$ shots for the reference-overlap circuit and $80\,000$ shots for the residual-overlap circuit, together with leakage-aware postprocessing.
We compared candidate pairs using the absolute error for each geometry, the larger of these two errors, and the signed error in $\Delta E_{\mathbf{R}_x-\mathbf{R}_0}$.
The magnitude of the latter was used when assessing the energy-difference accuracy.

The selected pair retained the existing top-weight $(m,k,l)=(1,21,24)$ support and identity spin-orbital ordering for $\mathbf{R}_x$.
For $\mathbf{R}_0$, it used the guarded $(1,21,22)$ construction without determinant addback and the selected spin-orbital mapping described in Supplementary Section~S2.4.
The corresponding Helios-1E results are reported in Fig.~\ref{fig:helios1e_1010_stateprep}.

\subsection{Parity and leakage error detection}
\label{sec:symmetry_error_detection}
The prepared chemical states have inherent symmetries \cite{bonetmonroig2018lowcost} that can be exploited for quantum error detection (QED).
In this work, we use the fixed numbers of spin-up and spin-down electrons, which give each spin register a known fermion-number parity.
For the register $\sigma\in\{\alpha,\beta\}$, the corresponding parity operator is
\begin{equation}
P_{\sigma}
=
\prod_{j\in R_{\sigma}}Z_j,
\label{eq:spin_register_parity}
\end{equation}
where $R_{\sigma}$ contains the qubits representing spin $\sigma$. $Z_j$ acts with a Pauli-Z on the $j$-th qubit and with the identity on the remaining qubits.
An ideal state with $N_{\sigma}$ electrons is an eigenstate of $P_{\sigma}$ with eigenvalue $(-1)^{N_{\sigma}}$.
We measured $P_{\alpha}$ and $P_{\beta}$ independently after state preparation and terminated runs with the wrong parity.
For the (10,10) calculations, $N_{\alpha}=6$ and $N_{\beta}=4$, so both expected parities were even.
The check detects errors that change either register parity, including an odd number of Pauli $X$ or $Y$ errors within that register.
It does not detect Pauli $Z$ errors, parity-preserving errors, or an even number of bit flips.

All Helios configurations additionally used native ternary measurement, which distinguishes computational outcomes from population leaked outside the qubit subspace.\cite{ransford2026helios}
Shots containing a leakage outcome were discarded before the overlap probabilities were evaluated.
Leakage postselection is separate from the coherent parity check (CPC) and measurement-based parity check (MBPC) described below.
A shot was retained only when it passed every enabled check.

\subsubsection{Coherent parity check (CPC)}
\label{sec:gate_model_parity_check}
The error-detection gadget to measure the parity of $D$ can be constructed following the proposal by Moore.\cite{moore1999quantumcircuitsfanoutparity}
The construction may use $A$ ancilla qubits, where $1 \leq A \leq D$.
The first ancilla stores the parity outcome, while the remaining $A-1$ ancillas are used as temporary branches to parallelize the interactions with the data register.
These additional ancilla qubits are uncomputed to $\ket{0}$ before the parity measurement.
Increasing $A$ trades additional ancilla qubits and gates for lower depth.
The number of two-qubit gates is
\begin{equation}
N_{\mathrm{2q}}^{\mathrm{CPC}}
=
D+2(A-1),
\label{eq:cpc_two_qubit_count}
\end{equation}
using $D$ two-qubit gates for the data ancilla interaction and $2(A-1)$ for the fan-out fan-in construction.
With all-to-all connectivity and parallel fan-out and fan-in, the two-qubit depth is
\begin{equation}
d_{\mathrm{2q}}^{\mathrm{CPC}}
=
\left\lceil\frac{D}{A}\right\rceil
+
2\left\lceil\log_2 A\right\rceil.
\label{eq:cpc_two_qubit_depth}
\end{equation}

Measuring the additional ancillas verifies that they returned to $\ket{0}$ during uncomputation.
Figure~\ref{fig:tups_parity_circuits}b shows the four-data-qubit example with $A=1$.
In this limit, no fan-out or fan-in is required.
For the reported (10,10) CPC calculations, each $D=10$ spin register used one parity ancilla.
The two checks can be applied in parallel, giving 22 total qubits, 20 additional two-qubit interactions, and a two-qubit depth of 10 for the parity-check layer.
\subsubsection{Measurement-based parity check (MBPC)}
\label{sec:measurement_based_parity_check}
Using measurement-based quantum computing, we can execute a parity measurement in constant depth.\cite{AnsisRosmanis_paper}
We show an example circuit in Fig.~\ref{fig:tups_parity_circuits}c.
The main idea is to replace the sequential interaction with the parity ancilla qubit with an entangled resource state that is consumed during circuit execution.
The measurement outcomes introduce Pauli-$Z$ byproducts that are removed using classically controlled corrections.
For the four-data-qubit example, let $y$ denote the parity ancilla and let $a,c\in\{0,1\}$ be the two resource-state measurement outcomes.
Before correction and omitting the final Hadamard gate on $y$, the implemented operation is
\begin{equation}
U_{a,c}
=
Z_y^c Z_{x_2}^a Z_{x_3}^a
\prod_{i=0}^{3}\operatorname{CZ}_{x_i,y}.
\label{eq:mbpc_uncorrected_operation}
\end{equation}
Applying the measurement-dependent correction
\begin{equation}
C_{a,c}
=
Z_y^c Z_{x_2}^a Z_{x_3}^a
\label{eq:mbpc_feedforward_correction}
\end{equation}
removes these byproducts,
\begin{equation}
C_{a,c}U_{a,c}
=
\prod_{i=0}^{3}\operatorname{CZ}_{x_i,y}.
\label{eq:mbpc_corrected_operation}
\end{equation}
The Hadamard gates on $y$ transform this into the parity-accumulation operation,
\begin{equation}
H_y C_{a,c}U_{a,c}H_y
=
\prod_{i=0}^{3}\operatorname{CX}_{x_i,y}.
\label{eq:mbpc_parity_operation}
\end{equation}
The bits $a$ and $c$ determine Pauli-frame corrections and are not themselves parity-failure outcomes.

For a register of $D$ data qubits, the measurement-based gadget requires
\begin{equation}
N_{\mathrm{2q}}^{\mathrm{MBPC}}
=
3(D-2),
\label{eq:mbpc_two_qubit_count}
\end{equation}
with
\begin{equation}
A^{\mathrm{MBPC}}
=
2D-5
\label{eq:mbpc_ancilla_count}
\end{equation}
ancilla qubits.
The two-qubit depth is three when the interactions are executed in parallel, independent of $D$.
For the full (10,10) implementation, the two $D=10$ spin-register checks can run in parallel using 30 ancillas and 20 data qubits.
This gives 50 qubits in total, 48 additional two-qubit interactions, and two-qubit depth three for the parity-check layer.

\subsection{Quantum execution and statistical analysis}

The Jordan--Wigner Hamiltonians and real sparse states were represented and contracted using zixy 0.2.2.
InQuanto was used to synthesize the sparse multiconfiguration circuits for $\ket{\Phi}$, $\ket{\Lambda}$, and the sMCS targets, while the pp-tUPS targets were constructed as described above.
The circuits were optimized to the Helios native gate set with pytket, with the global phase restored because it affects signed-overlap protocols.
The optimized pytket circuits were imported into guppy and compiled to HUGR programs for execution through the Quantinuum software stack.
For classical shadows we used code from \cite{lenihan2025excitationamplitudesamplinglow} to generate the Clifford tableaus and post-process the generated data, which was modified to allow for multi determinant mixed estimator reference states using InQuanto.

For the direct-overlap data in Fig.~\ref{fig:result_protocol_energy_histograms}a--c, the Guppy programs were sampled without noise using exact statevector simulation.

Each noiseless energy estimate used an effective budget of $N_{\mathrm{shot}}=10\,000$ shots.
For split protocols, CU, DST, SST, RHT, and RIST, $8000$ shots were assigned to the residual-overlap circuit and $2000$ to the reference-overlap circuit.
FCU and CS estimate both contractions from the same set of measurement outcomes and therefore use all $10\,000$ shots jointly.
For the noiseless distributions in Fig.~\ref{fig:result_protocol_energy_histograms}a--c, we first generated $40\,000$ raw outcomes per protocol, with CS using one shot per Clifford measurement circuit.
Split protocols used $32\,000$ residual and $8000$ reference outcomes.
We then formed $B=2000$ bootstrap energy estimates by sampling with replacement at the effective allocations above.
The plotted standard errors are the sample standard deviations of these bootstrap estimates.

The Helios-1E and Helios-1 estimates were evaluated directly from the finite-shot outcomes retained after postselection.
For CU, standard errors were propagated from the statistics of the all-zero probabilities.
For CS, uncertainties were obtained from 100 bootstrap resamples of the postselected outcomes.

For a vector of measured quantities $z$, covariance matrix $C_z$, and energy estimator $F(z)$, the leading large-sample variance was propagated as
\begin{equation}
\sigma_F^2=\nabla F^{T}C_z\nabla F.
\end{equation}
The covariance matrix includes the Bernoulli variance of the all-zero CU outcomes, the negative covariance between the mutually exclusive FCU events, and the protocol-specific covariances for the signed, swap-based, and classical-shadow estimators.
At the shot counts used here, the finite-sample bias of the nonlinear energy estimator was found to be negligible relative to the statistical uncertainty, and no bias correction was applied; see Supplementary Section~\ref{subsec:si_covariance_recipe}.
Reported error bars denote one standard deviation.
The $\mathbf{R}_x$ and $\mathbf{R}_0$ calculations were sampled independently, so
\begin{equation}
\sigma_{\Delta E_{\mathbf{R}_x-\mathbf{R}_0}}^2
=
\sigma_{E_{\mathbf{R}_x}}^2
+
\sigma_{E_{\mathbf{R}_0}}^2.
\end{equation}

The idealized shot requirements in Fig.~\ref{fig:result_protocol_energy_histograms}d were not obtained from the fixed $8000/2000$ allocation.
For each value of $m$, we evaluated the statevector overlaps and residual norm and minimized the leading variance over the independent circuit allocations.
Let $\epsilon$ be the target standard error.
For the PA baseline, non-identity Jordan--Wigner terms were sorted by decreasing coefficient magnitude and assigned greedily to the first qubit-wise-commuting group.
For $H=\sum_gH_g$, define $\sigma_g^2=\mel{\Psi}{H_g^2}{\Psi}-\mel{\Psi}{H_g}{\Psi}^2$, including within-group covariances.
If group $g$ receives $N_g$ shots, then $\epsilon_{\mathrm{PA}}^2=\sum_g\sigma_g^2/N_g$, and optimal allocation gives
\begin{equation}
N_{\mathrm{PA}}
=
\frac{\left(\sum_g\sigma_g\right)^2}{\epsilon^2}.
\label{eq:pa_shot_cost}
\end{equation}

For nonzero real $x$ and $y$, define
\begin{equation}
A(x,y)
=
\frac{\sqrt{1-x^2}}{|x|}
+
\frac{\sqrt{1-y^2}}{|y|}.
\end{equation}
Independent RHT and RIST measurements then require
\begin{equation}
N_{\mathrm{RHT}}=N_{\mathrm{RIST}}
=
\frac{r_{\Phi}^2}{\epsilon^2}
\left|\frac{h}{s}\right|^2
A(h,s)^2.
\label{eq:signed_overlap_shots}
\end{equation}
For real $h$ and $s$ and a known correction sign, independent CU measurements require
\begin{equation}
N_{\mathrm{CU}}
=
\frac{r_{\Phi}^2}{4\epsilon^2}
\left|\frac{h}{s}\right|^2
A(h,s)^2.
\label{eq:cu_shot_cost}
\end{equation}
The factor of four follows from estimating squared overlaps rather than signed binary expectations.
The corresponding SST and DST requirement is
\begin{equation}
N_{\mathrm{SST}}
=
N_{\mathrm{DST}}
=
\frac{r_{\Phi}^2}{4\epsilon^2}
\left|\frac{h}{s}\right|^2
A(|h|^2,|s|^2)^2.
\end{equation}
Exact orthogonality of the prepared reference and residual makes the two FCU events mutually exclusive, giving
\begin{equation}
N_{\mathrm{FCU}}
=
\frac{r_{\Phi}^2}{4\epsilon^2}
\frac{|h|^2+|s|^2}{|s|^4}.
\label{eq:fcu_shot_cost}
\end{equation}
Since $|h|^2+|s|^2\leq1$ under this condition, the shot-variance ordering is
\begin{equation}
N_{\mathrm{FCU}}
\leq
N_{\mathrm{CU}}
\leq
N_{\mathrm{SST}}
=
N_{\mathrm{DST}}.
\label{eq:absolute_overlap_shot_ordering}
\end{equation}
If the prepared residual has $c=\braket{\Phi|\Lambda}\neq0$, the unflagged all-zero FCU event instead measures $|h-cs|^2$.
Equations~\eqref{eq:fcu_shot_cost} and \eqref{eq:absolute_overlap_shot_ordering} therefore apply only to an orthogonal residual; truncating or otherwise approximating the states can introduce an additional systematic error.

For the two fixed real contractions entering the mixed estimator, global-Clifford CS gives, for large Hilbert-space dimension,
\begin{equation}
N_{\mathrm{CS}}
\approx
\frac{3r_{\Phi}^2}{\epsilon^2}
\frac{|h|^2+|s|^2}{|s|^4}.
\label{eq:cs_shot_cost}
\end{equation}
These expressions compare shot variance and do not include circuit duration, control overhead, or hardware runtime.
Full covariance derivations, optimal allocation rules, bootstrap details, and the nonorthogonal FCU case are given in Supplementary Section~\ref{SI-sec:si_mixed_estimator_details}.

\bmhead{Data availability}
The data supporting the findings of this study are available in the Supplementary Information, with additional data available from the corresponding author upon reasonable request.

\bmhead{Code availability}
Third-party software is identified in the Methods.

\clearpage
\putbib[references]
\end{bibunit}

\backmatter
\bmhead{Acknowledgments}
We are grateful to Ilyas Khan for his early confidence in this work and for the support that enabled it.
We appreciate Varun Ramdevan's support in establishing and advancing this collaboration and thank Kai Stoeber for sharing his expertise and for helpful scientific discussions.
We specially thank Harry Buhrman, the Quantinuum chemistry, hardware, and AI teams, Kentaro Yamamoto, Nathan Fitzpatrick, Robert Anderson, and George H. Booth for their insightful discussions and contributions throughout this work.
We thank Charlie Baldwin, Henrik Dreyer and Georgia Prokopiou for their comprehensive review of the manuscript and their valuable suggestions and insights.
\bmhead{Author contributions}
CL and DZM developed the direct ME method and overlap-measurement statistics.
KP developed the pp-tUPS calculations, GGD the sparse multiconfiguration-state calculations, and LE the quantum error-detection methods.
JG led the AI-assisted state and circuit optimization, with contributions from FF and ER.
CAG led the classical electronic structure and rate calculations, with contributions and perspective from GL, MM, CM, TMS and ES.
MCR, GGD, PK, KP, JG, CL, LE and DZM contributed to the initial exploration, implementation and hardware execution.
ES, TMS and DMR initiated the project and managed project resources.
DZM conceived the central methodology and coordinated the overall research.

\bmhead{Competing interests}
Some of the authors are employees of and hold equity interests in Quantinuum. The remaining authors declare no competing interests.

\clearpage
\makeatletter
\def\@extra@binfo{@SI}
\def\@extra@b@citeb{@SI}
\makeatother
\providecommand{\mainref}[1]{\ref{#1}}
\begin{bibunit}[sn-nature]
\begin{center}
{\Large Measurement-efficient energy estimator for quantum chemistry of photodynamic materials\par}
\vspace{1em}
{\large Connor Lenihan$^{1,\dagger}$, Kripa Panchagnula$^{1,\dagger}$, Luca Erhart$^{1,\dagger}$, Jem Guhit$^{2,\dagger}$, Carlo A. Gaggioli$^{1,\dagger}$, Gabriel Greene-Diniz$^{1,\dagger}$, Gabriel Laude$^1$, Megan Cerys Rowe$^1$, Fabian Finger$^2$, Pranav Kalidindi$^2$, Enrico Rinaldi$^2$, Mikko Muuronen$^3$, Chandrika Mulakala$^4$, Thomas M. Soini$^1$, Elvira Shishenina$^2$, David Muñoz Ramo$^1$, David Zsolt Manrique$^{1,*}$\par}
\vspace{0.75em}
{\small $^1$Quantinuum, Terrington House, 13--15 Hills Road, Cambridge, CB2 1NL, UK\par} {\small $^2$Quantinuum, Partnership House, Carlisle Place, London, SW1P 1BX, UK\par} {\small $^3$Johnson \& Johnson, Beerse, Belgium\par} {\small $^4$Johnson \& Johnson, South San Francisco, CA, USA\par}
\vspace{0.5em}
{\small $^\dagger$These authors contributed equally to this work.\par} {\small $^*$Corresponding author: david.zsolt.manrique@quantinuum.com\par}
\vspace{1.5em}
\end{center}

\numbered

\setcounter{section}{0}
\setcounter{figure}{0}
\setcounter{table}{0}
\setcounter{equation}{0}
\renewcommand{\thesection}{S\arabic{section}}
\renewcommand{\thefigure}{S\arabic{figure}}
\renewcommand{\thetable}{S\arabic{table}}
\renewcommand{\theequation}{S\arabic{equation}}
\renewcommand{\theHsection}{S\arabic{section}}
\renewcommand{\theHfigure}{S\arabic{figure}}
\renewcommand{\theHtable}{S\arabic{table}}
\renewcommand{\theHequation}{S\arabic{equation}}

\section*{Supplementary Information}
\suppressfloats[t]

\section{Classical rate modeling and electronic-structure calculations} \label{Classical rate modeling}

\subsection{DFT rate calculations} \label{dft_rates}

In this subsection we describe the DFT calculations of the rate of singlet oxygen formation on Ru-functionalized MOF.
We first optimized the Hf-MOF (Hf--DBB--Ru) using the UFF4MOF II force field as implemented in ADF see the Computational Details section \ref{comp_details}).
The picture of the optimized MOF is shown in Supplementary Fig.~\ref{fig:SI_MOF}.
After the optimization was done, we extracted one ligand from the MOF (which contains the Ru(bpy)$_3$ moiety acting as the photosensitizer, PS), and we capped the benzoate moieties with one H atom each, to mimic the effect of nearby Hf atoms (H$_2$DBB--Ru).
This molecular system is therefore charged +2 and is a closed-shell singlet in the ground state (denoted as \textbf{R$_0^S$}, see Supplementary Fig.~\ref{fig:S0_geom}).
We further fixed the positions of the carbon atoms of the COOH groups during all geometry optimizations, in order to maintain the rigidity of the linker as found in the MOF.
H$_2$DBB--Ru and the Hf--DBB--Ru MOF exhibited similar absorption and phosphorescence features,\cite{ni2018nanoscale} supporting the use of the molecular model throughout this work.

\begin{figure*}[t]
    \centering
    \includegraphics[width=0.8\textwidth]{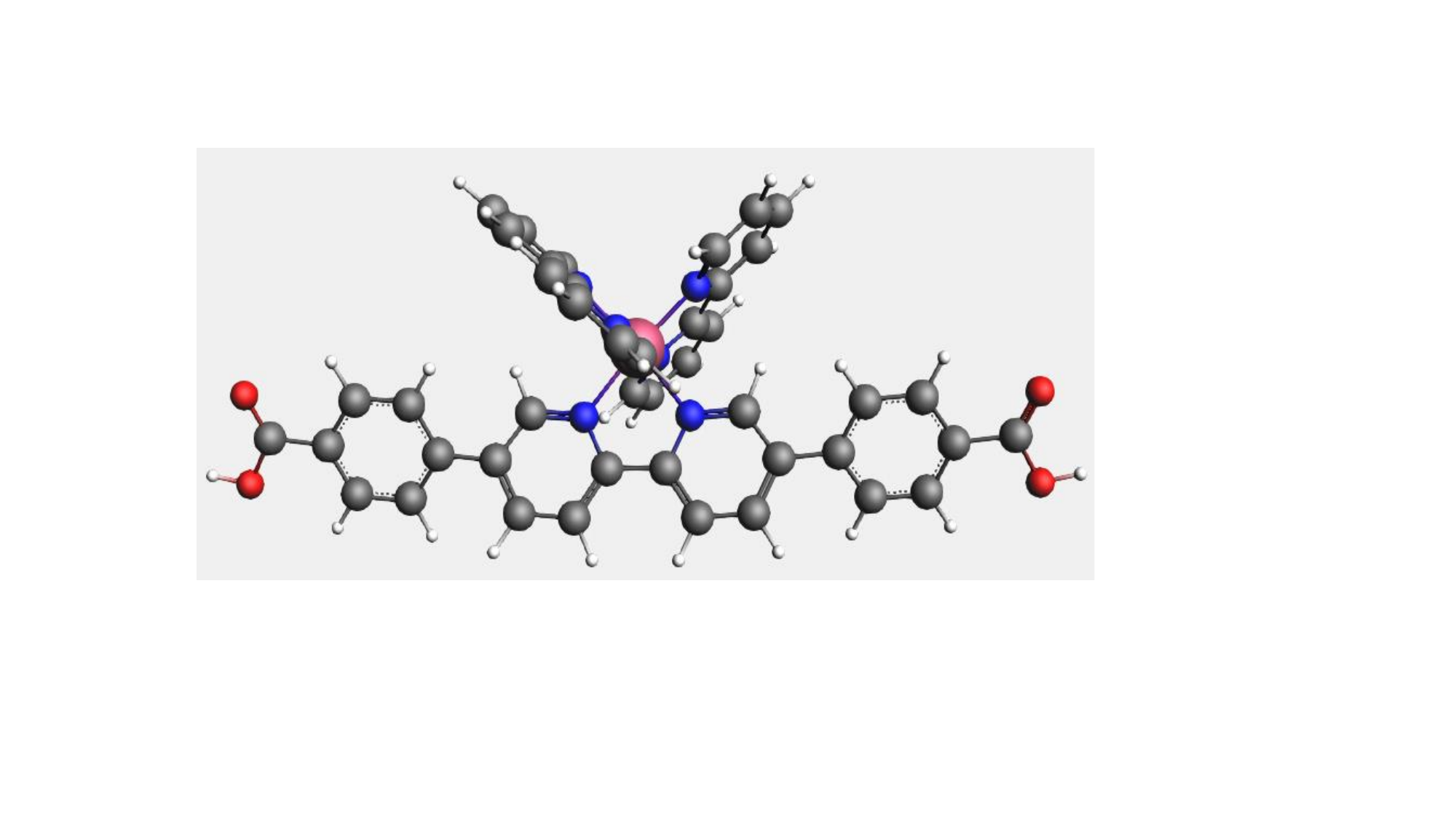}
    \caption{Ligand extracted from the optimized MOF, containing the Ru(bpy)$_3$ moiety acting as PS (H$_2$DBB--Ru).
    The carboxylate groups were capped with H atoms.
    Colour code: carbon in grey, hydrogen in white, oxygen in red, nitrogen in blue, and ruthenium in magenta.}
    \label{fig:S0_geom}
\end{figure*}

For the classical modeling of the rate of singlet oxygen production, we followed the divide-to-conquer algorithm developed by Bai and Barbatti \cite{bai2017divide,bai2017spatial} which uses the classical Marcus rate expression (Eq.~\eqref{eq:rate_marcus})
\begin{equation}\label{eq:rate_marcus}
k_{ij} = \frac{2\pi}{\hbar} |V_{ij}|^2
\frac{1}{\sqrt{4\pi \lambda k_B T}}
\exp\left( -\frac{\Delta E^{\ddagger}}{k_B T} \right)
\end{equation}

To calculate the rate between states $i$ and $j$, we need to calculate the diabatic couplings ($V_{ij}$) between these electronic states, the reorganization energy ($\lambda$) and the activation energy ($\Delta E^{\ddagger}$). In the following, all calculations are carried out in the triplet spin state, unless stated otherwise.

We start by analyzing the reorganization energy ($\lambda$), which is the difference in energy between the triplet state at \textbf{R$_0^S$} geometry (most stable geometry of H$_2$DBB-Ru in the singlet spin potential energy surface) and the triplet state at $\mathbf{R}_0$ geometry.
For the DFT study, we employed three density functional approximations, namely $\omega$B97X, B3LYP, and PBE functionals (see computational details for more information).
The optimized geometries of \textbf{R$_0^S$} and $\mathbf{R}_0$ (together with some bond lengths) are reported in Supplementary Fig.~\ref{fig:S0_S1_geom}.
We can see that their geometries are very similar, and therefore we would expect the reorganization energy to be rather small.
Indeed, we found the reorganization energy to be 0.034, 0.085 and 0.117 eV with $\omega$B97X, B3LYP, and PBE functionals respectively.
The variability of $\lambda$ with the functionals gives rise to a factor of two in the rate, which is acceptable for computational methods.

A previous work analyzed computationally the excited
(triplet) potential energy surface of Ru(bpy)$_3$, noticing
that the $\mathbf{R}_0$ state consists of a metal to ligand charge transfer state ($^3$MLCT, Ru $d$ to ligand $\pi^*$) and that there is a local minima (slightly higher in energy) in the triplet potential energy surface (PES) that has a metal centered ($^3$MC) transition (ruthenium $d$ to $d^*$), which then leads to the $^{3/1}$MECP (Minimum Energy Crossing Point) that allows relaxation into the singlet ground state (see Figure 3 in Ref. \cite{soupart2018dft}). This computational study is in line with the experimental Ru(bpy)$_3$ luminescence properties and highlights the complexity of the triplet PES.

Then, following the divide-to-conquer algorithm \cite{bai2017divide}, we optimized the $^{3/1}$MECP geometry, starting from the $^3$MC geometry, and the MECP geometry is shown in Supplementary Fig.~\ref{fig:MECP_geom}.
We then generated 20 images by linearly interpolating between $\mathbf{R}_0$, the $^3$MLCT minimum, and the MECP geometry, in order to locate $\mathbf{R}_x$.\cite{bai2017divide}
We remark that the linear interpolation is an approximation and may not provide the minimum energy path connecting $\mathbf{R}_0$ and MECP geometries.
However, the characterization of the whole singlet and triplet PESs would be computationally very demanding and we decided to proceed with the linear interpolation, as used in ref. \citenum{bai2017divide,bai2017spatial,manae2022quantum}.
The Mulliken spin densities on ruthenium for the MLCT, MC and MECP are 0.95, 1.82 and 1.77 respectively (calculated with $\omega$B97X).
We can therefore conclude that the MECP resembles the MC electronic structure.

We then analyzed the energies of MLCT, MC and MECP with different functionals, and we report the results in Supplementary Table~\ref{SI-tab:MLCT_MC_func}.
We can see that MLCT is the most stable structure with the PBE functional, with MC structure lying at 0.43 eV.
On the other hand, the hybrid B3LYP functional stabilizes more MC (and MECP) with respect to MLCT, with MC being at -0.31 eV than MLCT.
We also optimized the geometries of MLCT and MC with B3LYP, but the relative energies of the two structures did not change significantly.
The $\omega$B97X functional further stabilizes MC, going to -1.17 eV with respect to MLCT.
We can see that the functionals yield high variability in the relative energy of the different electronic structures, with changes up to 1.6 eV.

In order to calculate the diabatic couplings and activation energy (both needed to calculate the rate), we decided to move the O$_2$ molecule close to two carbon atoms (which are part of the ligand) in a parallel way with respect to the aromatic ring of one bipyridine.
We did this by performing a scan in which the two oxygen atoms were simultaneously moved close to these carbon atoms from 3.8 to 2.4 \AA {} (see Supplementary Fig.~\ref{fig:O2_scan} for a sketch of the scan).
We chose this position because it was shown that orbitals overlaps between $\mathbf{R}_0$ and O$_2$ would increase the rate of singlet oxygen production by maximizing the diabatic couplings,\cite{serrano2009singlet} and by inspecting the spin density in $\mathbf{R}_0$ (Supplementary Fig.~\ref{fig:T1_spin_density}) we can see that the carbon atoms we selected have high spin density.
Nevertheless, other $\mathbf{R}_0$-O$_2$ scans could be tried (that could potentially give rise to higher rates), but this analysis was outside the scope of this work.

We then looked at the singlet-triplet (S-T) splittings of all 20 images (between MLCT and MECP), with the data presented in Supplementary Fig.~\ref{fig:T_S_splittings}.
We can see that the S-T gap for image0 (i.e. MLCT) has a sizable variability with the functionals, specifically it is 2.29, 2.04 and 1.71 eV with $\omega$B97X, B3LYP, and PBE, respectively.
The S-T gap trend is similar among the three functionals, showing a gap for images up to 6-9 that is similar to the MLCT gap, and then the gap starts to reduce towards around zero at image21 (MECP geometry).
We then analyzed the energies of the triplet spin for all the images (setting image0 at energy zero), reported in Supplementary Fig.~\ref{fig:triplet_energies}.
In all three cases, we can see that the energy goes up until images 9-11, and then the energy goes down, reaching different energetics at the MECP, as reported in Supplementary Table~\ref{SI-tab:MLCT_MC_func}.
We also analyzed the Hirshfeld spin densities on ruthenium for all the images for the two hybrid functionals, with the results shown in Supplementary Fig.~\ref{fig:spin_images}.
We can see that in the first 5 to 9 images the spin density is about 0.3 to 1, indicating an MLCT configuration, with a jump in the spin density toward 1.6 for the last images, indicating a change of configuration from MLCT to MC.

We now discuss the rate calculations for the $^1\Delta_{g}$ oxygen production.
We were not able to study the rate of production of the $^1\Sigma_{g}$ state of oxygen, because Eq.~\eqref{eq:get_Rx} (required by the divide-to-conquer model to get the crossing structure $\mathbf{R}_x$) was never satisfied, mostly because the O$_2$ TDDFT (Time-Dependent Density Functional Theory) excitation energies for this electronic state are very far from experimental values (see Supplementary Section~\ref{subsec:tddft_excitations}).
Without being able to locate $\mathbf{R}_x$, we cannot calculate the activation energy and therefore the rate.
Hence, we focus on the calculations for the $^1\Delta_{g}$ oxygen production.

We then proceeded to calculate the diabatic couplings between the two singlet electronic configurations $^{1}[^{3}\!PS + (^{3}\Sigma_{g})O_{2}]$ and $^{1}[^{1}\!PS + (^{1}\Delta_{g})O_{2}]$, and the results (for the three functionals) are reported in Supplementary Fig.~\ref{fig:diabatic_couplings}.
We observe that the couplings increase when O$_2$ gets closer to the two carbon atoms of the ligand, although not monotonically.
For $\omega$B97X and B3LYP, the maximum couplings are around 8-12 cm$^{-1}$, while for PBE the maximum coupling is around 180 cm$^{-1}$.
We can see that the variability on the couplings is around 1 order of magnitude, which would give rise to two orders of magnitude difference in the rate (according to Eq.~\eqref{eq:rate_marcus}).

We then calculated the activation energies (as described in ref \cite{bai2017divide}), and the results are shown in Supplementary Fig.~\ref{fig:activation_energies}.
We can see that with the activation energies are in generally higher with B3LYP than $\omega$B97X (always around 0.3-0.4 eV higher, apart from the 2.4 \AA {} point), and with PBE we were only able to calculate 3 points (3.8, 3.6 and 3.4 \AA, with activation energy equal to 1.14 eV), because for the other points Eq.~\eqref{eq:get_Rx} was not satisfied.

Finally, by combining the reorganization energies, diabatic couplings and activation energies, we calculated the rates for the O$_2$ $^1\Delta_{g}$ production with the three functionals, and the results are reported in Supplementary Tables~\ref{SI-tab:wb97x_data_for_rates} and \ref{SI-tab:b3lyp_data_for_rates} and in Supplementary Fig.~\ref{SI-fig:rates}.
We notice that the highest rates are found with the $\omega$B97X functional, with the maximum being 47.6 s$^{-1}$ (at O$_2$-PS distance of 2.8 \AA).
The B3LYP rates are much lower in value, with the maximum being 1.3$\times 10^{-4}$ s$^{-1}$.
We can see that the two functionals give rise to a difference of more than 5 orders of magnitude in the rate values, which is a huge variability when performing in-silico predictions.
We also calculated the rates with PBE, although only for three values of O$_2$-PS distance, and the maximum rate is 7.2$\times 10^{-8}$ s$^{-1}$, other orders of magnitude smaller than with the other functionals.
However, shorter O$_2$-PS distances may give rise to higher rates, so the results with PBE are inconclusive.

The activation energy is the quantity that mostly modifies the rate (since it has an exponential effect on the rate, Eq.~\eqref{eq:rate_marcus}), and the difference in energy between $\mathbf{R}_x$ and $\mathbf{R}_0$ makes the greatest impact in the calculation of the rate.
There is another piece that takes part into the activation energy, and it is the energy required to bring O$_2$ close to the PS, but we calculated this quantity to be around 1-2 mHa, and therefore we focus our analysis with quantum algorithms on the difference in energy between $\mathbf{R}_x$ and $\mathbf{R}_0$.
We decided to select the $\mathbf{R}_x$ geometry that gives rise to the highest rate we found - to be used for subsequent active space methods and quantum computation - which is the $\mathbf{R}_x$ geometry of the 2.8 \AA{} point for $\omega$B97X in Supplementary Fig.~\ref{SI-fig:rates}.
The description on how to get this geometry is shown in Supplementary Section~\ref{subsec:rx_calculation}.
The geometry of $\mathbf{R}_x$ is reported in Supplementary Fig.~\ref{fig:geom_Rx}, and two Ru-N geometries are elongated with respect to MLCT, but shorter than the MECP ones.

After this analysis, we can conclude that DFT lacks predictive power on the calculation of singlet oxygen production rates, because of the high variability we found by changing the functional.

\begin{figure*}[!t]
    \centering
    \includegraphics[width=0.8\textwidth]{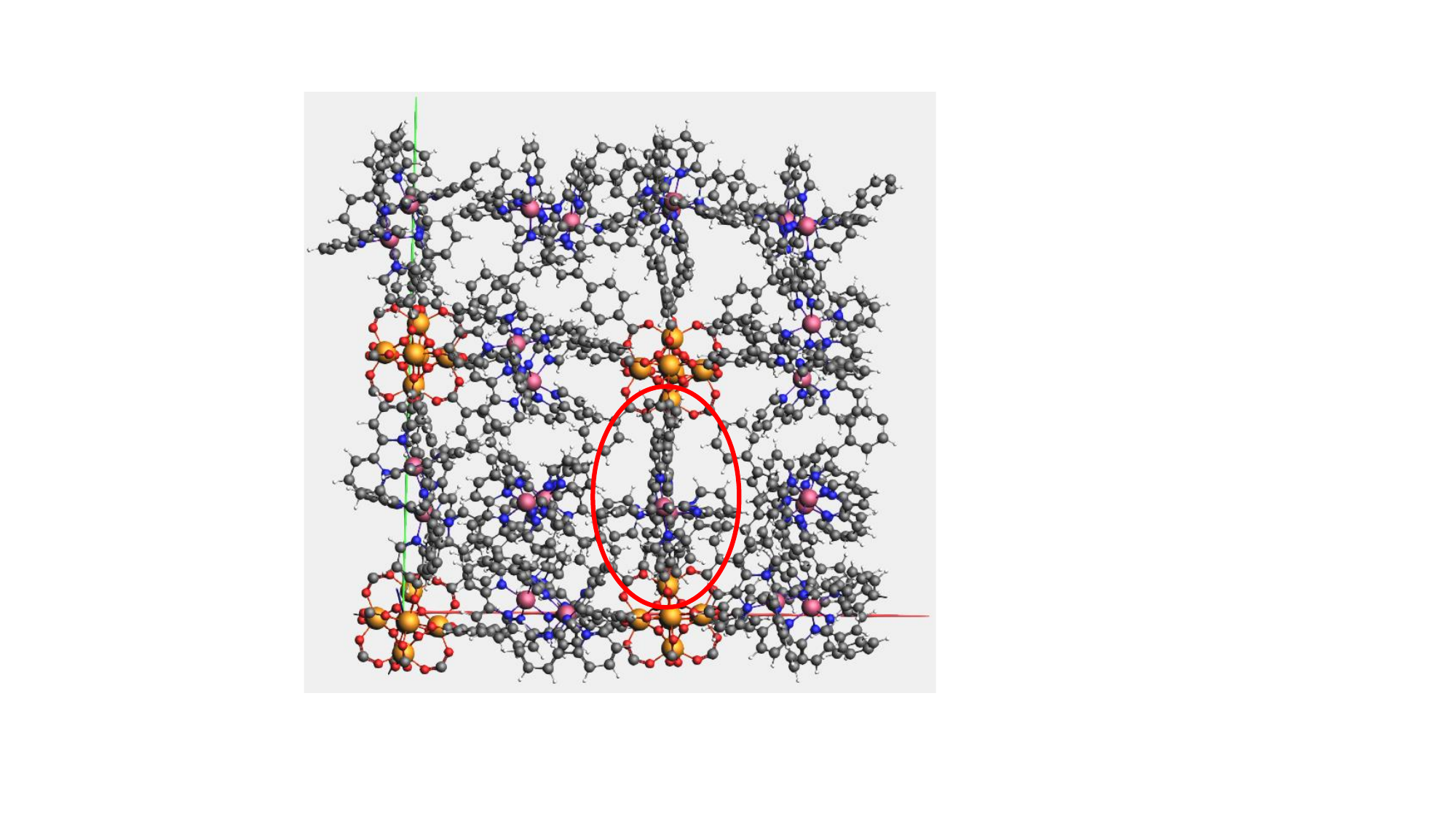}
    \caption{Optimized MOF using the UFF4MOF II force field as implemented in ADF.
    The extracted ligand used in all the calculations is circled in red.
    The optimized cell parameters are 38.4, 37.7 and 38.2 \AA, while the angles are 92.0, 91.5 and 88.9 degrees.
    The brute formula is: C$_{1056}$H$_{720}$Hf$_{24}$N$_{144}$O$_{128}$Ru$_{24}$.
    Colour code: carbon in grey, hydrogen in white, oxygen in red, nitrogen in blue, hafnium in orange and ruthenium in magenta.}
    \label{fig:SI_MOF}
\end{figure*}

\begin{figure*}[!t]
    \centering
    \includegraphics[width=0.8\textwidth]{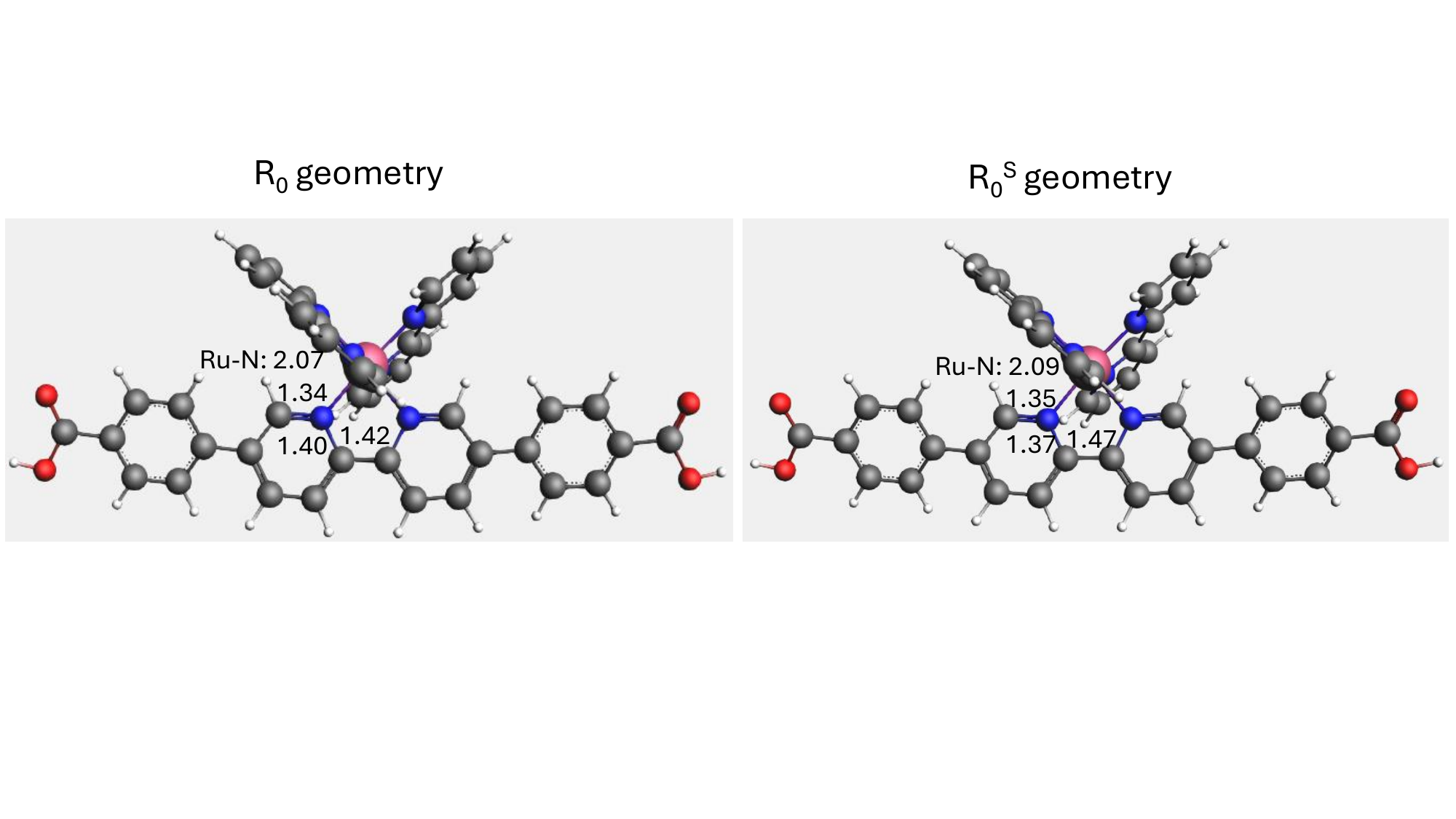}
    \caption{Comparison of $\mathbf{R}_0$ (left) and \textbf{R$_0^S$} geometries (right).
    Some bond lengths are also reported, highlighting the similarity between the two geometries.
    Colour code: carbon in grey, hydrogen in white, oxygen in red, nitrogen in blue, and ruthenium in magenta.}
    \label{fig:S0_S1_geom}
\end{figure*}

\begin{figure*}[!t]
    \centering
    \includegraphics[width=0.8\textwidth]{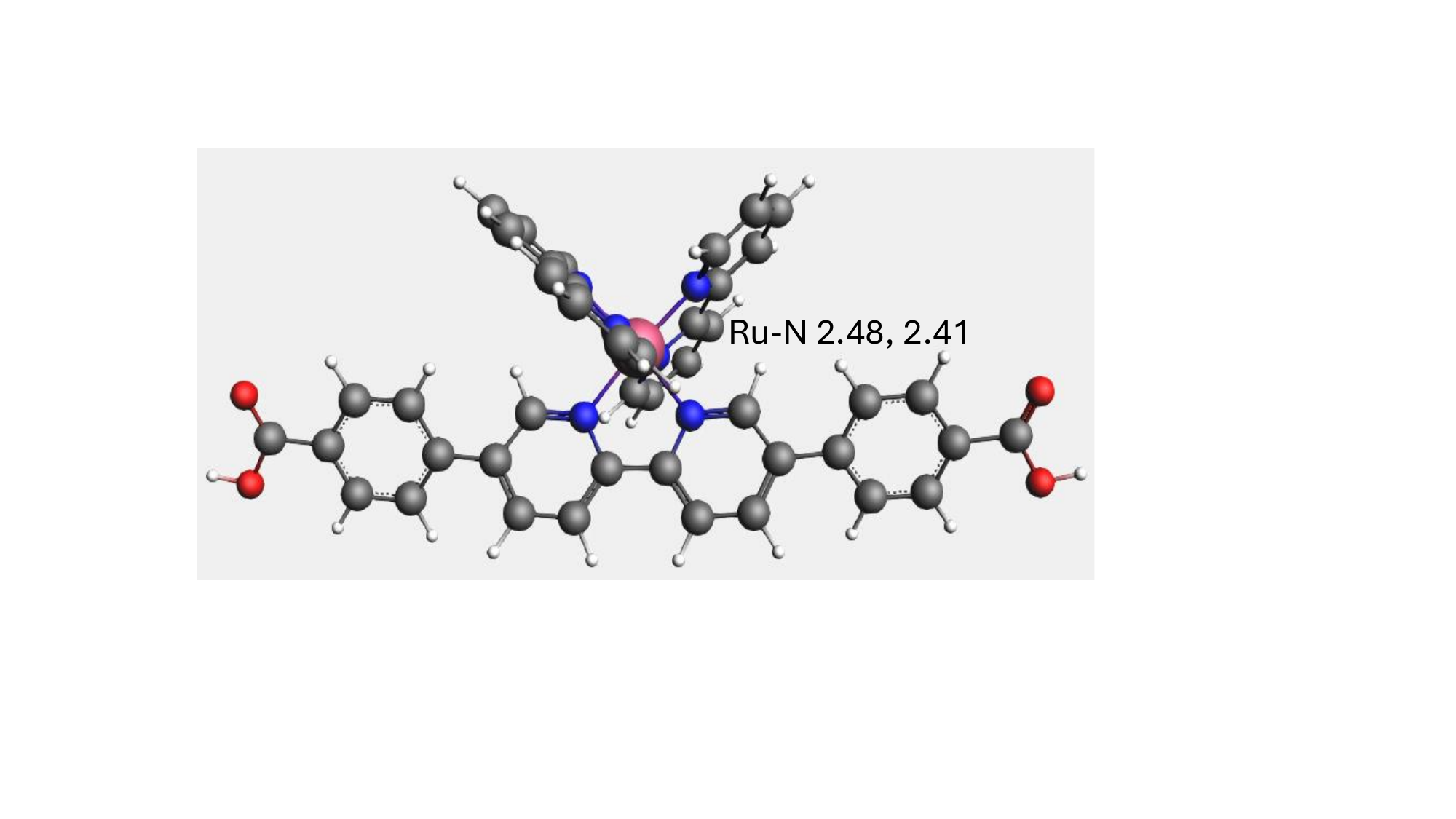}
    \caption{MECP geometry, with two Ru-N bond lengths reported (the ones that are most different with respect to $\mathbf{R}_0$).
    Colour code: carbon in grey, hydrogen in white, oxygen in red, nitrogen in blue, and ruthenium in magenta.}
    \label{fig:MECP_geom}
\end{figure*}

\begin{table}
\small
  \caption{DFT relative energies of MLCT, MC and MECP (setting the MLCT energy to zero).
  The evaluation was done as single points (SP) with the PBE, B3LYP and $\omega$B97X functionals, together with B3LYP geometry optimizations (OPT) of MLCT and MC structures.
  All energies in eV.}
\label{SI-tab:MLCT_MC_func}
    \begin{tabular}{ccc}
    \hline
    Functional & $\Delta$E MC-MLCT & $\Delta$E MECP-MLCT\\
    SP PBE  & 0.43 & 0.62\\
    SP B3LYP  & -0.31 & -0.05\\
    OPT B3LYP  & -0.29 & N/A\\
    SP $\omega$B97X  & -1.17 & -0.66\\
    \hline
    \end{tabular}
\end{table}

\begin{figure*}[!t]
    \centering
    \includegraphics[width=0.8\textwidth]{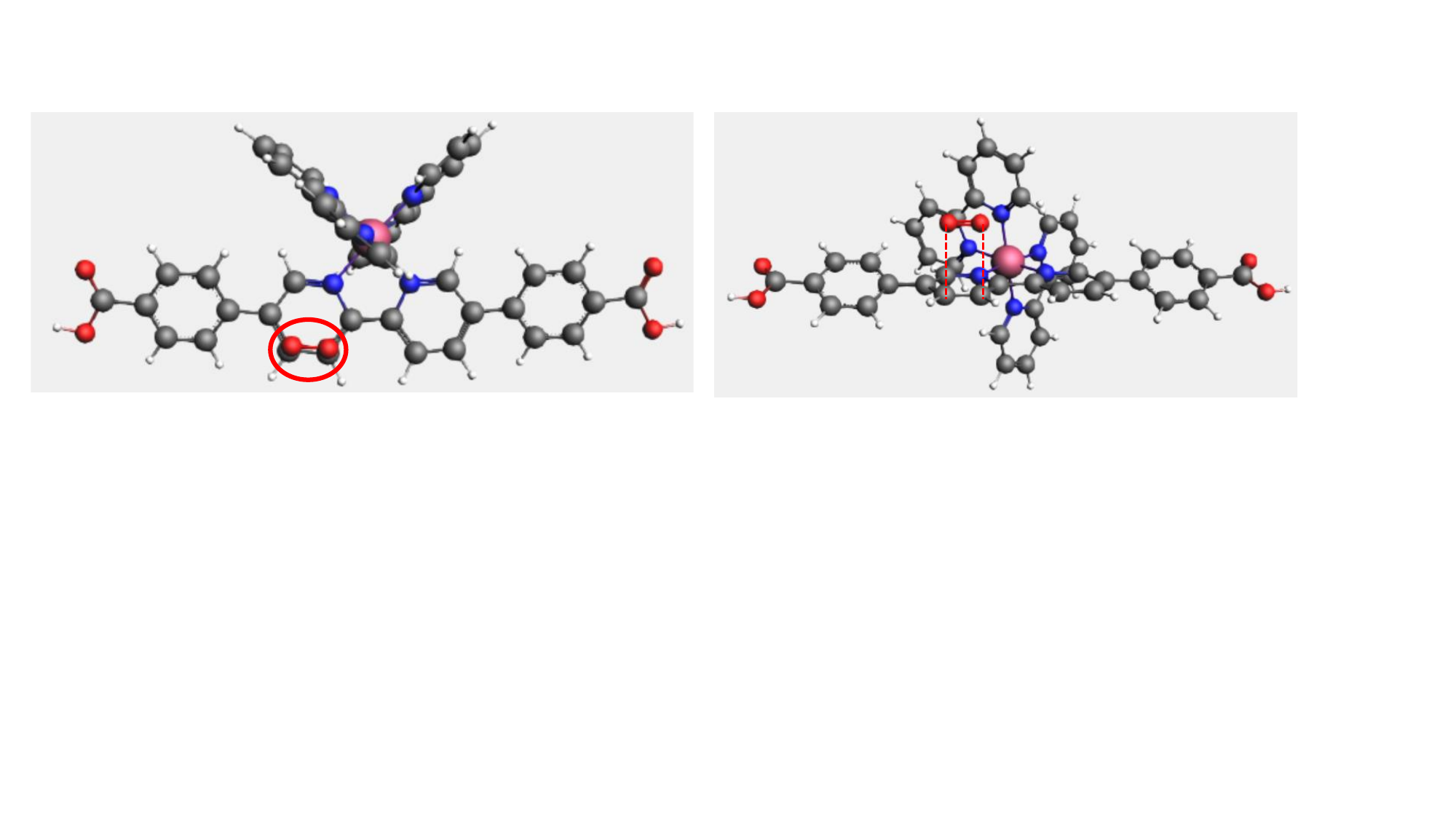}
    \caption{Two views of the scan of O$_2$ towards the aromatic ring of one bipyridine moiety.
    The two carbon atoms interacting with O$_2$ and the O$_2$ molecule are circled in red (left sketch) and are connected by dashed lines (right sketch).
    Colour code: carbon in grey, hydrogen in white, oxygen in red, nitrogen in blue, and ruthenium in magenta.}
    \label{fig:O2_scan}
\end{figure*}

\begin{figure*}[!t]
    \centering
    \includegraphics[width=0.8\textwidth]{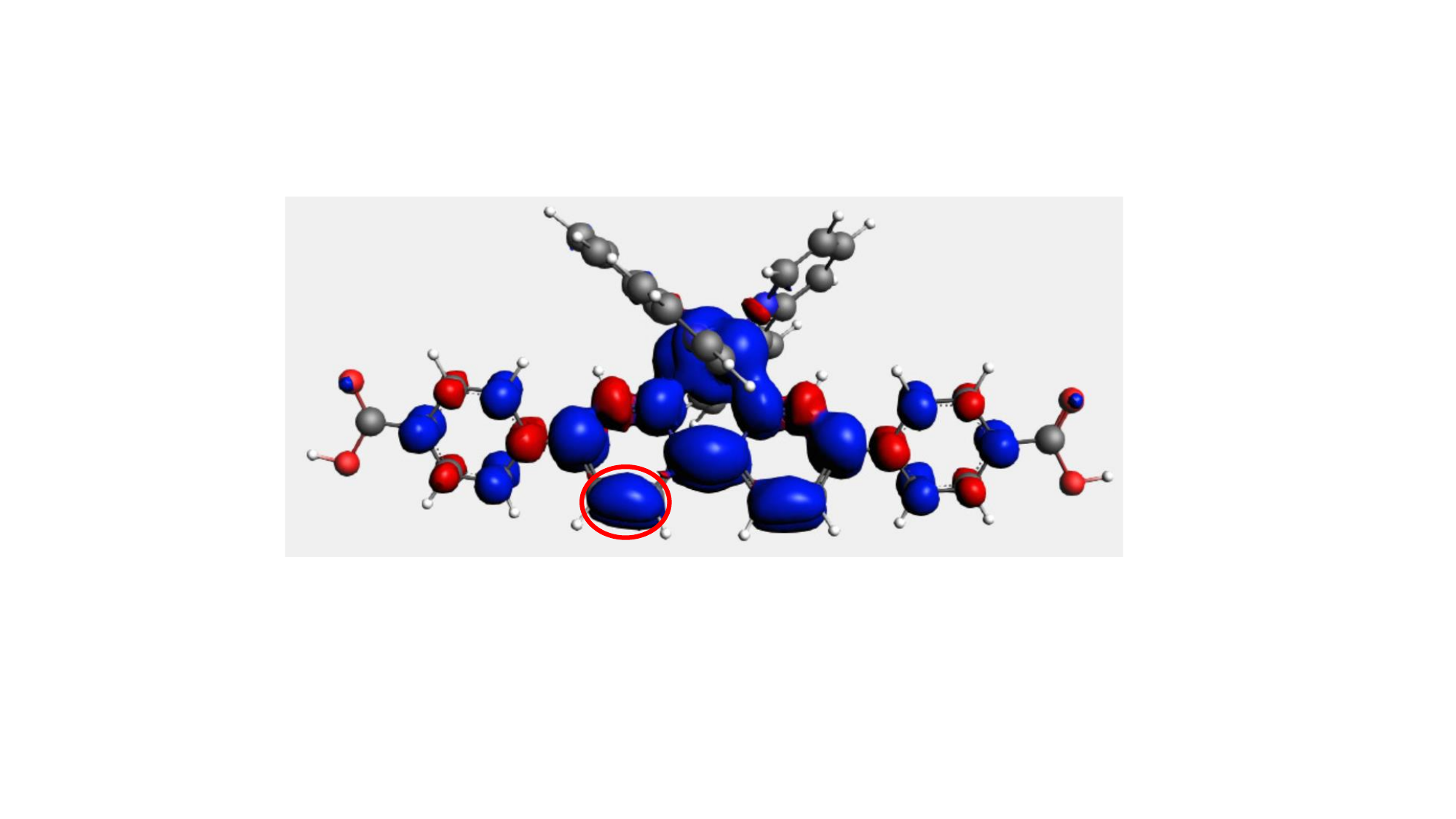}
    \caption{$\mathbf{R}_0$ geometry with the spin density isosurface in blue (0.001 a.u.).
    The carbon atoms that interacts with O$_2$ are circled in red, showing a sizable amount of spin density.
    Colour code: carbon in grey, hydrogen in white, oxygen in red, nitrogen in blue, and ruthenium in magenta.}
    \label{fig:T1_spin_density}
\end{figure*}

\begin{figure*}[!t]
    \centering
    \includegraphics[width=0.8\textwidth]{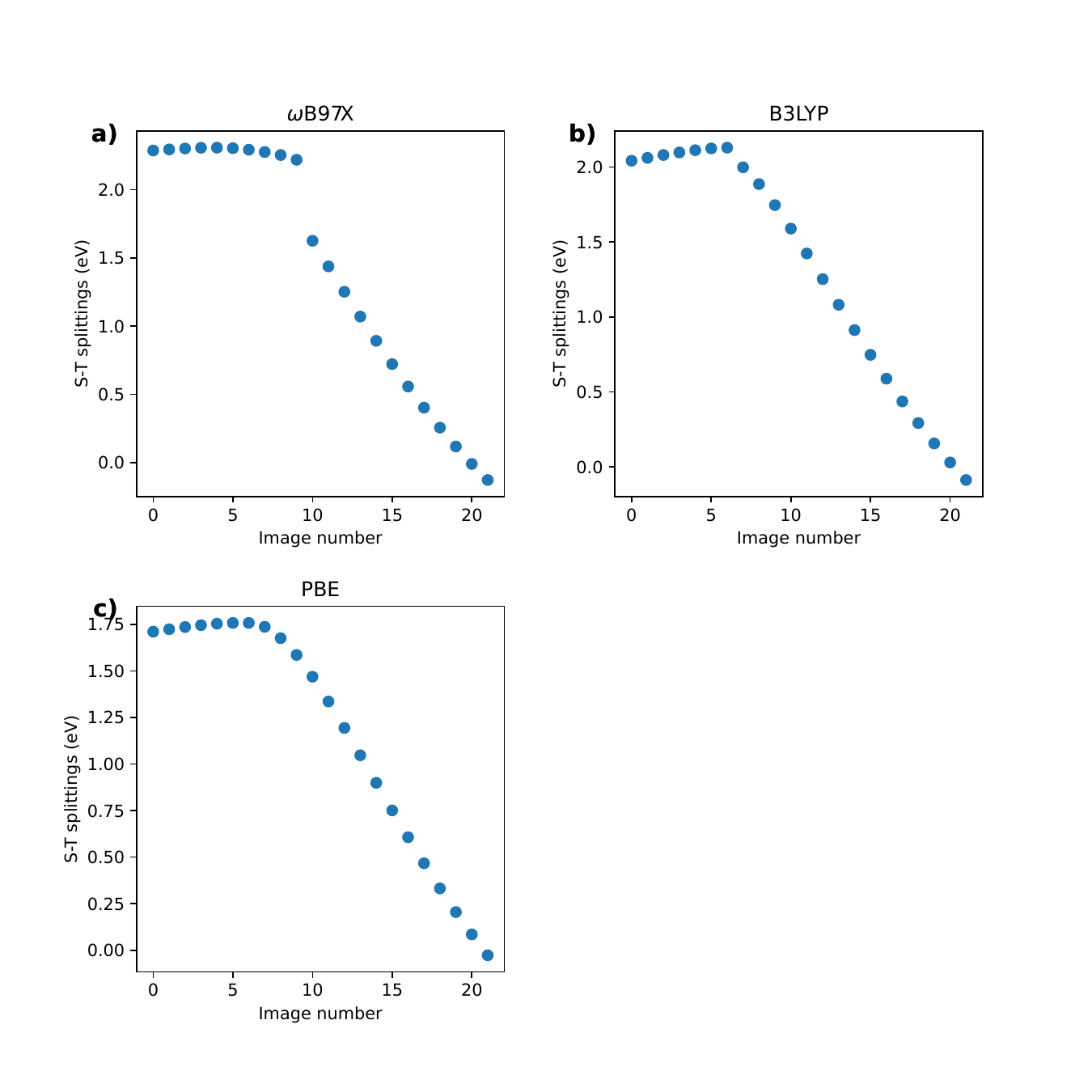}
    \caption{Singlet-triplet splittings (in eV) for the 21 images between $\mathbf{R}_0$ (image 0) and MECP (image 21).
    Calculations with the three functionals.
    All energies in eV.}
    \label{fig:T_S_splittings}
\end{figure*}

\begin{figure*}[!t]
    \centering
    \includegraphics[width=0.8\textwidth]{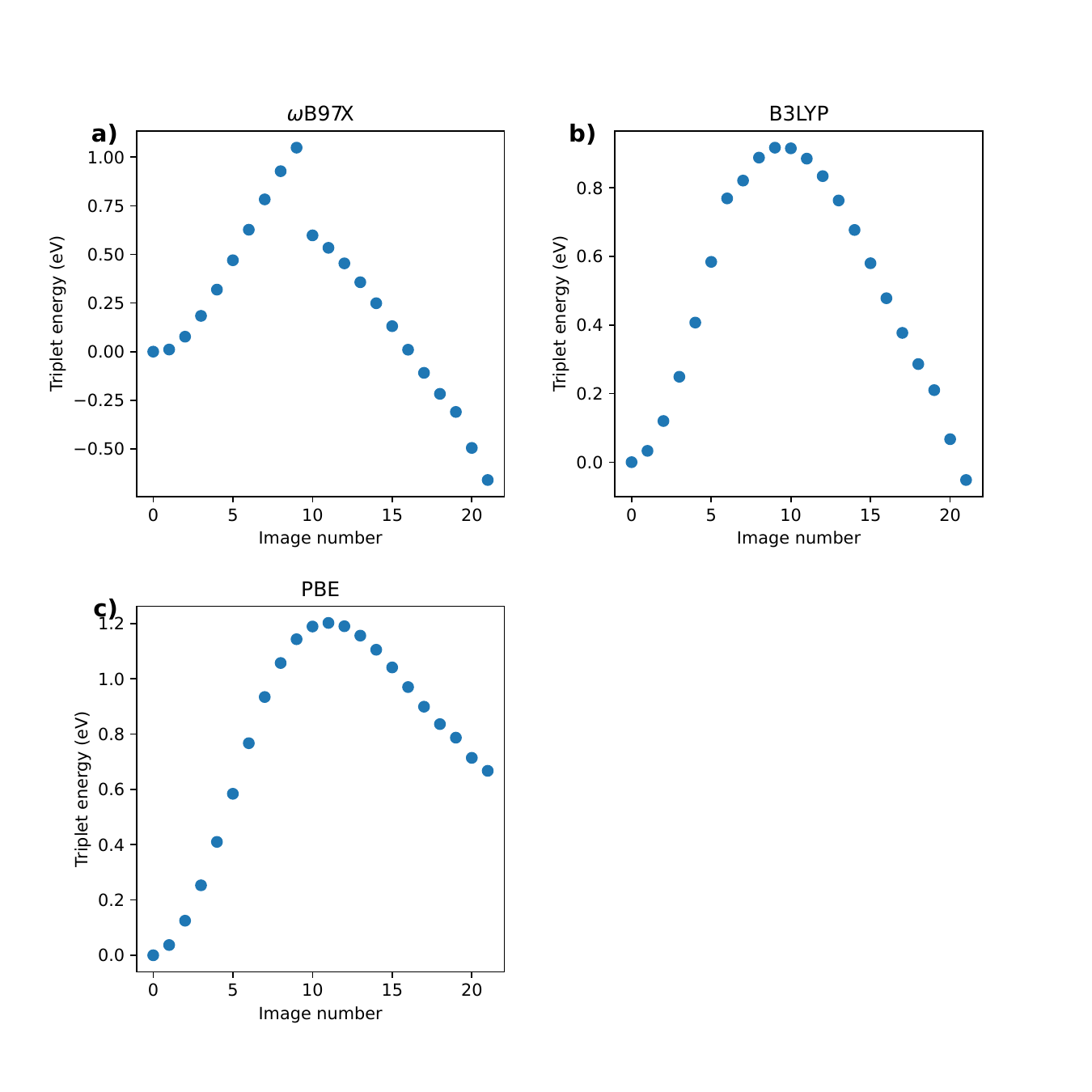}
    \caption{Energy of the triplet spin multiplicity (in eV) for the 21 images between $\mathbf{R}_0$ (image 0) and MECP (image 21).
    Calculations with the three functionals.
    All energies in eV.}
    \label{fig:triplet_energies}
\end{figure*}

\begin{figure*}[!t]
    \centering
    \includegraphics[width=0.8\textwidth]{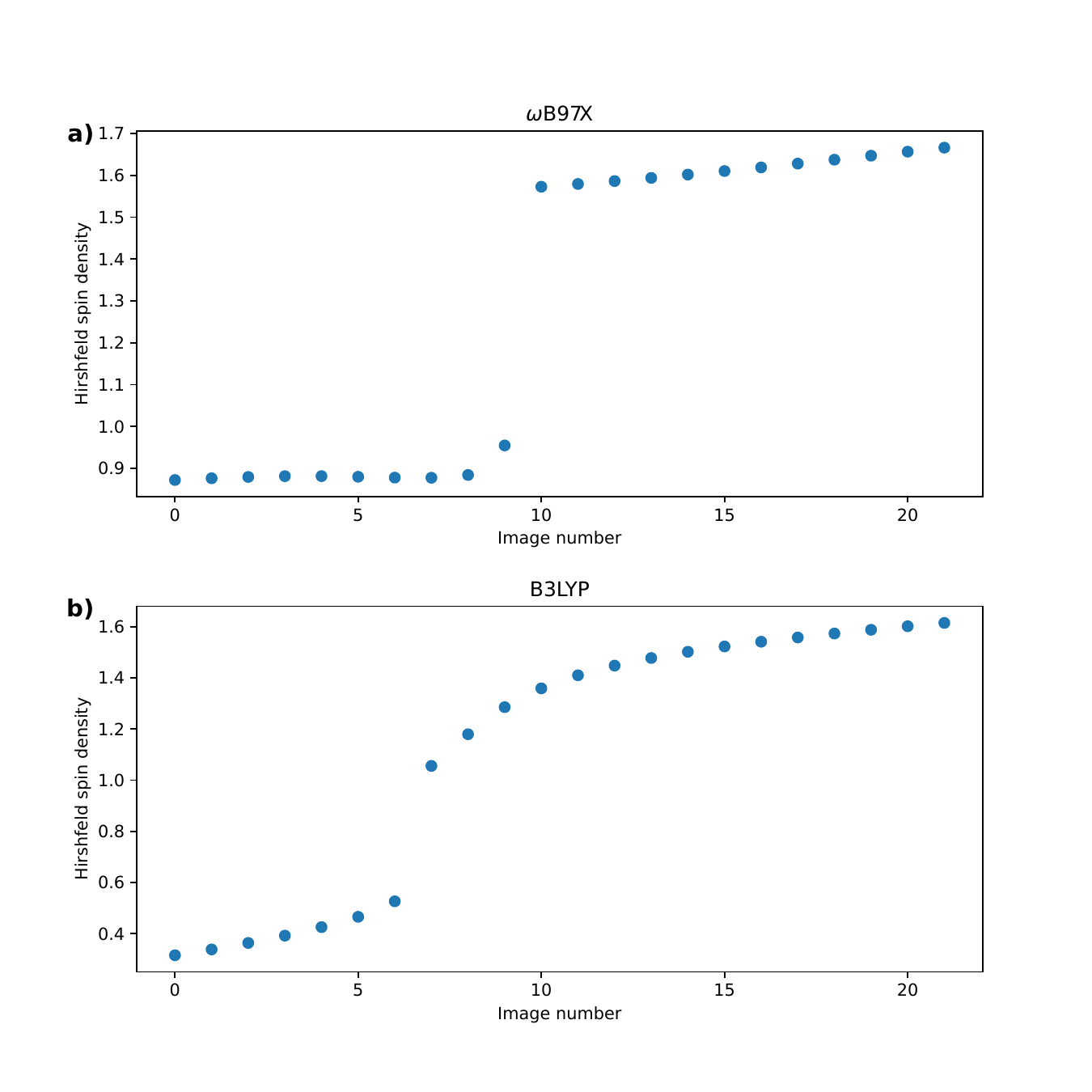}
    \caption{Hirshfeld spin densities for the 21 images between $\mathbf{R}_0$ (image 0) and MECP (image 21).
    Calculations with $\omega$B97X and B3LYP.}
    \label{fig:spin_images}
\end{figure*}

\begin{figure*}[!t]
    \centering
    \includegraphics[width=0.8\textwidth]{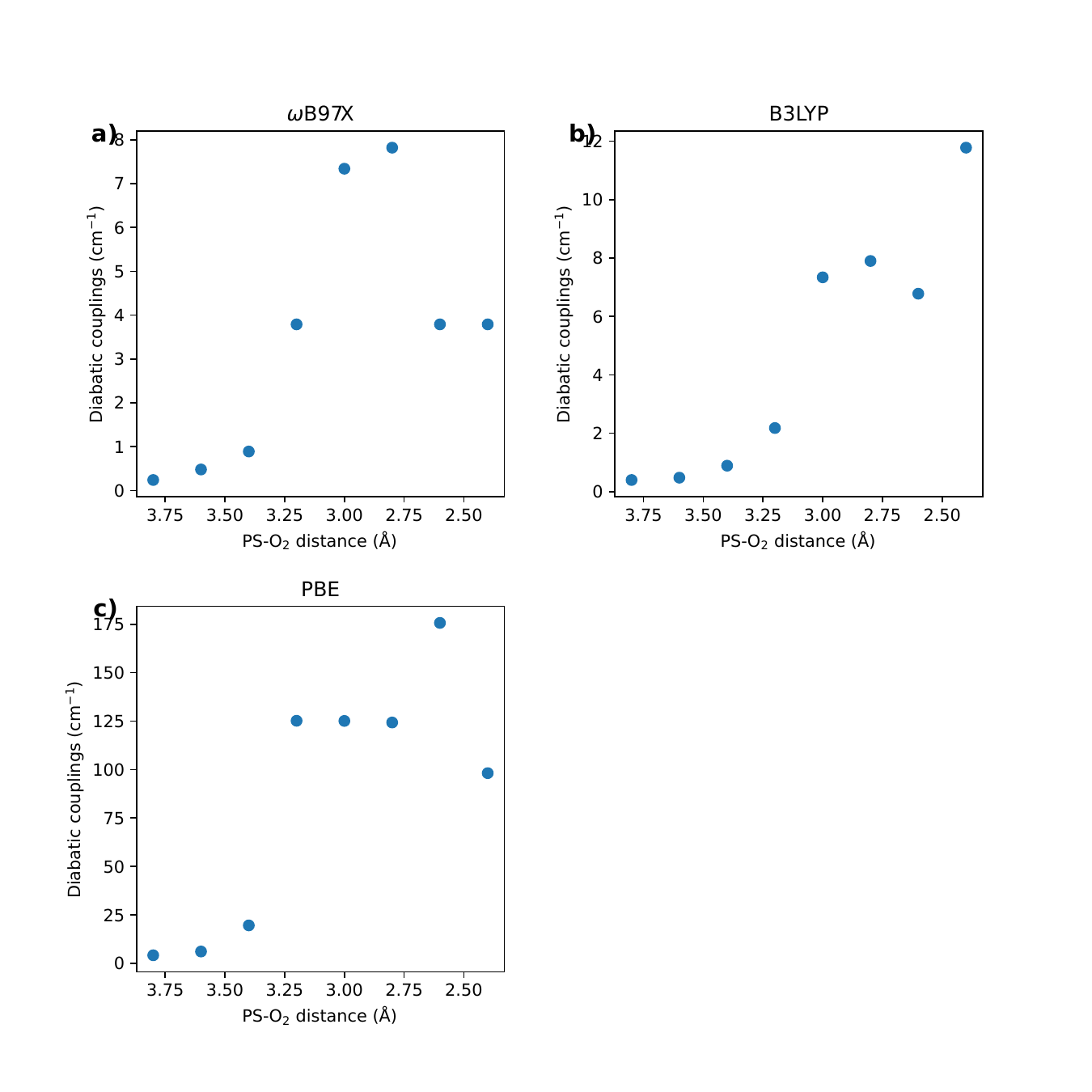}
    \caption{Diabatic couplings (in cm$^{-1}$) between the two singlet electronic configurations $^{1}[^{3}\!PS + (^{3}\Sigma_{g})O_{2}]$ and $^{1}[^{1}\!PS + (^{1}\Delta_{g})O_{2}]$, for all the PS-O$_2$ distances analyzed.
    Calculations with the three functionals.}
    \label{fig:diabatic_couplings}
\end{figure*}

\begin{figure*}[!t]
    \centering
    \includegraphics[width=0.8\textwidth]{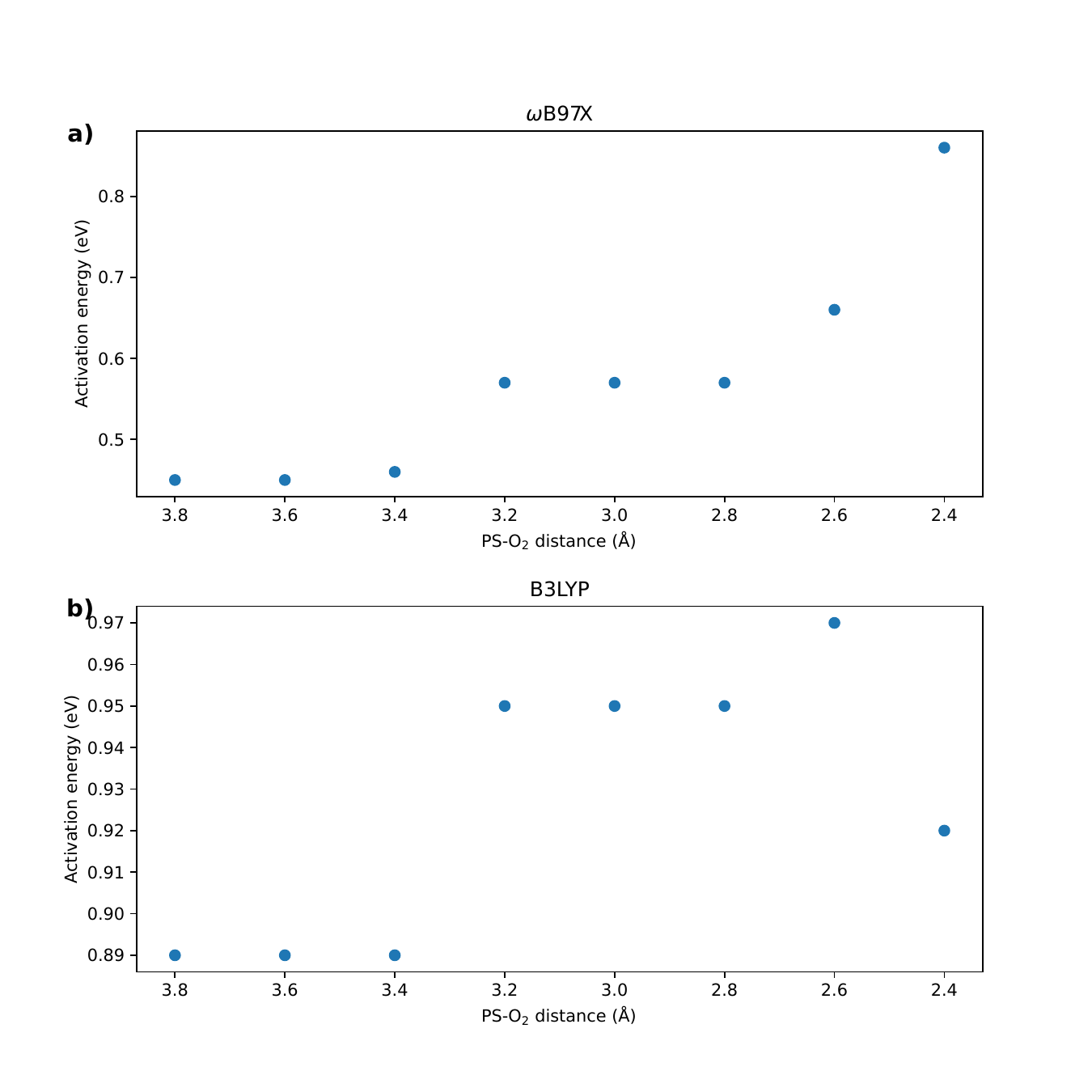}
    \caption{Activation energies (in eV) for all the PS-O$_2$ distances analyzed.
    Top) $\omega$B97X functional and Bottom) B3LYP functional.}
    \label{fig:activation_energies}
\end{figure*}

\begin{table}
\small
  \caption{Reorganization energy (eV), diabatic couplings (cm$^{-1}$), activation energies (eV) and rates (s$^{-1}$, for the $^1\Delta_{g}$ O$_2$ production) for all the calculated PS-O$_2$ distances (in \AA) with the $\omega$B97X functional.}
\label{SI-tab:wb97x_data_for_rates}
    \begin{tabular}{ccccc}
    \hline
    PS-O$_2$ & Reorganization & Diabatic & Activation & Rate\\
    distance & energy & coupling & energy & \\
    \hline3.8  &  0.03 & 0.24 & 0.45 & 3.9\\
    3.6  &  0.03 & 0.48 & 0.45 & 14.3\\
    3.4  &  0.03 & 0.89 & 0.46 & 42.4\\
    3.2  &  0.03 & 3.79 & 0.57 & 11.7\\
    3.0  &  0.03 & 7.34 & 0.57 & 42.2\\
    2.8  &  0.03 & 7.82 & 0.57 & 47.6\\
    2.6  &  0.03 & 3.79 & 0.66 & 3.3E-2\\
    2.4  &  0.03 & 3.79 & 0.86 & 2.2E-4\\
    \hline
    \end{tabular}
\end{table}

\begin{table}
\small
  \caption{Reorganization energy (eV), diabatic couplings (cm$^{-1}$), activation energies (eV) and rates (s$^{-1}$, for the $^1\Delta_{g}$ O$_2$ production) for all the calculated PS-O$_2$ distances (in \AA) with the B3LYP functional.}
\label{SI-tab:b3lyp_data_for_rates}
    \begin{tabular}{ccccc}
    \hline
    PS-O$_2$ & Reorganization & Diabatic & Activation & Rate\\
    distance & energy & coupling & energy & \\
    \hline3.8  &  0.08 & 0.40 & 0.89 & 5.03E-7\\
    3.6  &  0.08 & 0.48 & 0.89 & 6.77E-7\\
    3.4  &  0.08 & 0.89 & 0.89 & 2.03E-6\\
    3.2  &  0.08 & 2.18 & 0.95 & 1.38E-6\\
    3.0  &  0.08 & 7.34 & 0.95 & 1.53E-5\\
    2.8  &  0.08 & 7.90 & 0.95 & 1.76E-5\\
    2.6  &  0.08 & 6.78 & 0.98 & 4.93E-6\\
    2.4  &  0.08 & 11.78 & 0.92 & 1.30E-4\\
    \hline
    \end{tabular}
\end{table}

\begin{figure*}[!t]
    \centering
    \includegraphics[width=0.8\textwidth]{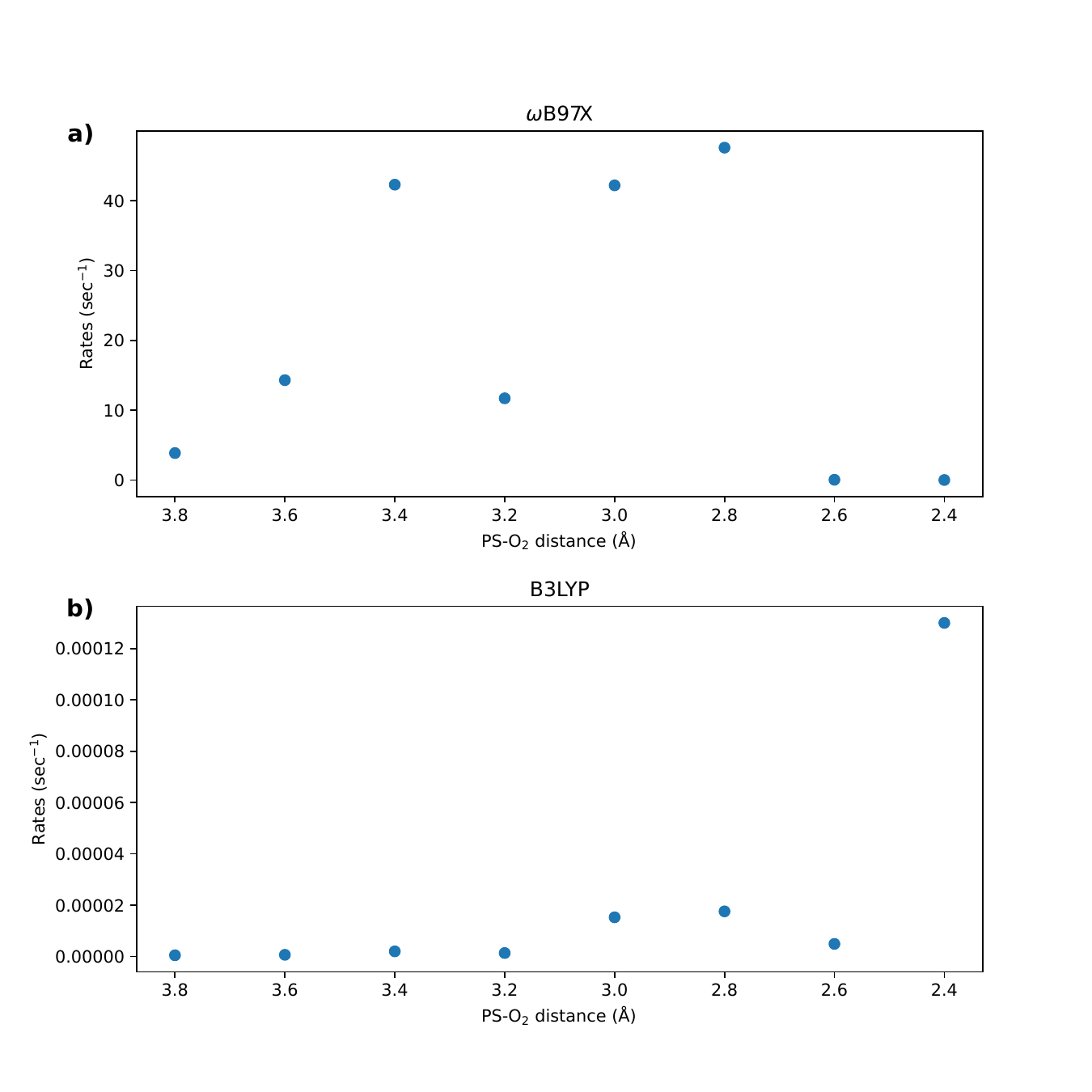}
    \caption{Rates of $^1\Delta_{g}$ O$_2$ production (in s$^{-1}$) with respect to the PS-O$_2$ distance.
    Top) $\omega$B97X functional and Bottom) B3LYP functional.}
    \label{SI-fig:rates}
\end{figure*}

\begin{figure*}[!t]
    \centering
    \includegraphics[width=0.8\textwidth]{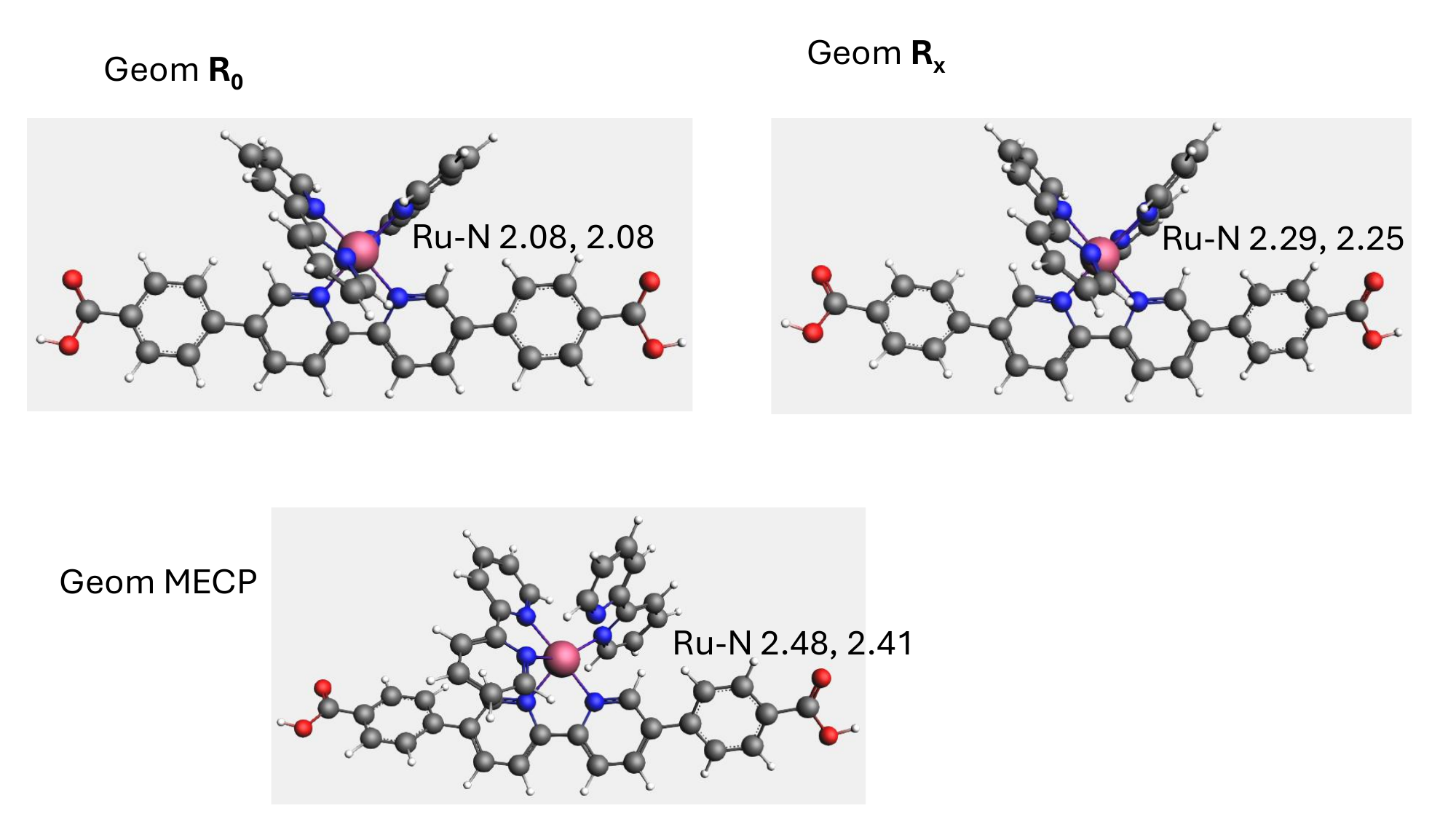}
    \caption{$\mathbf{R}_0$, $\mathbf{R}_x$ and MECP geometries, with the two Ru-N bond lengths that changes the most among the three structures.}
    \label{fig:geom_Rx}
\end{figure*}

\subsection{Tunnelling assessment}\label{subsec:tunnelling}

It is quite possible that nuclear quantum effects (e.g. tunnelling) could manifest in various photochemical processes. \cite{manae2022quantum, MilnikovNakamura2013, manae2024} For example, in photosensitization of thiothymine, it was shown that nuclear quantum effects could speed up the rate by multiple orders of magnitude. \cite{manae2022quantum} In order to see whether nuclear quantum effects would be important, we have chosen to evaluate the tunnelling onset temperature\cite{heller2022, manae2024} $T_0$, defined as
 \begin{align}\label{eq:tunnel-onset}
	T_0^3 = \frac{\hbar^2}{24 m k_B^3} \left(
	\frac{\kappa_T\kappa_S}{\kappa_T - \kappa_S}
	\right)^2,
\end{align}
where $\kappa_T, \kappa_S$ defines the slope of the triplet and singlet electronic states respectively at the minimum energy crossing point.
At temperatures below $T_0$, it can be assumed that tunnelling will begin to be important.
For the photosensitization of H$_2$DBB--Ru, we calculated the onset temperature to be $T_0 = 203 \text{ K}$, which is well below the temperature considered in this study (body temperature, $T = 309 \text{ K}$).
Given that $T \gg T_0$, we can safely assume (within the limits of DFT) that the classical rate expression from Marcus theory (Eq.~\eqref{eq:rate_marcus}) is sufficient for this study.

\subsection{TDDFT excitation energies}\label{subsec:tddft_excitations}

\begin{equation}
\label{eq:get_Rx}
\Delta E^{\text{PS}}_{T-S}(\mathbf{R}_x)
= \Delta E^{\text{PS}}_{T-S}(\mathbf{R}_0)
- \Delta E^{\text{PS--O2}}_{TT-SS}(\mathbf{R}_0, \mathbf{r}_0, D')
\end{equation}

In this equation (equation 14 in ref. \citenum{bai2017divide}), in order to get the $\mathbf{R}_x$ geometry, the triplet-singlet gap of each image has to be calculated, and the one that satisfies the equation will be the $\mathbf{R}_x$ geometry.
The first part in the right side of the equation is the singlet triplet gap at $\mathbf{R}_0$, which is 2.29 eV (with the $\omega$B97X functional).
The second part in the right side of the equation is the energy difference between the $^{1}[^{3}\!PS + (^{3}\Sigma_{g})O_{2}]$ (TT in Eq.~\eqref{eq:get_Rx}) and $^{1}[^{1}\!PS + (^{1}\Sigma_{g}/^{1}\Delta_{g})O_{2}]$ (SS in Eq.~\eqref{eq:get_Rx}) states for a fixed PS-O$_2$ distance (D$^{\prime}$) and fixed PS and O$_2$ geometries at their triplet equilibrium geometries ($\mathbf{R}_0$ and $\mathbf{r}_0$, respectively).
The problem for the $^{1}\Sigma_{g}$ state arises because the TDDFT calculated energy of the $^{1}[^{1}\!PS + (^{1}\Sigma_{g})O_{2}]$ state is 2.72 eV, while the energy of the $^{1}[^{3}\!PS + (^{3}\Sigma_{g})O_{2}]$ state is 2.45 eV (at O$_2$ distance of 2.8 \AA).
Therefore, the last piece of the equation is negative (-0.29 eV), and there is no image between $\mathbf{R}_0$ and MECP that has a S-T gap of 2.58 eV (2.29+0.29 eV), as seen in Supplementary Fig.~\ref{fig:T_S_splittings}.
Therefore, it was not possible to locate $\mathbf{R}_x$ for the $^{1}\Sigma_{g}$ state for any PS-O$_2$ distance.
In the following, we report the TDDFT excitation energies of O$_2$ (without PS) using the three functionals compared with experimental values (Table \ref{tab:O2_TDDFT}) .
We can see that while the $^{1}\Delta_{g}$ state is predicted to be at 1.20-1.3 eV (being 0.2-0.3 eV off with respect to experiment), the $^{1}\Sigma_{g}$ state is predicted to be at 2.2-2.4 eV, which is 0.6-0.8 eV off with respect to experiment.
The large inaccuracy of TDDFT in predicting the $^{1}\Sigma_{g}$ state of O$_2$ causes the inability to calculate activation energies for this state, as previously described.

\begin{table}
\small
  \caption{TDDFT excitation energies of O$_2$ (without PS) calculated with the three functionals and the experimental values.\cite{huber1979constants}
  All energies are in eV.}
\label{tab:O2_TDDFT}
    \begin{tabular}{ccccc}
    \hline
    \hline O$_2$ electronic state & $\omega$B97X & B3LYP & PBE & Experiment\\
    $^{1}\Delta_{g}$  &  1.20 & 1.28 & 1.30 & 0.98 \\
    $^{1}\Sigma_{g}$  &  2.25 & 2.36 & 2.29 & 1.64 \\
    \hline
    \end{tabular}
\end{table}

\subsection{Target geometry ($\mathbf{R}_x$) selection}\label{subsec:rx_calculation}

In this subsection, we show the calculations required to obtain $\mathbf{R}_x$, which is used together with $\mathbf{R}_0$ for the subsequent multiconfiguration and quantum computing calculations.
We invoke Eq.~\eqref{eq:get_Rx} to obtain $\mathbf{R}_x$.
The triplet-singlet gap at $\mathbf{R}_0$ is 2.29 eV (with the $\omega$B97X functional).
The second part in the right side of the equation is the energy difference between the $^{1}[^{3}\!PS + (^{3}\Sigma_{g})O_{2}]$ (TT in Eq.~\eqref{eq:get_Rx}) and $^{1}[^{1}\!PS + (^{1}\Delta_{g})O_{2}]$ (SS in Eq.~\eqref{eq:get_Rx}) states for a fixed PS-O$_2$ distance (D$^{\prime}$) and fixed PS and O$_2$ geometries at their triplet equilibrium geometries ($\mathbf{R}_0$ and $\mathbf{r}_0$, respectively).
This value is 0.86 eV (at PS-O$_2$ distance equal to 2.8 \AA), and therefore the T-S splitting at $\mathbf{R}_x$ has to be 1.43 eV (2.29-0.86 eV).
From Supplementary Fig.~\ref{fig:T_S_splittings} we can see that this gap occurs at image 11, and therefore we selected this geometry as $\mathbf{R}_x$ for the quantum computing calculations.

\subsection{SHCISCF and CASCI calculations}

In this subsection, we first report the orbitals from the (10,10) active-space calculations corresponding to the \textbf{R$_0$} and \textbf{R$_x$} geometries, shown in Supplementary Fig.~\ref{SI-fig:T1_and_Rx_orbs_1010}.
In table \ref{SI-tab:shciscf_data} we report energies and wavefunction analysis from the (6,6) SHCISCF calculations.
In Table \ref{SI-tab:casci_data} we report energies for all the active spaces analyzed using CASCI and CASCI+NEVPT2.

\begin{figure*}[!t]
    \centering
    \includegraphics[width=0.8\textwidth]{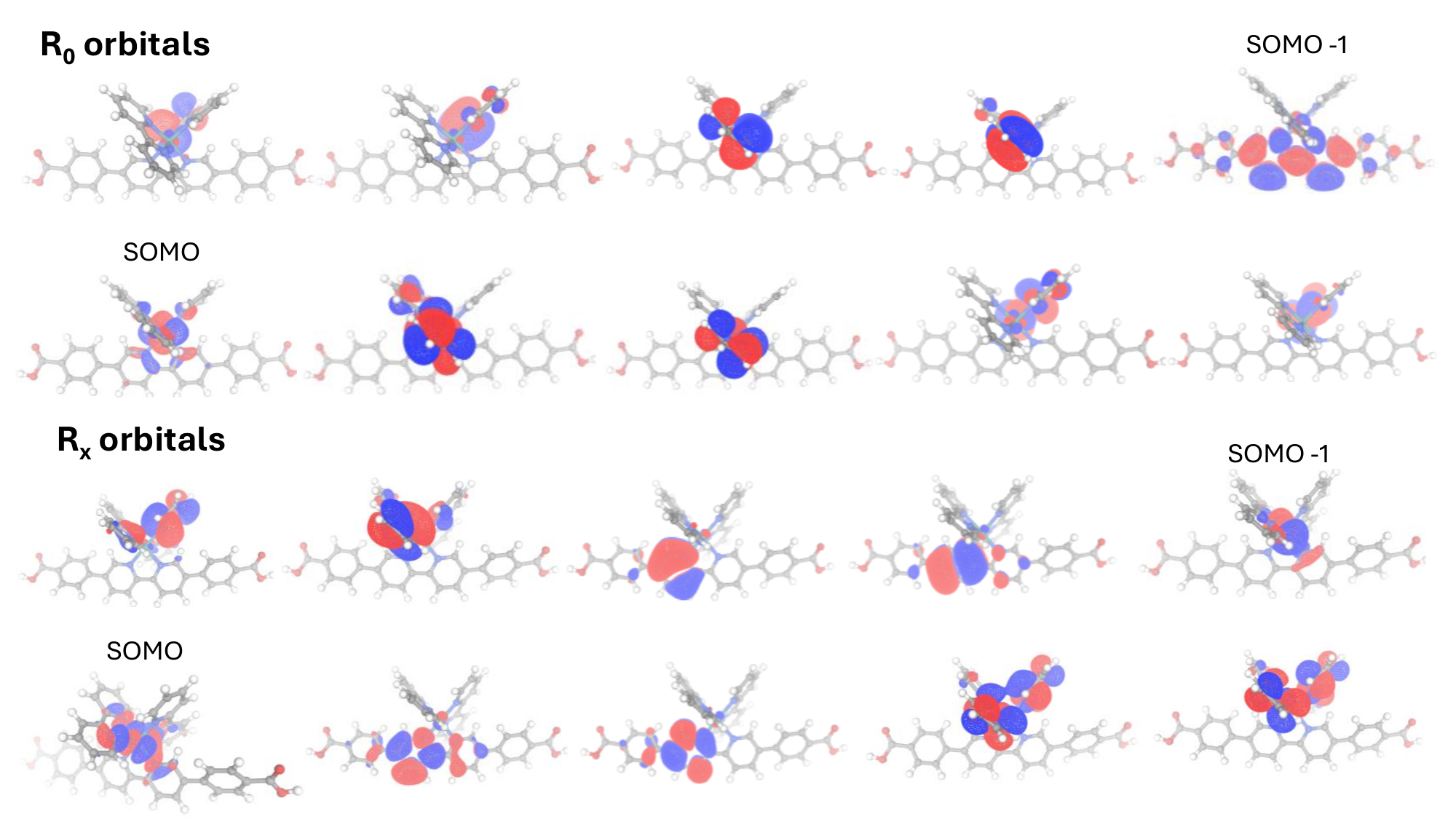}
    \caption{Natural orbitals obtained from the (10,10) SHCISCF calculations corresponding to the $\mathbf{R}_0$ (top) and $\mathbf{R}_x$ (bottom) geometries.
    The SOMO-1 and SOMO are highlighted (singly occupied orbitals).}
    \label{SI-fig:T1_and_Rx_orbs_1010}
\end{figure*}

\begin{table*}
\small
  \caption{Data from SHCISCF (6,6) calculations for $\mathbf{R}_0$ and $\mathbf{R}_x$.
  Molecule, Hartree-Fock energy, SHCISCF energy, difference in energy between SHCISCF and HF ($\Delta$E), active orbitals occupation numbers and first 5 determinants with the respective CI coefficients (wavefunction column). 2 stands for a doubly occupied orbital, 0 for a virtual orbital, a and b for singly occupied orbitals ($\alpha$ and $\beta$ spin, respectively).}
\label{SI-tab:shciscf_data}
  \begin{tabular*}{\textwidth}{@{\extracolsep{\fill}}cccccc}
    \hline
    Molecule & E HF (Ha) & E SHCISCF (Ha) & $\Delta$E (Ha) & Occ. numbers & Wavefunction\\
    \hline
    $\mathbf{R}_0$  & -2404.43803 & -2404.48009 & 0.042 & \begin{tabular}[c]{@{}l@{}}1.933,1.937,1.000, \\ 1.000,0.061,0.070\end{tabular} & \begin{tabular}[c]{@{}l@{}@{}@{}@{}} 22aa00 -0.967\\ 20aa20 0.138 \\ 02aa02 0.123 \\ abaaba 0.062 \\ baaaab 0.062 \end{tabular} \\
    \hline
    $\mathbf{R}_x$  & -2404.44278 & -2404.48636 & 0.044 & \begin{tabular}[c]{@{}l@{}}1.929,1.935,1.000, \\ 1.000,0.062,0.074\end{tabular} & \begin{tabular}[c]{@{}l@{}@{}@{}@{}} 22aa00 -0.966\\ 20aa20 -0.146 \\ 02aa02 -0.127 \\ baaaab 0.075 \\ abaaba 0.075 \end{tabular}\\
    \hline
    \end{tabular*}
\end{table*}

\begin{table*}
\small
  \caption{CASCI energies (in Ha) and CASCI+NEVPT2 energies (in Ha) for different active spaces calculated using the SHCISCF orbitals of the respective active space.
  Absolute energies for both $\mathbf{R}_0$ and $\mathbf{R}_x$ are reported.
  The energy difference $\Delta E_{\mathbf{R}_x-\mathbf{R}_0}$ is reported for CASCI and CASCI+NEVPT2 energies (in Ha).}
\label{SI-tab:casci_data}
  \centering
  \begin{tabular}{llrrr}
    \hline
    Method & Active space & $E_{\mathbf{R}_0}$ (Ha) & $E_{\mathbf{R}_x}$ (Ha) & $\Delta E_{\mathbf{R}_x-\mathbf{R}_0}$ (Ha) \\
    \hline
    CASCI & (6,6) & -2404.481168 & -2404.487426 & -0.0063 \\
    CASCI & (8,8) & -2404.499432 & -2404.505779 & -0.0063 \\
    CASCI & (10,10) & -2404.524434 & -2404.524315 & 0.0001 \\
    CASCI+NEVPT2 & (6,6) & -2412.281304 & -2412.245936 & 0.0354 \\
    CASCI+NEVPT2 & (8,8) & -2412.272536 & -2412.237553 & 0.0350 \\
    CASCI+NEVPT2 & (10,10) & -2412.276388 & -2412.229338 & 0.0470 \\
    \hline
  \end{tabular}
\end{table*}

\subsection{Electronic-structure computational details} \label{comp_details}

All DFT calculations were performed with the ADF 2025.1 package.\cite{baerends2025amsterdam}
All geometry optimizations were carried out with the PBE functional \cite{perdew1996generalized} and the TZP (for Ru) and DZP (for the other atoms) basis set with the D3 dispersion correction and BJ damping factor.\cite{grimme2010consistent,grimme2011effect}
The scalar relativistic Zero Order Regular Approximation (ZORA) hamiltonian has been employed \cite{van1996relativistic,van1999geometry}, together with Core Large and NumericalQuality Good.
The carbon atoms of the COOH groups were frozen during the optimizations.
For DFT single point calculations, we employed three functionals, namely $\omega$B97X, \cite{chai2008systematic} B3LYP, \cite{stephens1994ab} and PBE functionals. We used the same computational details as for the geometry optimizations, except for the use of Core None.
Excitation energies and diabatic couplings were calculated with the same details as the DFT single points, together with the keywords TDA (Tamm-Dancoff Approximation), SFTDDFT (Spin-Flip TDDFT), FOCDFT (Fragment Orbital Constrained DFT) and ExcitonTransfer.
Two fragments were considered for these calculations, which were the PS and O$_2$ molecules.
The DBB-Ru MOF was optimized using the UFF4MOF II force field as implemented in ADF.\cite{coupry2016extension}
SHCISCF calculations were performed with the Dice implementation \cite{smith2017cheap} in the PySCF package \cite{sun2018pyscf,sun2020recent} using the crenbl basis set and the crenbl electron core potential (ECP) for Ru \cite{fernandez1985ab,ermler1991ab} and the 6-31G** basis set for the other atoms.
Three sweep iterations with $\epsilon$=[0.01,0.001,0.0001] were used.
CASCI and NEVPT2 calculations were carried out with PySCF and same basis set as SHCISCF.
All hamiltonians used for subsequent quantum computing calculations have been generated with the InQuanto software platform.~\cite{inquanto}

\FloatBarrier

\section{Mixed-estimator truncation and state preparation}
\label{SI-sec:mixed_estimator_truncation}

\subsection{The \texorpdfstring{$(m,k,l)$}{(m,k,l)} state-size convention}

We use the $(m,k,l)$ convention of the main text, where $m$ and $k$ are the numbers of determinants in the reference and residual states, while $l$ is the number of pp-tUPS layers or sMCS determinants.
Let $\ket{\Psi}=\sum_{x\in S_\Psi}c_x\ket{x}$ be the exact CASCI eigenstate, i.e. the exact eigenstate within the chosen active space.
The set $S_m\subset S_\Psi$ contains its $m$ largest-amplitude determinants.
The reference state is
\begin{equation}
\ket{\Phi_m}=\sum_{x\in S_m}a_x^{(m)}\ket{x},
\end{equation}
where the coefficients $a_x^{(m)}$ are obtained by diagonalizing $H$ in the subspace spanned by $S_m$.
We then define
\begin{equation}
Q_{\Phi_m}=I-\ket{\Phi_m}\bra{\Phi_m},
\end{equation}
\begin{equation}
r_{\Phi_m}=\norm{Q_{\Phi_m}H\ket{\Phi_m}},
\end{equation}
\begin{equation}
\ket{\Lambda_m}=\frac{Q_{\Phi_m}H\ket{\Phi_m}}{r_{\Phi_m}}.
\end{equation}
The set $S_{m,k}\subset S_{\Lambda_m}$ contains the $k$ largest-amplitude determinants of $\ket{\Lambda_m}$, while $S_l\subset S_\Psi$ contains the $l$ largest-amplitude determinants of $\ket{\Psi}$.
Their normalized truncations are
\begin{equation}
\ket{\Lambda_{m,k}}
=
\frac{\sum_{x\in S_{m,k}}\braket{x|\Lambda_m}\ket{x}}
{\sqrt{\sum_{x\in S_{m,k}}|\braket{x|\Lambda_m}|^2}},
\label{eq:chi_k_def_correct}
\end{equation}
\begin{equation}
\ket{\Psi_l}
=
\frac{\sum_{x\in S_l}c_x\ket{x}}
{\sqrt{\sum_{x\in S_l}|c_x|^2}}.
\end{equation}
With $E_{\Phi_m}=\mel{\Phi_m}{H}{\Phi_m}$, $h_{m,k,l}=\braket{\Psi_l|\Lambda_{m,k}}$, and $s_{m,l}=\braket{\Psi_l|\Phi_m}$, the truncated mixed estimator is
\begin{equation}
E_{\text{ME}}^{(m,k,l)}
=
E_{\Phi_m}
+
r_{\Phi_m}\frac{h_{m,k,l}}{s_{m,l}}.
\end{equation}
For fixed $m$ and $l$, define $\delta_{m,k}=\norm{\ket{\Lambda_{m,k}}-\ket{\Lambda_m}}$.
The error introduced by residual truncation obeys
\begin{equation}
\left|E_{\text{ME}}^{(m,k,l)}-E_{\text{ME}}^{(m,|S_{\Lambda_m}|,l)}\right|
\leq
\frac{r_{\Phi_m}}{|s_{m,l}|}\delta_{m,k}.
\end{equation}
Thus $r_{\Phi_m}\delta_{m,k}/|s_{m,l}|<\epsilon$ is a conservative stopping condition for a target energy accuracy $\epsilon$.
The distance $\delta_{m,k}$ is obtained classically from the cumulative coefficient weight discarded from $\ket{\Lambda_m}$.
This bound assumes the worst possible alignment with the target state, while the scans below evaluate the actual mixed-estimator error and can therefore support smaller values of $k$.
For a fixed reference size $m$, taking $k=|S_{\Lambda_m}|$ and $l=|S_\Psi|$ recovers the exact residual and CASCI target states for sMCS preparation.
In general practice $|S_\Psi|$ is the complete active space, however in practice we included lower number of determinants than that, that had no numerically significant impact on the statevector calculations.

\subsection{sMCS state-preparation convergence}

\begin{figure}
        \centering
        \includegraphics[width=0.95\linewidth]{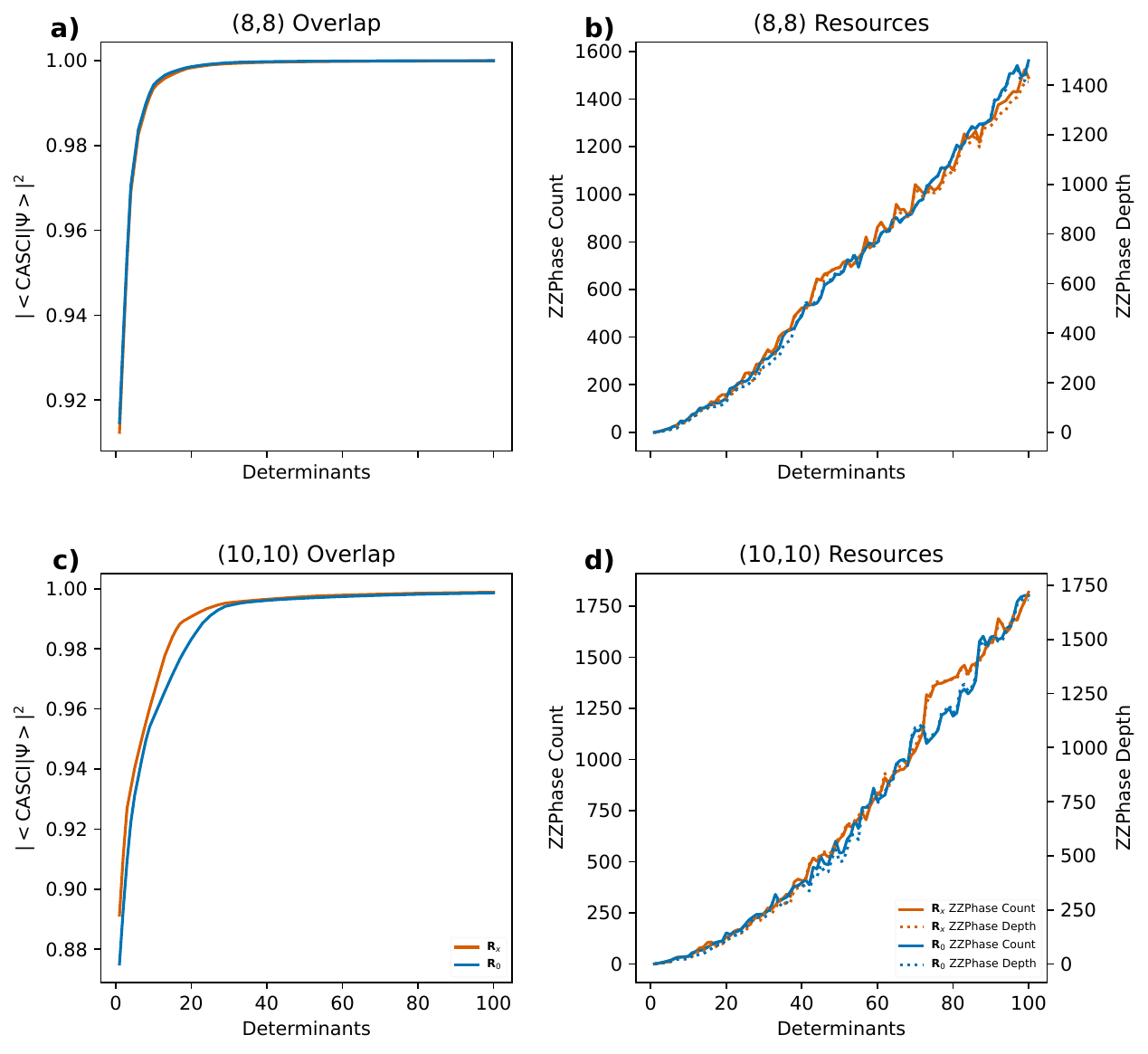}
        \caption{sMCS state-preparation quality and circuit resources for the (8,8) and (10,10) ASs.
        \textbf{a)}, \textbf{b)} show the squared overlap between the (8,8) CASCI and sMCS states and the corresponding two-qubit sMCS resources.
        \textbf{c)}, \textbf{d)} are analogous for the (10,10) sMCS states.
        Orange and blue correspond to the $\mathbf{R}_x$ and $\mathbf{R}_0$ geometries.
        Two-qubit gate count is shown as a solid line, with dotted corresponding to depth.}
        \label{fig:mc_cas881010_stateprep}
    \end{figure}

Supplementary Fig.~\ref{fig:mc_cas881010_stateprep} shows that the sMCS overlap with the CASCI state increases monotonically with the number of retained determinants, while the compiled ZZPhase-gate count and depth also increase. Convergence is slower in the (10,10) active space than in the (8,8) active space.

\FloatBarrier

\subsection{Mixed-estimator truncation scans}

    \begin{figure}
        \centering
        \includegraphics[width=0.95\linewidth]{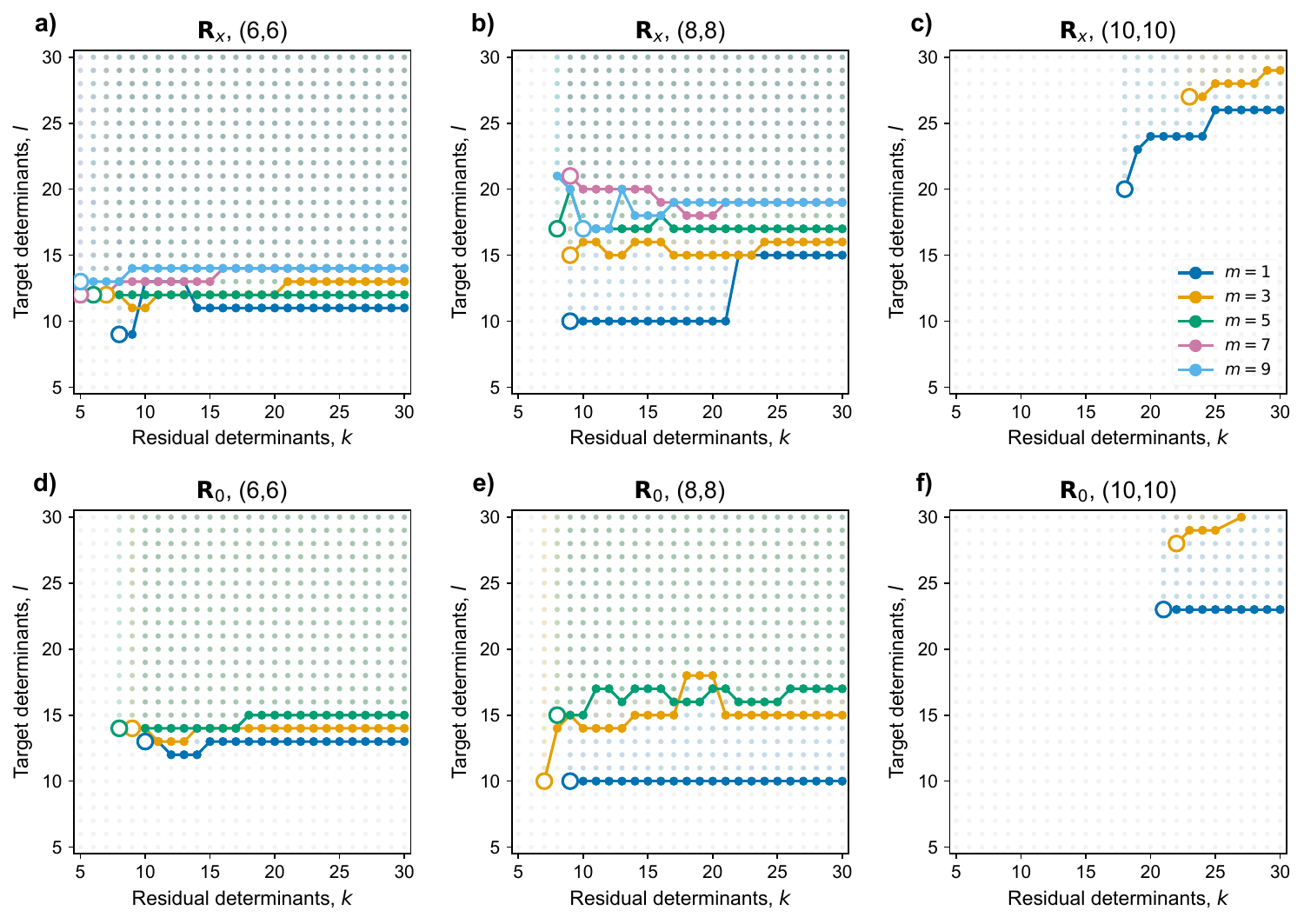}
        \caption{Mixed-estimator truncation landscapes for $\mathbf{R}_x$ (top) and $\mathbf{R}_0$ (bottom) in the (6,6), (8,8), and (10,10) active spaces from left to right.
        The axes give the residual-state size $k$ and sMCS target-state size $l$.
        The coloured curves mark the $1.6$-mHa error contour for each reference size $m$, and open circles identify the smallest $k+l$ on each contour.}
        \label{fig:mixed_estimator_truncation}
\end{figure}
\FloatBarrier

Supplementary Fig.~\ref{fig:mixed_estimator_truncation} shows that the residual- and target-state sizes required to meet the energy criterion depend jointly on the reference size and need not vary monotonically.
The determinant supports therefore cannot generally be selected using independent truncation thresholds.
This coupled landscape motivates joint optimization of the state supports and their compiled circuits.

\subsection{Joint support and circuit optimization}

For the (10,10) active space, we used AI-assisted search to optimize the determinant supports and circuit compilation together, as detailed under ``AI-assisted state and circuit optimization'' in Methods.
Rather than using the common $(m,k,l)=(1,21,24)$ choice for both geometries, the search retained this support for $\mathbf{R}_x$ but selected a smaller $(1,21,22)$ support for $\mathbf{R}_0$, formed from the union of the reference- and residual-state determinant supports.
This reduced $l$ from 24 to 22 for $\mathbf{R}_0$.
The selected circuit used the identity spin-orbital mapping at $\mathbf{R}_x$.
At $\mathbf{R}_0$, the selected spin-orbital permutation was
\[
(13,9,3,5,18,17,8,10,16,19,4,12,15,14,11,6,0,1,7,2).
\]
This permutation was recorded under the internal candidate identifier \texttt{seed\_006}.

We compared the resulting sparse multiconfiguration circuits with the tUPS circuits for the common $(1,21,24)$ choice.
Each mixed-estimator energy requires two compute--uncompute circuits, one for the reference overlap and one for the residual overlap.
The target state $\ket{\Psi_l}$ is prepared in both circuits, while the single-determinant reference $\ket{\Phi_1}$ requires no ZZPhase gates.
The state-preparation cost for each geometry is therefore $2G(\Psi_l)+G(\Lambda_{1,k})$, with the depth counted over the same circuit blocks.

Summed over both geometries, the compiled ZZPhase-gate count decreased from 857 to 755, a reduction of approximately $12\%$.
For $\mathbf{R}_x$, the count decreased from 424 to 417 and the two-qubit depth from 357 to 344.
For $\mathbf{R}_0$, the count decreased from 433 to 338 and the depth from 351 to 262.
The corresponding resource and Helios-1E emulator comparison is shown in Supplementary Fig.~\ref{fig:helios1e_1010_stateprep}.

\begin{figure}
    \centering
    \includegraphics[width=0.95\linewidth]{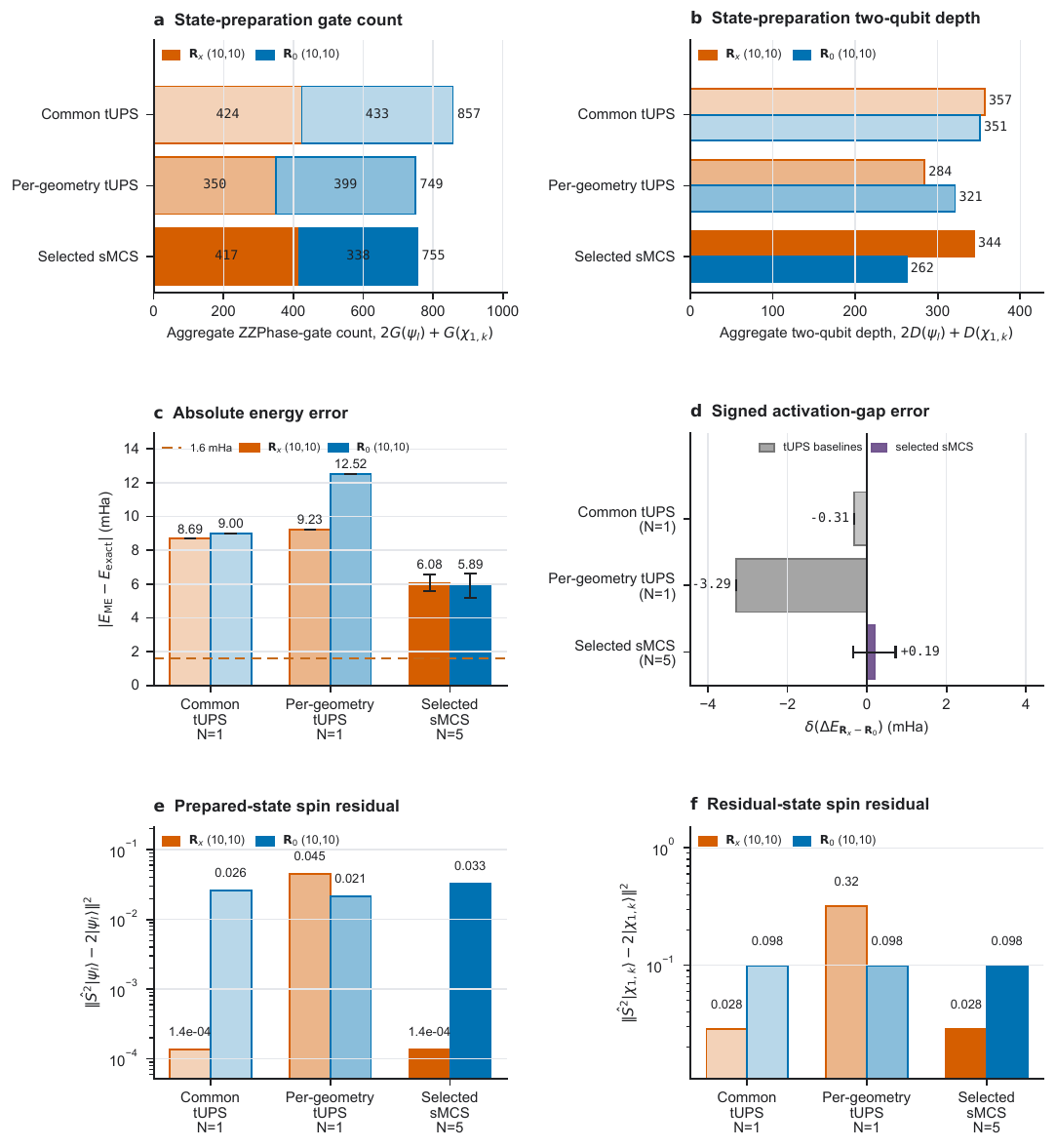}
    \caption{Helios-1E comparison of state-preparation strategies for the (10,10) active space.
    \textbf{a}, Aggregate compiled ZZPhase-gate count for the $\mathbf{R}_x$ and $\mathbf{R}_0$ mixed-estimator circuits.
    \textbf{b}, Corresponding compiled two-qubit depths.
    \textbf{c}, Absolute mixed-estimator energy errors relative to CASCI; the dashed line marks $1.6$~mHa.
    \textbf{d}, Signed error in the energy difference $E_{\mathbf{R}_x}-E_{\mathbf{R}_0}$.
    \textbf{e}, Prepared-state spin residual.
    \textbf{f}, Residual-state spin residual.
    Red and teal denote $\mathbf{R}_x$ and $\mathbf{R}_0$, respectively.
    The selected sparse multiconfiguration results show the mean and one standard deviation over five emulator repeats in which both geometries were evaluated within the same campaign.
    The tUPS baselines are single emulator evaluations and therefore do not include statistical error bars.}
    \label{fig:helios1e_1010_stateprep}
\end{figure}
\FloatBarrier

On Helios-1E, the selected sparse multiconfiguration pair gave absolute energy errors of $6.08\pm0.50$~mHa for $\mathbf{R}_x$ and $5.89\pm0.71$~mHa for $\mathbf{R}_0$, where the uncertainties denote one standard deviation across the five emulator repeats.
The signed error in $E_{\mathbf{R}_x}-E_{\mathbf{R}_0}$ was $0.19\pm0.53$~mHa and was therefore consistent with zero within the observed emulator variability.
The individual-state errors were numerically lower than those from the single-run common and geometry-specific pp-tUPS baselines; however, the absence of repeated baseline measurements prevents a formal statistical comparison.
Together with the reduction in compiled state-preparation resources, these results motivated carrying this state pair forward to the subsequent symmetry-based error-detection study.

\section{Overlap-measurement circuits}
\label{SI-sec:overlap_measurement_circuits}

The protocols parameterized by $\ket{\Omega}$ are applied separately with $\ket{\Omega}=\ket{\Phi}$ and $\ket{\Omega}=\ket{\Lambda}$ to obtain the two contractions entering the mixed estimator.
FCU and CS instead obtain both contractions from the same set of measurement outcomes.
Throughout, $U_{\alpha}\ket{0}^{\otimes n}=\ket{\alpha}$ denotes preparation of state $\ket{\alpha}$.

\subsection{Real signed-overlap circuits}

The real Hadamard test (RHT) and real interferometric-swap test (RIST) return an ancilla bit $a\in\{0,1\}$.
For either circuit, the single-shot estimator is
\begin{equation}
X_{\Omega}=(-1)^a,
\qquad
\mathbb{E}[X_{\Omega}]=\operatorname{Re}\braket{\Psi|\Omega}.
\end{equation}
Running the circuit separately for $\Omega=\Phi$ and $\Omega=\Lambda$ therefore estimates $s$ and $h$, including their signs.

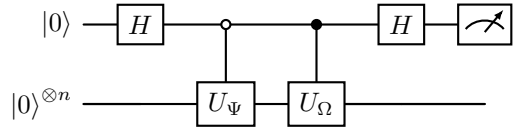
\begin{figure}[htbp]
\centering
\begin{quantikz}[column sep=0.45cm]
\lstick{$\ket{0}$} & \gate{H} & \octrl{1} & \ctrl{1} & \gate{H} & \meter{} \\
\lstick{$\ket{0}^{\otimes n}$} & \qw & \gate{U_{\Psi}} & \gate{U_{\Omega}} & \qw & \qw
\end{quantikz}
\caption{
Hadamard-test circuit for real-overlap estimation.
The state-preparation boxes denote $U_{\Psi}\ket{0}^{\otimes n}=\ket{\Psi}$ and $U_{\Omega}\ket{0}^{\otimes n}=\ket{\Omega}$, where $\Omega=\Phi$ or $\Omega=\Lambda$.
}
\label{fig:hadamard_circuit}
\end{figure}

\begin{figure}[htbp]
\centering
\begin{quantikz}[column sep=0.45cm]
\lstick{$\ket{0}$} & \gate{H} & \octrl{1} & \ctrl{2} & \ctrl{1} & \gate{H} & \meter{} \\
\lstick{$\ket{0}^{\otimes n}$} & \qw & \gate{U_{\Psi}} & \qw & \swap{1} & \qw & \qw \\
\lstick{$\ket{0}^{\otimes n}$} & \qw & \qw & \gate{U_{\Omega}} & \targX{} & \qw & \qw
\end{quantikz}
\caption{
Interferometric swap-test circuit for estimating $\mathrm{Re}\braket{\Psi|\Omega}$.
The open-controlled branch prepares $U_{\Psi}\ket{0}^{\otimes n}=\ket{\Psi}$ on the first register, the closed-controlled branch prepares $U_{\Omega}\ket{0}^{\otimes n}=\ket{\Omega}$ on the second register, and the ancilla-controlled SWAP brings the two branches onto the same register before the final Hadamard and measurement.
}
\label{fig:interferometric_swap_circuit}
\end{figure}

\FloatBarrier

The fan-out implementation in Supplementary Fig.~\ref{fig:fanout_cswap_circuit} has the same ancilla outcome and therefore uses the same estimator $X_{\Omega}=(-1)^a$.
The additional control qubits allow the controlled-SWAP operations to be applied in parallel, reducing the controlled-SWAP-layer depth and, when parallel gates are supported, the circuit duration.

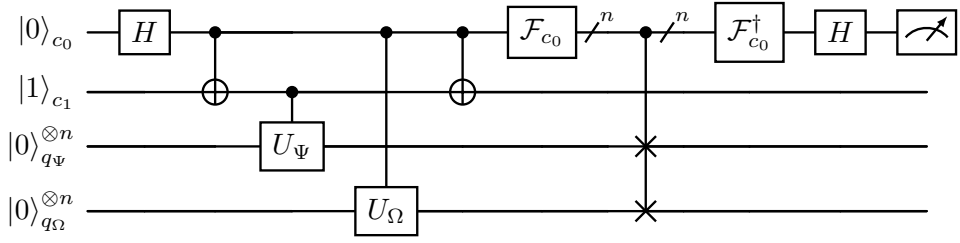
\begin{figure}[htbp]
\centering
\resizebox{\columnwidth}{!}{%
\begin{quantikz}[column sep=0.38cm,row sep=0.20cm]
\lstick{$\ket{0}_{c_0}$} & \gate{H} & \ctrl{1} & \qw & \ctrl{3} & \ctrl{1} & \gate{\mathcal{F}_{c_0}} & \qwbundle{n} & \ctrl{2} & \qwbundle{n} & \gate{\mathcal{F}_{c_0}^{\dagger}} & \gate{H} & \meter{} \\
\lstick{$\ket{1}_{c_1}$} & \qw & \targ{} & \ctrl{1} & \qw & \targ{} & \qw & \qw & \qw & \qw & \qw & \qw & \qw \\
\lstick{$\ket{0}^{\otimes n}_{q_\Psi}$} & \qw & \qw & \gate{U_{\Psi}} & \qw & \qw & \qw & \qw & \swap{1} & \qw & \qw & \qw & \qw \\
\lstick{$\ket{0}^{\otimes n}_{q_\Omega}$} & \qw & \qw & \qw & \gate{U_{\Omega}} & \qw & \qw & \qw & \targX{} & \qw & \qw & \qw & \qw
\end{quantikz}
}
\caption{
Interferometric swap-test circuit with separate branch-control fanout and fanout-parallelized controlled-SWAP.
The first CX gate prepares $c_1=\bar{c}_0$, so $c_1$ controls $U_\Psi$ while $c_0$ controls $U_\Omega$; the second CX gate uncomputes $c_1$.
The operation $\mathcal{F}_{c_0}$ is drawn as a one-wire to $n$-wire fanout, $\mathcal{F}_{c_0}^{\dagger}$ recombines the bundle, and the grouped controlled-SWAP column denotes $\mathrm{CSWAP}_{\parallel}$ between corresponding qubits of the $q_\Psi$ and $q_\Omega$ registers.
}
\label{fig:fanout_cswap_circuit}
\end{figure}

\FloatBarrier

\subsection{Absolute-squared-overlap circuits}

Compute--uncompute (CU), the standard swap test (SST), and the destructive swap test (DST) estimate $|\braket{\Psi|\Omega}|^2$.
Let $b$ be the CU system-register outcome, $a$ the SST ancilla outcome, and $u,v\in\{0,1\}^n$ the two DST register outcomes after the Bell-basis rotation.
Their single-shot estimators are
\begin{equation}
X_{\Omega}^{\mathrm{CU}}
=
\begin{cases}
1, & b=0^n,\\
0, & \text{otherwise}.
\end{cases}
\end{equation}
\begin{equation}
X_{\Omega}^{\mathrm{SST}}=(-1)^a,
\end{equation}
\begin{equation}
X_{\Omega}^{\mathrm{DST}}=(-1)^{\sum_{j=1}^{n}u_jv_j}.
\end{equation}
To see how the swap-test estimators arise, define
\begin{equation}
p_{\Omega}=|\braket{\Psi|\Omega}|^2.
\end{equation}
For SST, the ancilla probabilities are
\begin{equation}
P(a=0)=\frac{1+p_{\Omega}}{2},
\qquad
P(a=1)=\frac{1-p_{\Omega}}{2}.
\end{equation}
Consequently,
\begin{equation}
\mathbb{E}[X_{\Omega}^{\mathrm{SST}}]
=
P(a=0)-P(a=1)
=
p_{\Omega}.
\end{equation}
For DST, the Bell-basis rotation diagonalizes the SWAP operator on each qubit pair.
The outcome $(u_j,v_j)=(1,1)$ is the antisymmetric outcome and contributes eigenvalue $-1$, while the other three outcomes contribute $+1$.
The product over all pairs is therefore $(-1)^{\sum_j u_jv_j}$, and
\begin{equation}
\mathbb{E}[X_{\Omega}^{\mathrm{DST}}]
=
\mel{\Psi\otimes\Omega}{\operatorname{SWAP}}{\Psi\otimes\Omega}
=
|\braket{\Psi|\Omega}|^2
=
p_{\Omega}.
\end{equation}
The CU all-zero indicator likewise has expectation $p_{\Omega}$.
Two independently sampled circuits with $\Omega=\Phi$ and $\Omega=\Lambda$ therefore give $|s|^2$ and $|h|^2$, respectively.

Although $p_{\Omega}\geq0$, individual SST and DST outcomes are $\pm1$, and their finite-shot sample mean can be negative when the overlap is poorly resolved.

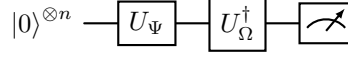
\begin{figure}[htbp]
\centering
\begin{quantikz}[column sep=0.45cm]
\lstick{$\ket{0}^{\otimes n}$} & \gate{U_{\Psi}} & \gate{U_{\Omega}^{\dagger}} & \meter{}
\end{quantikz}
\caption{
Compute-uncompute circuit for estimating $|\braket{\Psi|\Omega}|^2$ from the all-zero measurement probability.
Here $U_{\Psi}\ket{0}^{\otimes n}=\ket{\Psi}$ and $U_{\Omega}\ket{0}^{\otimes n}=\ket{\Omega}$, with $\Omega=\Phi$ or $\Omega=\Lambda$.
}
\label{fig:compute_uncompute_circuit}
\end{figure}

\begin{figure}[htbp]
\centering
\begin{quantikz}[column sep=0.45cm]
\lstick{$\ket{0}$} & \gate{H} & \ctrl{1} & \gate{H} & \meter{} \\
\lstick{$\ket{0}^{\otimes n}$} & \gate{U_{\Psi}} & \swap{1} & \qw & \qw \\
\lstick{$\ket{0}^{\otimes n}$} & \gate{U_{\Omega}} & \targX{} & \qw & \qw
\end{quantikz}
\caption{
Standard swap-test circuit for estimating $|\braket{\Psi|\Omega}|^2$.
The state-preparation boxes denote $U_{\Psi}\ket{0}^{\otimes n}=\ket{\Psi}$ and $U_{\Omega}\ket{0}^{\otimes n}=\ket{\Omega}$, where $\Omega=\Phi$ or $\Omega=\Lambda$.
}
\label{fig:swap_test_circuit}
\end{figure}
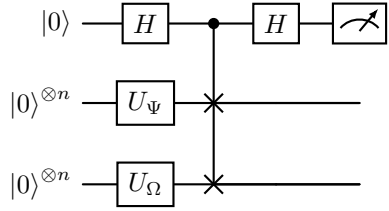

\begin{figure}[htbp]
\centering
\begin{quantikz}[column sep=0.45cm]
\lstick{$\ket{0}^{\otimes n}$} & \gate{U_{\Psi}} & \ctrl{1} & \gate{H^{\otimes n}} & \meter{} \\
\lstick{$\ket{0}^{\otimes n}$} & \gate{U_{\Omega}} & \targ{} & \qw & \meter{}
\end{quantikz}
\caption{
Destructive-swap-test circuit for estimating $|\braket{\Psi|\Omega}|^2$ by Bell-basis measurement of paired qubits.
The state-preparation boxes denote $U_{\Psi}\ket{0}^{\otimes n}=\ket{\Psi}$ and $U_{\Omega}\ket{0}^{\otimes n}=\ket{\Omega}$, where $\Omega=\Phi$ or $\Omega=\Lambda$.
}
\label{fig:destructive_swap_circuit}
\end{figure}
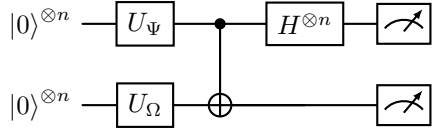

\FloatBarrier

When $\braket{\Phi|\Lambda}=0$, flagged compute--uncompute (FCU), shown in main-text Fig.~\mainref{fig:four_qubit_fcu_circuit}, estimates $|s|^2$ and $|h|^2$ from two disjoint events in the outcomes of the same circuit.
If $f_{\Phi}$ is the flag outcome and $o_{\Lambda}$ is the system outcome after uncomputing $\ket{\Lambda}$, the binary estimators are

\begin{align}
    X_q
&=
\begin{cases}
1, & f_{\Phi}=1,\\
0, & \text{otherwise},
\end{cases}\\
X_p
&=
\begin{cases}
1, & f_{\Phi}=0\text{ and }o_{\Lambda}=0^n,\\
0, & \text{otherwise},
\end{cases}
\end{align}

with $\mathbb{E}[X_q]=|s|^2$ and $\mathbb{E}[X_p]=|h|^2$.

\subsection{Classical-shadow measurements}

For classical shadows (CS), each shot applies a sampled global Clifford $U_i$ to $\ket{\tau}=\left(\ket{0}^{\otimes n}+\ket{\Psi}\right)/\sqrt{2}$ and returns a computational-basis outcome $b$.
With $d=2^n$, the record $(U_i,b)$ gives the single-shot shadow
\begin{equation}
\hat{\rho}_{\tau}=(d+1)U_i^{\dagger}\ket{b}\bra{b}U_i-I.
\end{equation}
For $O_x=\ket{0^n}\bra{x}+\ket{x}\bra{0^n}$, the corresponding real-amplitude estimator is
\begin{equation}
\hat{c}_x=\operatorname{Tr}\left(O_x\hat{\rho}_{\tau}\right),
\qquad
\mathbb{E}[\hat{c}_x]=\braket{x|\Psi}.
\end{equation}
The same shadow records are contracted with the sparse coefficients of $\ket{\Phi}$ and $\ket{\Lambda}$ to reconstruct $s$ and $h$.

\begin{figure}[htbp]
\centering
\begin{quantikz}[column sep=0.45cm]
\lstick{$\ket{0}^{\otimes n}$} & \gate{U_{\tau}} & \gate{U_i} & \meter{}
\end{quantikz}
\caption{
Classical-shadow measurement circuit.
The state-preparation box denotes $U_{\tau}\ket{0}^{\otimes n}=\ket{\tau}$ with $\ket{\tau}=\left(\ket{0}^{\otimes n}+\ket{\Psi}\right)/\sqrt{2}$, and $U_i$ is a randomly sampled Clifford circuit.
}
\label{fig:classical_shadows_circuit}
\end{figure}
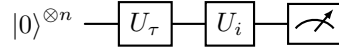

\FloatBarrier

\section{Mixed-estimator statistical analysis}
\label{SI-sec:si_mixed_estimator_details}

\subsection{Covariance propagation for mixed-estimator measurements}
\label{subsec:si_covariance_recipe}

Suppose a collection of independently sampled quantum circuits is used to estimate a vector of quantities $z$.
Its components may include real or imaginary parts of overlaps, squared overlap magnitudes, or expectation values of grouped Pauli operators.
Let $X_\ell^{(c)}$ denote the random outcome of shot $\ell$ from circuit $c$, which is sampled $N_c$ times.
The measurement protocol defines a function $g_j^{(c)}$ that maps each outcome $X^{(c)}$ to a single-shot estimator of $z_j$.
We assume explicitly that this estimator is unbiased,
\begin{equation}
\mathbb{E}\left[g_j^{(c)}(X^{(c)})\right]=z_j.
\end{equation}
The corresponding sample mean is
\begin{equation}
\hat z_j=
\frac{1}{N_c}
\sum_{\ell=1}^{N_c}
g_j^{(c)}\left(X_\ell^{(c)}\right).
\end{equation}
Let $C_z=\operatorname{Cov}(\hat z)$ denote the covariance matrix after the chosen shot allocation.
For two quantities estimated from the same circuit,
\begin{equation}
\left[C_z\right]_{jk}
=
\operatorname{Cov}(\hat z_j,\hat z_k)
=
\frac{\operatorname{Cov}\left(g_j^{(c)}(X^{(c)}),g_k^{(c)}(X^{(c)})\right)}{N_c}.
\end{equation}
Quantities estimated from independently sampled circuits have zero covariance, so $C_z$ is block diagonal with one block for each circuit.
If a reported quantity combines outputs from several circuits, those outputs are included as separate components of $z$ and combined through the final estimator.

Given a total budget $N_{\mathrm{shot}}$, the circuit shot counts satisfy
\begin{equation}
\sum_c N_c=N_{\mathrm{shot}},
\end{equation}
and are chosen to minimize the leading variance of the final estimator.
At fixed allocation fractions $N_c/N_{\mathrm{shot}}$, the covariance matrix scales as $C_z=\mathcal{O}(N_{\mathrm{shot}}^{-1})$.

For any smooth scalar estimator $F(z)$, such as the mixed estimator, the multivariate delta method gives the leading large-shot variance
\begin{equation}
\sigma_F^2
=
\operatorname{Var}[F(\hat z)]
=
\nabla F(z)^{T}
C_z
\nabla F(z)
+
\mathcal{O}\left(N_{\mathrm{shot}}^{-2}\right).
\label{eq:variance_from_gradient}
\end{equation}
The measurement protocol and the shot allocation determine the entries of $C_z$.
Even though every component of $\hat z$ is unbiased, a nonlinear estimator $F(\hat z)$ is generally biased because averaging and nonlinear functions do not commute.
This statistical bias is the difference between the average finite-shot estimate and the exact estimator value,
\begin{equation}
\operatorname{Bias}[F(\hat z)]
=
\mathbb{E}[F(\hat z)]-F(z)
\end{equation}
and is distinct from hardware-noise and model-error biases.
A Taylor expansion gives the leading statistical bias
\begin{equation}
\operatorname{Bias}[F(\hat z)]
=
\frac{1}{2}
\operatorname{Tr}\left(H_F C_z\right)
+
\mathcal{O}\left(N_{\mathrm{shot}}^{-2}\right),
\end{equation}
where $H_F$ is the Hessian of $F$ evaluated at $z$, and all shot counts are scaled together.
If needed, the bias can be estimated from the same covariance matrix or by resampling.

\subsection{Grouped Pauli-averaging baseline}
\label{SI-sec:pauli_averaging}

To establish the expected sampling cost of direct energy measurement, we used grouped Pauli averaging (PA) as a baseline.
For each geometry and active space, we calculated the group variances exactly on the corresponding CASCI state and then determined the ideal shot count required for a per-state standard error of $1.6$~mHa.
Thus, the reported counts are variance-based estimates rather than Monte Carlo estimates and do not include hardware noise or circuit errors.

The identity contribution was removed from the Jordan--Wigner Hamiltonian because it shifts the energy deterministically and has zero measurement variance.
The remaining Pauli terms were sorted by decreasing coefficient magnitude and partitioned into qubit-wise-commuting (QWC) groups.
We used pytket's \texttt{NonConflictingSets} partition strategy with the
\texttt{Lazy} graph-colouring method.

For group $g$, we evaluated
\begin{equation}
\sigma_g^2
=
\mel{\Psi}{H_g^2}{\Psi}
-
\mel{\Psi}{H_g}{\Psi}^2
=
\norm{H_g\ket{\Psi}}^2
-
\mel{\Psi}{H_g}{\Psi}^2
\end{equation}
using sparse Zixy operator application and inner products.
This includes all within-group covariances.

If $N_g$ shots are assigned to group $g$, the total PA variance is
\begin{equation}
\epsilon_{\mathrm{PA}}^2
=
\sum_g\frac{\sigma_g^2}{N_g},
\end{equation}
which is with variance-minimizing allocation is, $N_g\propto\sigma_g$, giving
\begin{equation}
N_{\mathrm{PA}}
=
\frac{\left(\sum_g\sigma_g\right)^2}{\epsilon^2}.
\label{SI-eq:pauli_averaging_shots}
\end{equation}
Supplementary Table~\ref{SI-tab:pauli_averaging_shots} reports this result for a CASCI target state and standard error of $\epsilon=1.6$~mHa.

\begin{table}[t]
\centering
\small
\caption{Grouped Pauli-averaging results.
$N_P$ is the number of nonidentity Pauli terms, $N_g$ is the number of coefficient-sorted QWC groups, and the shot count is $N_{\mathrm{PA}}$ for $\epsilon=1.6$~mHa.}
\label{SI-tab:pauli_averaging_shots}
\begin{tabular}{@{}lccrrr@{}}
\toprule
Geometry & Active space & Qubits & $N_P$ & $N_g$ & Shots \\
\midrule
$\mathbf{R}_x$ & $(6,6)$   & 12 & 1818  & 596  & $694\,194$  \\
$\mathbf{R}_0$ & $(6,6)$   & 12 & 1818  & 599  & $798\,146$  \\
$\mathbf{R}_x$ & $(8,8)$   & 16 & 5792  & 1877 & $1\,205\,657$ \\
$\mathbf{R}_0$ & $(8,8)$   & 16 & 5792  & 1840 & $962\,712$  \\
$\mathbf{R}_x$ & $(10,10)$ & 20 & 14250 & 4636 & $3\,754\,694$ \\
$\mathbf{R}_0$ & $(10,10)$ & 20 & 14250 & 4630 & $2\,522\,182$ \\
\bottomrule
\end{tabular}
\end{table}

\subsection{Real signed overlaps}

This case covers circuits whose outcomes give unbiased signed estimates of real overlaps,
\begin{equation}
h=\operatorname{Re}\braket{\Psi|\Lambda},
\qquad
s=\operatorname{Re}\braket{\Psi|\Phi}.
\end{equation}
For the real-valued states considered here, these overlaps determine the mixed estimator.
Therefore
\begin{equation}
z
=
\begin{bmatrix}
h\\
s
\end{bmatrix}
\end{equation}
and
\begin{equation}
F(z)
=
E_{\Phi}
+
r_{\Phi}\frac{h}{s}.
\end{equation}
The gradient and Hessian are
\begin{equation}
\nabla F
=
r_{\Phi}
\begin{bmatrix}
s^{-1}\\
-hs^{-2}
\end{bmatrix},
\end{equation}
\begin{equation}
H_{F}
=
r_{\Phi}
\begin{bmatrix}
0 & -s^{-2}\\
-s^{-2} & 2hs^{-3}
\end{bmatrix}.
\end{equation}

The relevant circuits are the real Hadamard test and the real interferometric-swap test shown in Supplementary Figs.~\ref{fig:hadamard_circuit} and \ref{fig:interferometric_swap_circuit}, including the fan-out RIST implementation in Supplementary Fig.~\ref{fig:fanout_cswap_circuit}.
The independent single-shot estimators from the two overlap circuits, $X_h,X_s\in\{-1,1\}$, satisfy
\begin{equation}
\mathbb{E}[X_h]=h,
\qquad
\mathbb{E}[X_s]=s,
\end{equation}
with variances
\begin{equation}
\operatorname{Var}(X_h)=1-h^2,
\qquad
\operatorname{Var}(X_s)=1-s^2.
\end{equation}
Since $h$ and $s$ are measured with independent circuits using $N_h$ and $N_s$ shots, respectively,
\begin{equation}
C_{z}
=
\begin{bmatrix}
\frac{1-h^2}{N_h} & 0\\
0 & \frac{1-s^2}{N_s}
\end{bmatrix},
\end{equation}
and Eq.~\eqref{eq:variance_from_gradient} gives
\begin{equation}
\sigma_F^2
=
r_{\Phi}^2
\left(
\frac{1-h^2}{s^2N_h}
+
\frac{h^2(1-s^2)}{s^4N_s}
\right)
\end{equation}
For nonzero real $x$ and $y$, define
\begin{equation*}
A(x,y)
=
\frac{\sqrt{1-x^2}}{|x|}
+
\frac{\sqrt{1-y^2}}{|y|}.
\end{equation*}
With the optimal allocation of $N_{\mathrm{shot}}=N_h+N_s$, the variance becomes
\begin{equation}
\sigma_{\mathrm{RHT}}^2
=
\sigma_{\mathrm{RIST}}^2
=
\frac{r_{\Phi}^2}{N_{\mathrm{shot}}}
\left|\frac{h}{s}\right|^2
A(h,s)^2.
\end{equation}

\subsection{Absolute-squared overlaps}

This case covers circuits whose outcomes give unbiased estimates of absolute-squared overlaps,
\begin{equation}
p=|h|^2,
\qquad
q=|s|^2.
\end{equation}
For the ground-state mixed estimator, the sign of the correction is known.
After choosing the reference phase so that $s>0$, the correction is negative, so the absolute-squared overlaps are sufficient.
Therefore
\begin{equation}
z
=
\begin{bmatrix}
p\\
q
\end{bmatrix}
\end{equation}
and
\begin{equation}
F(z)
=
E_{\Phi}
-
r_{\Phi}
\sqrt{\frac{p}{q}}.
\end{equation}
The gradient and Hessian directly follow as
\begin{equation}
\nabla F
=
-\frac{r_{\Phi}}{2}
\begin{bmatrix}
\left(pq\right)^{-1/2}\\
-p^{1/2}q^{-3/2}
\end{bmatrix}.
\end{equation}
\begin{equation}
H_F
=
-\frac{r_{\Phi}}{4}
\begin{bmatrix}
-p^{-3/2}q^{-1/2} & -p^{-1/2}q^{-3/2}\\
-p^{-1/2}q^{-3/2} & 3p^{1/2}q^{-5/2}
\end{bmatrix}.
\end{equation}

The relevant circuits are compute--uncompute, the swap test, and the destructive-swap test shown in Supplementary Figs.~\ref{fig:compute_uncompute_circuit}--\ref{fig:destructive_swap_circuit}.
For compute--uncompute, the independent random variables from the outcome of the two overlap circuits, $X_p,X_q\in\{0,1\}$, satisfy
\begin{equation}
\mathbb{E}[X_p]=p,
\qquad
\mathbb{E}[X_q]=q,
\end{equation}
and the variances
\begin{equation}
\operatorname{Var}(X_p)=p(1-p),
\qquad
\operatorname{Var}(X_q)=q(1-q).
\end{equation}
For swap-test or destructive-swap parity estimators, $X_p,X_q\in\{-1,1\}$ satisfy the same expectation equations, but the variances are
\begin{equation}
\operatorname{Var}(X_p)=1-p^2,
\qquad
\operatorname{Var}(X_q)=1-q^2.
\end{equation}
Since $p$ and $q$ are measured with independent circuits using $N_p$ and $N_q$ shots, respectively,
\begin{equation}
C_z
=
\begin{bmatrix}
\frac{\operatorname{Var}(X_p)}{N_p} & 0\\
0 & \frac{\operatorname{Var}(X_q)}{N_q}
\end{bmatrix},
\end{equation}
Equation~\eqref{eq:variance_from_gradient} gives
\begin{equation}
\sigma_F^2
=
\frac{r_{\Phi}^2}{4}
\left(
\frac{\operatorname{Var}(X_p)}{pqN_p}
+
\frac{p\,\operatorname{Var}(X_q)}{q^3N_q}
\right)
\end{equation}
and the optimal allocation of $N_{\mathrm{shot}}=N_p+N_q$ gives
\begin{equation}
\sigma_F^2
=
\frac{r_{\Phi}^2}{4N_{\mathrm{shot}}}
\frac{p}{q}
\left(
\frac{\sqrt{\operatorname{Var}(X_p)}}{p}
+
\frac{\sqrt{\operatorname{Var}(X_q)}}{q}
\right)^2.
\end{equation}
Using $p/q=|h/s|^2$ and the same $A(x,y)$ defined above, for compute-uncompute this becomes
\begin{equation}
\sigma_{\mathrm{CU}}^2
=
\frac{r_{\Phi}^2}{4N_{\mathrm{shot}}}
\left|\frac{h}{s}\right|^2
A(h,s)^2.
\end{equation}
For swap-test or destructive-swap parity estimators this becomes
\begin{equation}
\sigma_{\mathrm{SST}}^2
=
\sigma_{\mathrm{DST}}^2
=
\frac{r_{\Phi}^2}{4N_{\mathrm{shot}}}
\left|\frac{h}{s}\right|^2
A\left(|h|^2,|s|^2\right)^2.
\end{equation}

\subsection{Flagged compute--uncompute}

Flagged compute--uncompute, shown in main-text Fig.~\mainref{fig:four_qubit_fcu_circuit}, is a special absolute-squared-overlap protocol that requires
\begin{equation}
\braket{\Phi|\Lambda}=0.
\end{equation}
After preparing $\ket{\Psi}$, the circuit flags the exact $\ket{\Phi}$ component and uncomputes the unflagged branch against $\ket{\Lambda}$.
A single circuit therefore estimates both overlaps.
The two relevant disjoint events are the $\Phi$ flag event and the $\Lambda$ uncompute event,
\begin{equation}
f_{\Phi}=1,
\qquad
o_{\Lambda}=0^n,\ f_{\Phi}=0,
\end{equation}
where $o_{\Lambda}$ is the output bit string after uncomputing $\ket{\Lambda}$.
These events have probabilities
\begin{equation}
p=P\left(o_{\Lambda}=0^n,\ f_{\Phi}=0\right)=|h|^2,
\end{equation}
\begin{equation}
q=P\left(f_{\Phi}=1\right)=|s|^2.
\end{equation}
Therefore
\begin{equation}
z
=
\begin{bmatrix}
p\\
q
\end{bmatrix}
\end{equation}
and, using the same ground-state sign convention as above,
\begin{equation}
F(z)
=
E_{\Phi}
-
r_{\Phi}
\sqrt{\frac{p}{q}}.
\end{equation}
The gradient and Hessian directly follow as
\begin{equation}
\nabla F
=
-\frac{r_{\Phi}}{2}
\begin{bmatrix}
\left(pq\right)^{-1/2}\\
-p^{1/2}q^{-3/2}
\end{bmatrix}.
\end{equation}
\begin{equation}
H_F
=
-\frac{r_{\Phi}}{4}
\begin{bmatrix}
-p^{-3/2}q^{-1/2} & -p^{-1/2}q^{-3/2}\\
-p^{-1/2}q^{-3/2} & 3p^{1/2}q^{-5/2}
\end{bmatrix}.
\end{equation}

Let $X_p$ and $X_q$ be the two binary variables from one flagged compute-uncompute shot.
The possible records are
\begin{equation}
\left(X_p,X_q\right)
=
\begin{cases}
\left(1,0\right), & o_{\Lambda}=0^n,\ f_{\Phi}=0,\\
\left(0,1\right), & f_{\Phi}=1,\\
\left(0,0\right), & \mathrm{otherwise},\\
\left(1,1\right), & \mathrm{never}.
\end{cases}
\end{equation}
The corresponding probabilities are $p$, $q$, and $1-p-q$.
Thus
\begin{equation}
\mathbb{E}[X_p]=p,
\qquad
\mathbb{E}[X_q]=q,
\qquad
X_pX_q=0.
\end{equation}
The last identity holds shot by shot because the flag event and the uncompute event are mutually exclusive.
In contrast to two independent compute-uncompute circuits, the two estimates are therefore negatively correlated.
The individual Bernoulli variances are
\begin{equation}
\operatorname{Var}(X_p)=p(1-p),
\qquad
\operatorname{Var}(X_q)=q(1-q),
\end{equation}
while the covariance is
\begin{equation}
\operatorname{Cov}(X_p,X_q)
=
\mathbb{E}[X_pX_q]
-
\mathbb{E}[X_p]\mathbb{E}[X_q]
=
-pq.
\end{equation}
For $N_{\mathrm{shot}}$ shots, this gives
\begin{equation}
C_z
=
\frac{1}{N_{\mathrm{shot}}}
\begin{bmatrix}
p(1-p) & -pq\\
-pq & q(1-q)
\end{bmatrix}.
\end{equation}
There is no separate allocation between numerator and denominator shots in this protocol, because both probabilities are estimated from the same shot record.
Using Eq.~\eqref{eq:variance_from_gradient} gives
\begin{equation}
\sigma_F^2
=
\nabla F^T
C_z
\nabla F
=
\frac{r_{\Phi}^2}{4N_{\mathrm{shot}}}
\frac{p+q}{q^2}.
\end{equation}
In terms of $h$ and $s$, the FCU variance is therefore
\begin{equation}
\sigma_{\mathrm{FCU}}^2
=
\frac{r_{\Phi}^2}{4N_{\mathrm{shot}}}
\frac{|h|^2+|s|^2}{|s|^4}.
\end{equation}

\subsection{Classical-shadow overlap reconstruction}

\subsubsection{Linear contraction into overlaps}

This case covers protocols where a measured real-amplitude vector is first contracted into the final overlap vector.
Classical shadows are the main example.
We assume real state coefficients throughout, so $h$ and $s$ are real.
Write the states as
\begin{equation}
\begin{aligned}
\ket{\Psi}
&=
\sum_{x\in S_\Psi} c_x\ket{x},
\\
\ket{\Lambda}
&=
\sum_{x\in S_\Lambda} b_x\ket{x},
\\
\ket{\Phi}
&=
\sum_{x\in S_\Phi} a_x\ket{x}.
\end{aligned}
\end{equation}
The coefficients are taken to be zero outside their own supports.
Let $S=S_\Lambda\cup S_\Phi$ be the set of amplitudes of $\ket{\Psi}$ that are needed for the two overlaps.
Then
\begin{equation}
h
=
\sum_{x\in S} b_xc_x,
\qquad
s
=
\sum_{x\in S} a_xc_x.
\end{equation}
Let $\hat c_x$ be the measured estimate of $c_x$.
The measured overlaps are then
\begin{equation}
\hat h
=
\sum_{x\in S} b_x\hat c_x,
\qquad
\hat s
=
\sum_{x\in S} a_x\hat c_x.
\end{equation}
Thus the covariance matrix for $\hat z=[\hat h,\hat s]^T$ is obtained directly from the covariance of the amplitude estimates.
Its entries are
\begin{equation}
\begin{aligned}
\left[C_z\right]_{hh}
&=
\sum_{x,y\in S} b_xb_y\operatorname{Cov}(\hat c_x,\hat c_y),
\\
\left[C_z\right]_{hs}
&=
\sum_{x,y\in S} b_xa_y\operatorname{Cov}(\hat c_x,\hat c_y),
\\
\left[C_z\right]_{ss}
&=
\sum_{x,y\in S} a_xa_y\operatorname{Cov}(\hat c_x,\hat c_y).
\end{aligned}
\end{equation}
After this contraction, the scalar estimator is the real mixed estimator $F(z)$.

\subsubsection{Classical-shadow covariance for the mixed estimator}

This case covers the real-amplitude classical-shadow estimator shown in Supplementary Fig.~\ref{fig:classical_shadows_circuit}.
Let
\begin{equation}
\ket{\Psi}
=
\sum_{x\in S_\Psi} c_x\ket{x},
\qquad
\sum_{x\in S_\Psi}c_x^2=1,
\end{equation}
where $S_\Psi$ is the support of the prepared state.
The requested amplitude set is $S=S_\Lambda\cup S_\Phi$, as in the previous subsection, and $|S|$ is its size.
The coefficient $c_x$ is zero when $x\notin S_\Psi$.
The shadow state is
\begin{equation}
\ket{\tau}
=
\frac{1}{\sqrt{2}}
\left(
\ket{0}^{\otimes n}
+
\ket{\Psi}
\right),
\end{equation}
where $\ket{0}^{\otimes n}$ is chosen outside the relevant supports, and $d=2^n$.
For every $x\in S$, define
\begin{equation}
O_x
=
\ket{0^n}\bra{x}
+
\ket{x}\bra{0^n}.
\end{equation}
These operators extract the real amplitudes because
\begin{equation}
\mel{\tau}{O_x}{\tau}
=
c_x.
\end{equation}

A single shadow sample applies a Clifford $U_i$, measures a bit string $b$, and forms the shadow
\begin{equation}
\hat{\rho}_{\tau}
=
(d+1)U_i^{\dagger}\ket{b}\bra{b}U_i-I.
\end{equation}
The amplitude estimator is
\begin{equation}
\hat{c}_x
=
\operatorname{Tr}
\left(
O_x\hat{\rho}_{\tau}
\right),
\end{equation}
and it is unbiased
\begin{equation}
\mathbb{E}[\hat{c}_x]
=
c_x,
\end{equation} where $\mathbb{E}$ refers to the expectation value with all possible shadows.
In order to calculate $C_z$ we need to calculate the covariance
\begin{equation}
\begin{aligned}
\operatorname{Cov}(\hat c_x,\hat c_y)
&=
\frac{1}{N_{\mathrm{shot}}}
\Bigg(
\mathbb{E}\left[
\operatorname{Tr}(O_x\hat{\rho}_{\tau})
\operatorname{Tr}(O_y\hat{\rho}_{\tau})
\right]
\\
&\quad
-
\mathbb{E}\left[
\operatorname{Tr}(O_x\hat{\rho}_{\tau})
\right]
\mathbb{E}\left[
\operatorname{Tr}(O_y\hat{\rho}_{\tau})
\right]
\Bigg)
\\
&=
\frac{1}{N_{\mathrm{shot}}}
\left(
\operatorname{Tr}\left(
(O_x\otimes O_y)
\mathbb{E}\left[
\hat{\rho}_{\tau}\otimes\hat{\rho}_{\tau}
\right]
\right)
-
c_xc_y
\right).
\end{aligned}
\end{equation}
The tensor moment $\mathbb{E}[\hat{\rho}_{\tau}\otimes\hat{\rho}_{\tau}]$ is evaluated using the 3-design properties of the Clifford group, as detailed in Supplementary Section~\ref{subsec:shadow_tensor_evaluation}.
The result is
\begin{equation}
\begin{aligned}
\mathbb{E}\left[
\hat{\rho}_{\tau}\otimes\hat{\rho}_{\tau}
\right]
&=
\frac{d+1}{d+2}
\left(
F
+
F(\rho_{\tau}\otimes\mathbb{I})
+
(\rho_{\tau}\otimes\mathbb{I})F
\right)
\\
&\quad
-
\frac{1}{d+2}
\left(
\mathbb{I}\otimes\mathbb{I}
+
\rho_{\tau}\otimes\mathbb{I}
+
\mathbb{I}\otimes\rho_{\tau}
\right),
\end{aligned}
\end{equation}
where $d=2^n$ and $F$ is the swap operator on the two copies of the Hilbert space, that is $F\ket{a}\otimes\ket{b}=\ket{b}\otimes\ket{a}$.
By using \begin{equation}
\begin{aligned}
\operatorname{Tr}\left((A\otimes B)F\right)
&=
\sum_{a,b}
\left(\bra{a}\otimes\bra{b}\right)
(A\otimes B)
\left(\ket{b}\otimes\ket{a}\right)
\\
&=
\sum_{a,b}
\bra{a}A\ket{b}
\bra{b}B\ket{a}
\\
&=
\operatorname{Tr}(AB)
\end{aligned}
\end{equation} the trace can be evaluated as
\begin{equation}
\begin{aligned}
&\operatorname{Tr}\left(
(O_x\otimes O_y)
\mathbb{E}\left[
\hat{\rho}_{\tau}\otimes\hat{\rho}_{\tau}
\right]
\right)
\\
&\quad =
\frac{d+1}{d+2}
\left(
\operatorname{Tr}(O_xO_y)
+
\operatorname{Tr}(O_xO_y\rho_{\tau})
+
\operatorname{Tr}(O_x\rho_{\tau}O_y)
\right)
\\
&\quad =
\frac{d+1}{d+2}
\left(
3\delta_{xy}+c_xc_y
\right),
\end{aligned}
\end{equation}
where we used that
\begin{equation}
O_xO_y=
\delta_{xy}\ket{0}\bra{0}
+
\ket{x}\bra{y},
\end{equation}
therefore $\operatorname{Tr}(O_xO_y)=2\delta_{xy}$.
The other terms are
\begin{equation}
\begin{aligned}
\operatorname{Tr}(O_xO_y\rho_{\tau})
&=
\mel{\tau}{O_xO_y}{\tau}
\\
&=
\delta_{xy}\abs{\braket{0|\tau}}^2
+
\braket{\tau|x}\braket{y|\tau}
\\
&=
\frac{1}{2}\left(\delta_{xy}+c_xc_y\right).
\end{aligned}
\end{equation}
By substituting, covariance
\begin{equation}
\operatorname{Cov}(\hat c_x,\hat c_y)
=
\frac{1}{N_{\mathrm{shot}}}
\left(
\frac{3(d+1)}{d+2}\delta_{xy}
-
\frac{1}{d+2}c_xc_y
\right).
\end{equation}

Contracting the amplitude covariance gives

\begin{equation}
\begin{aligned}
\left[C_z\right]_{hh}
&=
\frac{1}{N_{\mathrm{shot}}}
\left(
\frac{3(d+1)}{d+2}
-
\frac{h^2}{d+2}
\right) \approx \frac{3}{N_{\mathrm{shot}}},
\\
\left[C_z\right]_{hs}
&=
\frac{1}{N_{\mathrm{shot}}}
\left(
\frac{3(d+1)}{d+2}\braket{\Lambda|\Phi}
-
\frac{hs}{d+2}
\right) \approx 0,
\\
\left[C_z\right]_{ss}
&=
\frac{1}{N_{\mathrm{shot}}}
\left(
\frac{3(d+1)}{d+2}
-
\frac{s^2}{d+2}
\right) \approx \frac{3}{N_{\mathrm{shot}}}.
\end{aligned}
\end{equation}
In the large-$d$ limit, the resulting mixed-estimator variance is
\begin{equation}
\sigma_{\mathrm{CS}}^2
\approx
\frac{3r_{\Phi}^2}{N_{\mathrm{shot}}}
\frac{|h|^2+|s|^2}{|s|^4}.
\label{eq:cs_variance}
\end{equation}

\subsubsection{Evaluation of $\mathbb{E}[\hat{\rho}\otimes\hat{\rho}]$}
\label{subsec:shadow_tensor_evaluation}
Following the classical shadow protocol, the expectation value of the tensor is defined as the expectation over sampling $U$ from the Clifford group and $\ket{b}$ measured from the state $U\rho U^{\dagger}$, that is
\begin{align}
&\mathbb{E}[\hat{\rho}\otimes\hat{\rho}] = \label{eqn:tensor_shadow_exp}\nonumber\\
&\mathbb{E}_U\Bigg[\sum_b \operatorname{Tr}(U\rho U^{\dagger} \ket{b}\bra{b}) \Bigg\{(2^n +1)^2(U^{\dagger}\ket{b}\bra{b}U) \otimes (U^{\dagger}\ket{b}\bra{b}U)\\
&- (2^n + 1)  \bigg((U^{\dagger}\ket{b}\bra{b}U) \otimes \mathbb{I} + \mathbb{I} \otimes  (U^{\dagger}\ket{b}\bra{b}U)  \bigg) + \mathbb{I}\otimes \mathbb{I} \Bigg\}\Bigg].\nonumber
\end{align}
As $\mathbb{E}_U \big[ \sum_b \operatorname{Tr}(U\rho U^{\dagger} \ket{b}\bra{b}) (U^{\dagger}\ket{b}\bra{b}U)\big] = \frac{\rho + \mathbb{I}}{(2^n + 1)}$ (eq. S35 in ref. \cite{Huang2020}) we only need to focus on the first term, which is
\begin{align}
    &\mathbb{E}_U \Bigg[\sum_b \operatorname{Tr}(U\rho U^{\dagger} \ket{b}\bra{b})(U^{\dagger}\ket{b}\bra{b}U) \otimes (U^{\dagger}\ket{b}\bra{b}U)\nonumber \Bigg]\\
    &= \sum_b \mathbb{E}_U\Bigg[ \operatorname{Tr}_3\bigg( (\mathbb{I}\otimes\mathbb{I}\otimes\rho) (U^{\dagger}\ket{0}\bra{0}U)^{\otimes3} \bigg) \Bigg] \label{eq:first_term_worked_out_clifford_bijection}\\
    &=  2^n\operatorname{Tr}_3\bigg( (\mathbb{I}\otimes\mathbb{I}\otimes\rho) \mathbb{E}_U\Big[(U^{\dagger}\ket{0}\bra{0}U)^{\otimes3}\Big] \bigg)\nonumber,
\end{align}
where $\operatorname{Tr}_3$ means a partial trace over the third constituent of the tensor product space and we have used the fact there exists a Clifford group element $C$ such that $\ket{b} = C\ket{0}$ and multiplication by a group element is a bijection on a group.
Then using eq. S77 from ref. \cite{Huang2020} we get that
\begin{equation}
    \mathbb{E}_U\Big[(U^{\dagger}\ket{0}\bra{0}U)^{\otimes3}\Big] = \binom{2^n + 2}{3}^{-1} P_{\vee^3},
\end{equation}
with $P_{\vee^3}$ the projector onto the totally symmetric subspace of $\mathbb{C}^d\otimes\mathbb{C}^d\otimes\mathbb{C}^d$.
$P_{\vee^3}$ can be expanded as the sum over all permutations of the tensor subspaces \cite{Huang2020}
\begin{equation}
    P_{\vee^3} = \frac{1}{6}\sum_{\pi\in S_3}W_{\pi},
\end{equation}
so that
\begin{equation}
    \mathbb{E}_U\Big[(U^{\dagger}\ket{0}\bra{0}U)^{\otimes3}\Big] = \frac{1}{2^n(2^n+1)(2^n+2)}\sum_{\pi\in S_3}{W_{\pi}}
\end{equation}
and by substituting into Eq.~\eqref{eq:first_term_worked_out_clifford_bijection}
\begin{equation}
2^n\operatorname{Tr}_3\bigg( (\mathbb{I}\otimes\mathbb{I}\otimes\rho) \mathbb{E}_U\Big[(U^{\dagger}\ket{0}\bra{0}U)^{\otimes3}\Big] \bigg) = \frac{1}{(2^n+1)(2^n+2)}\sum_{\pi}\operatorname{Tr}_3 \bigg( (\mathbb{I}\otimes\mathbb{I}\otimes\rho)W_{\pi} \bigg).
\end{equation}
There are six permutations of the three subspaces of the tensor product, the identity, three swaps and two cyclic permutations and we can compute $\operatorname{Tr}_3 \bigg( (\mathbb{I}\otimes\mathbb{I}\otimes\rho)W_{\pi} \bigg)$ for each by representing the tensor product operator in a basis and tracing over the final element:
\begin{align}
    \operatorname{Tr}_3 \bigg( (\mathbb{I}\otimes\mathbb{I}\otimes\rho)W_{\mathrm{id}} \bigg) &= \mathbb{I}\otimes \mathbb{I}\\
    \operatorname{Tr}_3 \bigg( (\mathbb{I}\otimes\mathbb{I}\otimes\rho)W_{(12)} \bigg) &= F\\
    \operatorname{Tr}_3 \bigg( (\mathbb{I}\otimes\mathbb{I}\otimes\rho)W_{\mathrm{(13)}} \bigg) &= \rho\otimes \mathbb{I}\\
    \operatorname{Tr}_3 \bigg( (\mathbb{I}\otimes\mathbb{I}\otimes\rho)W_{\mathrm{(23)}} \bigg) &= \mathbb{I}\otimes \rho \\
    \operatorname{Tr}_3 \bigg( (\mathbb{I}\otimes\mathbb{I}\otimes\rho)W_{\mathrm{(231)}} \bigg) &= F(\rho\otimes \mathbb{I})\\
    \operatorname{Tr}_3 \bigg( (\mathbb{I}\otimes\mathbb{I}\otimes\rho)W_{\mathrm{(312)}} \bigg) &= (\rho\otimes \mathbb{I})F,
\end{align}
where $F$ is the swap operator on the tensor product of two spaces $F\ket{\Phi}\otimes \ket{\Psi} = \ket{\Psi}\otimes \ket{\Phi}$.
Thus we have that
    \begin{align}
        &2^n\operatorname{Tr}_3\bigg( (\mathbb{I}\otimes\mathbb{I}\otimes\rho) \mathbb{E}_U\Big[(U^{\dagger}\ket{0}\bra{0}U)^{\otimes3}\Big] \bigg) =\\ \nonumber
        &\frac{1}{(2^n+1)(2^n+2)}\bigg(\mathbb{I}\otimes \mathbb{I}+F+\rho\otimes \mathbb{I}+\mathbb{I}\otimes \rho+F(\rho\otimes \mathbb{I})+(\rho\otimes \mathbb{I})F \bigg)
    \end{align}
Now we substitute this result back into Eq.~\eqref{eqn:tensor_shadow_exp} to get our final expression
\begin{align}
    &\mathbb{E}[\hat\rho\otimes \hat\rho] = \\ \nonumber
    &=\frac{(2^n+1)^2}{(2^n+1)(2^n+2)}\bigg(\mathbb{I}\otimes \mathbb{I}+F+\rho\otimes \mathbb{I}+\mathbb{I}\otimes \rho+F(\rho\otimes \mathbb{I})+(\rho\otimes \mathbb{I})F \bigg)-\\ \nonumber
    &(2^n+1)\bigg(\frac{\rho\otimes \mathbb{I} + \mathbb{I}\otimes \mathbb{I}}{(2^n + 1)}+\frac{\mathbb{I}\otimes \rho + \mathbb{I}\otimes \mathbb{I}}{(2^n + 1)}\bigg)+\mathbb{I}\otimes \mathbb{I}\\ \nonumber
    &=\frac{(2^n+1)}{(2^n+2)}\bigg(\mathbb{I}\otimes \mathbb{I}+F+\rho\otimes \mathbb{I}+\mathbb{I}\otimes \rho+F(\rho\otimes \mathbb{I})+(\rho\otimes \mathbb{I})F \bigg)- \\ \nonumber
    &\bigg( \rho\otimes\mathbb{I} + \mathbb{I}\otimes \rho + \mathbb{I}\otimes \mathbb{I} \bigg).
\end{align}

\subsection{Protocol variance comparison}\label{SI-sec:protocol_variance_comparison}

For the same total number of shots, $|h|^2+|s|^2\le 1$ follows from $\braket{\Phi|\Lambda}=0$.
Comparison with the independent compute-uncompute variance gives
\begin{equation}
\sigma_{\mathrm{CU}}^2
-
\sigma_{\mathrm{FCU}}^2
\ge 0.
\end{equation}
Thus, whenever the flagged readout is available, it is no worse in shot variance than two separately allocated compute-uncompute measurements.
It is strictly better if $|h|>0$ and $\ket{\Psi}$ has weight outside the span of $\ket{\Phi}$ and $\ket{\Lambda}$.
For $0<|h|<1$ and $0<|s|<1$, the swap-parity coefficient is larger term by term than the compute-uncompute coefficient, so
\begin{equation}
\sigma_{\mathrm{FCU}}^2
\le
\sigma_{\mathrm{CU}}^2
\le
\sigma_{\mathrm{SST}}^2
=
\sigma_{\mathrm{DST}}^2.
\end{equation}
The signed-overlap results also give $\sigma_{\mathrm{RHT}}^2=\sigma_{\mathrm{RIST}}^2=4\sigma_{\mathrm{CU}}^2$, while the large-$d$ classical-shadow result in Eq.~\eqref{eq:cs_variance} gives $\sigma_{\mathrm{CS}}^2\approx12\sigma_{\mathrm{FCU}}^2$.
Under $|h|^2+|s|^2\le1$, FCU, CU, RHT, RIST, and large-$d$ CS have standard-error bounds proportional to $r_{\Phi}/(\sqrt{N_{\mathrm{shot}}}|s|^2)$.

In the limit $s\to1$, orthogonality requires $|h|\to0$.
However, $r_{\Phi}=\norm{Q_{\Phi}H\ket{\Phi}}$ also depends on the reference and vanishes when the reference is the exact eigenstate.
For an exact target, $r_{\Phi}h=(E-E_{\Phi})s$, so these quantities cannot be varied independently.
The leading forms at fixed total shot count are therefore
\begin{equation}
\sigma_{\mathrm{FCU}}^2
\sim
\frac{r_{\Phi}^2}{4N_{\mathrm{shot}}}.
\end{equation}
\begin{equation}
\sigma_{\mathrm{CU}}^2
\sim
\frac{r_{\Phi}^2}{4N_{\mathrm{shot}}}.
\end{equation}
\begin{equation}
\sigma_{\mathrm{RHT}}^2
=
\sigma_{\mathrm{RIST}}^2
\sim
\frac{r_{\Phi}^2}{N_{\mathrm{shot}}}.
\end{equation}
\begin{equation}
\sigma_{\mathrm{CS}}^2
\sim
\frac{3r_{\Phi}^2}{N_{\mathrm{shot}}},
\qquad d\gg1.
\end{equation}
\begin{equation}
\sigma_{\mathrm{SST}}^2
=
\sigma_{\mathrm{DST}}^2
\sim
\frac{r_{\Phi}^2}{4N_{\mathrm{shot}}|h|^2}.
\end{equation}
The $1/|h|^2$ SST/DST factor diverges, but the behaviour of the absolute variance depends on the simultaneous scaling of $r_{\Phi}^2/|h|^2$ and cannot be inferred from $h\to0$ alone.
Setting each variance equal to $\epsilon^2$ gives the conditional leading shot requirements quoted in the main text.
These comparisons describe shot variance only; circuit depth, control overhead, and hardware runtime may change the practical ordering.

\subsection{Circuit duration and time to precision}
\label{SI-sec:runtime_comparison}

To compare time to precision, let $p=p_{\Lambda}$ and $q=p_{\Phi}$, and let $\tau_{\Psi}$, $\tau_{\Phi}$, and $\tau_{\Lambda}$ denote the durations of the corresponding state-preparation blocks, with the same duration assumed for each inverse.
We denote by $\tau_{\mathrm{flag}}$ the duration of the all-zero flag, flag measurement, and feed-forward.
Common state-independent readout and reset times may be absorbed into these durations.
For a maximally parallel balanced fan-in construction, the flag operation has duration $\tau_{\mathrm{flag}}=O(\log n)$, using $O(n)$ work qubits and entangling gates.
A serial construction with fewer work qubits instead has duration $O(n)$.

The CU residual- and reference-overlap circuits have durations $\tau_{\Psi}+\tau_{\Lambda}$ and $\tau_{\Psi}+\tau_{\Phi}$, respectively.
Large-shot propagation gives
\begin{equation}
\sigma_{\mathrm{CU}}^2
=
\frac{r_{\Phi}^2(1-p)}{4qN_p}
+
\frac{r_{\Phi}^2p(1-q)}{4q^2N_q}.
\end{equation}
The corresponding total execution time is
\begin{equation}
T_{\mathrm{CU}}
=
N_p\left(\tau_{\Psi}+\tau_{\Lambda}\right)
+
N_q\left(\tau_{\Psi}+\tau_{\Phi}\right).
\end{equation}
We minimize $T_{\mathrm{CU}}$ over $N_p$ and $N_q$ subject to $\sigma_{\mathrm{CU}}=\epsilon$ using the cost function
\begin{equation}
\mathcal{L}
=
T_{\mathrm{CU}}
+
\lambda\left(\sigma_{\mathrm{CU}}^2-\epsilon^2\right),
\end{equation}
where $\lambda$ is a Lagrange multiplier.
This gives
\begin{equation}
T_{\mathrm{CU}}^*
=
\frac{r_{\Phi}^2}{4\epsilon^2}
\left(
\sqrt{\frac{1-p}{q}\left(\tau_{\Psi}+\tau_{\Lambda}\right)}
+
\sqrt{\frac{p(1-q)}{q^2}\left(\tau_{\Psi}+\tau_{\Phi}\right)}
\right)^2.
\end{equation}

For dynamic FCU, the sequence $U_{\Psi}$, $U_{\Phi}^{\dagger}$, and the flag operation is always executed, whereas $U_{\Phi}$ and $U_{\Lambda}^{\dagger}$ are required only for the unflagged outcome, which occurs with probability $1-q$.
Its average duration is therefore
\begin{equation}
\overline{\tau}_{\mathrm{FCU}}
=
\tau_{\Psi}+\tau_{\Phi}+\tau_{\mathrm{flag}}
+
(1-q)\left(\tau_{\Phi}+\tau_{\Lambda}\right).
\end{equation}
Propagation of the joint FCU record, including $\operatorname{Cov}(X_p,X_q)=-pq$, gives
\begin{equation}
T_{\mathrm{FCU}}^*
=
\frac{r_{\Phi}^2(p+q)}{4q^2\epsilon^2}
\overline{\tau}_{\mathrm{FCU}}.
\end{equation}
After cancelling the common factor $r_{\Phi}^2/(4q^2\epsilon^2)$, dynamic FCU is faster than optimally allocated CU precisely when
\begin{equation}
\begin{split}
&(p+q)
\left[
\tau_{\Psi}+\tau_{\Phi}+\tau_{\mathrm{flag}}
+(1-q)\left(\tau_{\Phi}+\tau_{\Lambda}\right)
\right]
\\
&<
\left[
\sqrt{q(1-p)\left(\tau_{\Psi}+\tau_{\Lambda}\right)}
+
\sqrt{p(1-q)\left(\tau_{\Psi}+\tau_{\Phi}\right)}
\right]^2.
\end{split}
\label{eq:fcu_runtime_crossover}
\end{equation}
In the high-quality-reference limit $q\to1$, orthogonality requires $p\leq1-q\to0$, and the crossover reduces to
\begin{equation}
\tau_{\Phi}+\tau_{\mathrm{flag}}<\tau_{\Lambda}.
\end{equation}
Thus dynamic FCU can be faster when preparing the residual is more expensive than preparing and discriminating the reference state.
A static FCU circuit instead has duration $\tau_{\Psi}+2\tau_{\Phi}+\tau_{\Lambda}+\tau_{\mathrm{flag}}$ and does not gain this early-termination advantage.
The actual dynamic crossover remains dependent on compilation, flag construction, and backend scheduling.

The same construction gives ideal per-shot durations for the other overlap protocols.
Let $\tau_{c\alpha}$ denote controlled preparation of $\ket{\alpha}$.
For $\Omega\in\{\Phi,\Lambda\}$, RHT has duration $\tau_{c\Psi}+\tau_{c\Omega}$.
With separate branch controls and maximally parallel fan-out, RIST has duration
\begin{equation}
\tau_{\Omega}^{\mathrm{RIST}}
=
\max\left(\tau_{c\Psi},\tau_{c\Omega}\right)
+2\tau_{\mathrm{fan}}(n)
+\tau_{\mathrm{CSWAP}},
\end{equation}
where $\tau_{\mathrm{fan}}(n)$ is proportional to $\log n$ using a number of ancillary controls proportional to $n$.
Without fan-out, the controlled-SWAP duration is proportional to $n$.
SST similarly has
\begin{equation}
\tau_{\Omega}^{\mathrm{SST}}
=
\max\left(\tau_{\Psi},\tau_{\Omega}\right)
+2\tau_{\mathrm{fan}}(n)
+\tau_{\mathrm{CSWAP}}.
\end{equation}
DST removes the shared control and has ideal duration
\begin{equation}
\tau_{\Omega}^{\mathrm{DST}}
=
\max\left(\tau_{\Psi},\tau_{\Omega}\right)
+\tau_{\mathrm{CX}}
+\tau_H,
\end{equation}
because all Bell-basis pairs can be processed in parallel.
A CS record has duration $\tau_{\tau}+\tau_{\mathrm{Cliff}}(n)$, where the Clifford contribution depends on the sampled ensemble and its compilation.
Each qubit-wise-commuting PA group has duration $\tau_{\Psi}+\tau_{\mathrm{basis},g}$, but requires its own allocated measurements.

Combining these durations with the leading shot requirements gives
\begin{equation}
T_{\mathrm{RHT}}^*=T_{\mathrm{RIST}}^*
\propto
\frac{r_{\Phi}^2}{\epsilon^2}\tau_h,
\end{equation}
\begin{equation}
T_{\mathrm{CS}}^*
\propto
\frac{3r_{\Phi}^2}{\epsilon^2}\tau_{\mathrm{CS}},
\end{equation}
and
\begin{equation}
T_{\mathrm{SST}}^*=T_{\mathrm{DST}}^*
\propto
\frac{r_{\Phi}^2}{4\epsilon^2|h|^2}\tau_h,
\end{equation}
where $\tau_h$ is the residual-overlap circuit duration for the corresponding protocol.
Thus the shallow Bell layer does not generally make DST competitive when $|h|$ is small, while CS becomes more favourable when its record is reused across several quantities.

\subsection{Shot allocation and bootstrap uncertainties}

The following resampling procedure was used for the noiseless distributions in Fig.~\mainref{fig:result_protocol_energy_histograms}a--c.
The direct-protocol sampling and resampling used base random seeds 0 and $10^6$ for $\mathbf{R}_x$ and $\mathbf{R}_0$, respectively, with fixed offsets for the separate protocol records.
The Helios-1E and Helios-1 estimates were instead evaluated directly from the finite-shot records returned by the backend, with uncertainties propagated from their protocol-specific covariance matrices.

For each protocol the energy was evaluated as
\begin{equation}
E = E_\Phi + r_\Phi R ,
\end{equation}
where $R$ is the estimated ratio $h/s$.
For the squared-overlap protocols, CU, DST, SST, and FCU, the ratio is computed as
\begin{equation}
R = -\sqrt{\frac{p}{q}},
\end{equation}
with $p = |\braket{\Psi|\Lambda}|^2$, $q = |\braket{\Psi|\Phi}|^2$ .
For the real-overlap protocols, RHT and RIST, the ratio is computed as
\begin{equation}
R = \frac{h}{s},
\end{equation}
with $h = \operatorname{Re}\braket{\Psi|\Lambda}$, $s = \operatorname{Re}\braket{\Psi|\Phi}$.
FCU is not split: a single circuit estimates both $p$ and $q$.
The raw pool contains
\begin{equation}
N_{\mathrm{raw}}^{\mathrm{FCU}} = 40\,000
\end{equation}
shots.
Each bootstrap energy is computed from
\begin{equation}
N_{\mathrm{boot}}^{\mathrm{FCU}} = 10\,000
\end{equation}
shots sampled with replacement from the raw FCU pool.
For CS, every shot contributes to both $h$ and $s$, so
\begin{equation}
N_{\mathrm{raw}}^{\mathrm{CS}} = 40\,000,\qquad
N_{\mathrm{boot}}^{\mathrm{CS}} = 10\,000,
\end{equation}
as for FCU.
CU, DST, SST, RHT, and RIST are split estimators, for which the $\Lambda$ and $\Phi$ circuits are sampled separately.
The raw allocation is
\begin{equation}
N_{\Lambda,\mathrm{raw}} = 32\,000,\qquad
N_{\Phi,\mathrm{raw}} = 8000,
\end{equation}
so each split protocol has $40\,000$ raw shots in total.
Each bootstrap energy uses
\begin{equation}
N_{\Lambda,\mathrm{boot}} = 8000,\qquad
N_{\Phi,\mathrm{boot}} = 2000,
\end{equation}
shots sampled with replacement from the corresponding raw $\Lambda$ and $\Phi$ pools.
We generated $B=2000$ bootstrap energies for each protocol, $E_1,\ldots,E_B$.
The reported bootstrap standard error is
\begin{equation}
\sigma_E =
\sqrt{\frac{1}{B-1}\sum_{b=1}^{B}
\left(E_b-\bar E\right)^2},
\end{equation}
where
\begin{equation}
\bar E = \frac{1}{B}\sum_{b=1}^{B} E_b .
\end{equation}

\section{Hardware Results}
    \begin{table}
        \centering
        \resizebox{\columnwidth}{!}{%
        \begin{tabular}{ccccccc}
             \hline CAS&Geometry&State Prep&$(m,k,l)$&Error Detection&$E$ (Ha)&Discarded Shots\\
             \hline (6,6)&$\mathbf{R}_x$&pp-tUPS&$(1,10,2)$&Leakage&-2404.488339$\pm$0.001133&(664,112)\\
             (6,6)&$\mathbf{R}_0$&pp-tUPS&$(1,10,2)$&Leakage&-2404.484198$\pm$0.00114&(666,148)\\
             (6,6)&$\mathbf{R}_x$&pp-tUPS&$(3,10,2)$&Leakage&-2404.485820$\pm$0.00079&(596,126)\\
             (6,6)&$\mathbf{R}_0$&pp-tUPS&$(3,10,2)$&Leakage&-2404.481158$\pm$0.000772&(633,121)\\
             (6,6)&$\mathbf{R}_x$&sMCS&$(1,10,13)$&Leakage&-2404.488276$\pm$0.001048&(406,54)\\
             (6,6)&$\mathbf{R}_0$&sMCS&$(1,10,13)$&Leakage&-2404.4781649$\pm$0.001068&(417,59)\\
             (6,6)&$\mathbf{R}_x$&sMCS&$(3,10,13)$&Leakage&-2404.488073$\pm$0.000751&(308,53)\\
             (6,6)&$\mathbf{R}_0$&sMCS&$(3,10,13)$&Leakage&-2404.481199$\pm$0.000730&(352,56)\\
            (10,10)&$\mathbf{R}_x$&AI-sMCS&$(1,21,24)$&Leakage&-2404.517851$\pm$0.001724&$(1148,153)$\\
            (10,10)&$\mathbf{R}_0$&AI-sMCS&$(1,21,22)$&Leakage&
            -2404.514175$\pm$0.001658&$(998,118)$\\
            (10,10)&$\mathbf{R}_x$&AI-sMCS&$(1,21,24)$&Leakage + CPC&
            -2404.518388$\pm$0.001734&$(1696,271)$\\
            (10,10)&$\mathbf{R}_0$&AI-sMCS&$(1,21,22)$&Leakage + CPC&
            -2404.511333$\pm$0.001691&$(1387,241)$\\
            (10,10)&$\mathbf{R}_x$&AI-sMCS&$(1,21,24)$&Leakage + MBPC&
            -2404.517853$\pm$0.001734&$(1697,297)$\\
            (10,10)&$\mathbf{R}_0$&AI-sMCS&$(1,21,22)$&Leakage + MBPC&
            -2404.513668$\pm$0.001687&$(1494,261)$\\
             \hline
        \end{tabular}
        }
        \caption{Helios-1 measured CU ME energies.
        All Helios results use ternary leakage detection and postselection, with $N_s=2000$ and $N_h=8000$ submitted shots per geometry on hardware.
        The discarded shots column indicates the number of shots discarded for the $(h,s)$ circuits respectively due to the leakage and error detection when applied.
        Uncertainties are one standard deviation.}
        \label{SI-tab:helios1_cu_me_energies}
    \end{table}

    \begin{table}
        \centering
        \resizebox{\columnwidth}{!}{%
        \begin{tabular}{ccccccc}
             \hline CAS&Geometry&State Prep&$(m,k,l)$&Error Detection&$E$ (Ha)&Discarded Shots\\
             \hline (6,6)&$\mathbf{R}_x$&pp-tUPS&$(1,10,2)$&Leakage&-2404.492325$\pm$0.004523&1062\\
             (6,6)&$\mathbf{R}_0$&pp-tUPS&$(1,10,2)$&Leakage&-2404.489726$\pm$0.005614&1121\\
             (6,6)&$\mathbf{R}_x$&sMCS&$(1,10,13)$&Leakage&-2404.480324$\pm$0.004205&766\\
             (6,6)&$\mathbf{R}_0$&sMCS&$(1,10,13)$&Leakage&-2404.478163$\pm$0.004221&891\\
             \hline
        \end{tabular}
        }
        \caption{Helios-1 measured CS ME energies.
        All Helios results use leakage detection and postselection, with $2000$ separate Clifford measurement circuits and $5$ submitted shots per circuit on hardware.
        The discarded shots column indicates the number of shots discarded for the circuits due to the leakage detection.
        Uncertainties are one standard deviation.}
        \label{SI-tab:helios1_cs_me_energies}
    \end{table}

    CU mixed-estimator energies for the (6,6) and (10,10) active spaces are reported in Supplementary Table~\ref{SI-tab:helios1_cu_me_energies}.
    The corresponding (6,6) CS energies are reported in Supplementary Table~\ref{SI-tab:helios1_cs_me_energies}.
    These data are plotted in Figs.~\ref{fig:tups_mc_cu_cs} and \ref{fig:result_active_space}.
    All measurements used ternary leakage detection and postselection.
    CU used $8000$ and $2000$ shots for the $h$ and $s$ circuits, respectively, while CS used $2000$ Clifford circuits with five shots per circuit.
    The corresponding discarded-shot counts are also reported.

\section{Optimized geometries XYZ files}

Here we report the optimized geometries (XYZ) together with the PBE electronic energies from the geometry optimisations.\\

S0 geometry\\
87\\
E= -21.50228054 a.u.\\
O        22.173499       7.005746      20.405809\\
O        20.995754       5.855936      18.833590\\
C        21.581092       6.853747      19.199705\\
O        23.553835      21.948202       7.288453\\
O        22.320360      20.691032       5.846326\\
C        22.850272      20.851328       6.925982\\
Ru       22.213102      12.462587      11.492880\\
N        22.376688      12.588045      9.4296162\\
C        23.650555      12.659518       8.9382257\\
C        23.879493      12.802580       7.5691670\\
C        22.804540      12.864237       6.6915005\\
C        21.511033      12.771755       7.2022657\\
C        21.337240      12.632799       8.571037\\
H        24.896696      12.868931       7.1856869\\
H        22.974677      12.978786       5.619672\\
H        20.638643      12.805458       6.547911\\
H        20.345454      12.559828       9.01612\\
N        22.306596      14.524266      11.818014\\
N        20.141040      12.318088      11.602980\\
N        24.272633      12.465173      11.237180\\
N        21.932018      10.407761      11.397314\\
N        22.183441      12.528916      13.58241\\
C        24.709579      12.585964       9.947686\\
C        22.242409      14.890851     13.134748\\
C        22.429596      15.466887     10.861129\\
C        19.640561      11.046983      11.643532\\
C        19.292882      13.361973      11.699465\\
C        25.176751      12.395917     12.235295\\
C        20.640197       9.981223     11.536340\\
C        22.915740       9.493523     11.27237\\
C        22.148747      13.792424      14.104642\\
C        22.115914      11.459287      14.400419\\
C        26.0740485      12.62433       9.657824\\
C        22.272431      16.247791    13.469989\\
C        22.463954      16.844013      11.125965\\
H        22.516643      15.099883       9.839514\\
C        18.268792      10.824458     11.768658\\
C        17.921657      13.195106     11.825406\\
H        19.746107      14.352416      11.672380\\
C        26.543376      12.427878    12.000456\\
H        24.770950      12.311373      13.24305\\
C        20.339617       8.6191803      11.580112\\
C        22.669875       8.1290091      11.308661\\
H        23.924423       9.8880012      11.153530\\
C        22.035567      13.958087    15.488385\\
C        21.996667      11.552090      15.7946145\\
H        22.182090      10.485313     13.917655\\
C        27.001015      12.542486     10.689218\\
H        26.415795      12.716312      8.627837\\
C        22.371765      17.214564      12.479610\\
H        22.207578      16.556783      14.512745\\
C        17.399084      11.903873      11.859784\\
H        17.875489       9.809067     11.79265\\
H        17.275706      14.071207     11.897429\\
H        27.236912      12.365156     12.840182\\
C        21.359803       7.682713     11.471697\\
H        19.309213      8.286544     11.698680\\
H        23.500245       7.42829      11.21096\\
C        21.954026      12.85384      16.324952\\
H        21.991470      14.95781      15.919273\\
H        28.070187      12.56969      10.47169\\
H        22.353390      18.27059      12.753528\\
H        16.324819      11.73858      11.957384\\
H        21.824029      12.99931      17.39846\\
H        21.134421       6.61552      11.50905\\
C        22.591585      17.84682      10.04770\\
C        21.968331      17.67328       8.797630\\
C        22.067762      18.648430       7.809711\\
C        22.794574      19.825210       8.033536\\
C        23.429707      19.999001       9.269454\\
C        23.328242      19.025692      10.26066\\
H        21.356670      16.788445       8.60945\\
H        21.566113      18.525894      6.84801\\
H        23.090679      11.208103     18.25982\\
H        24.010175      20.902924       9.456534\\
H        23.856938      19.175041      11.203959\\
C        21.917145      10.350461      16.65105\\
C        21.228443       9.194561      16.23704\\
C        21.135298       8.083965      17.07027\\
C        21.730077       8.086385      18.33893\\
C        22.426982       9.228090      18.75319\\
C        22.517237      10.342655      17.92251\\
H        20.717278       9.180397      15.27210\\
H        20.583520       7.194366     16.76104\\
H        22.905287       9.243452     19.73294\\
H        23.528895      22.556307       6.504779\\
H        22.019036       6.154740      20.89204\\

T1 geometry\\
87\\
E= -21.43198963 a.u.\\
O        21.910791       7.086404      20.486871\\
O        21.2189065       5.77919      18.758630\\
C        21.581092       6.853747     19.199705\\
O        23.254764      22.059728       7.367881\\
O        22.579372      20.589766       5.768734\\
C        22.850272      20.851328       6.925982\\
Ru       22.122152      12.465127      11.483233\\
N        22.248555      12.587913       9.405561\\
C        23.516125      12.685464       8.899362\\
C        23.720249      12.860905       7.528388\\
C        22.631093      12.917209       6.670174\\
C        21.345222      12.786880       7.196900\\
C        21.195202      12.623781       8.565330\\
H        24.729756      12.956805       7.130841\\
H        22.782524      13.056429       5.598415\\
H        20.463277      12.811645       6.555470\\
H        20.211819      12.527971       9.02535\\
N        22.257361      14.508890      11.782909\\
N        20.050625      12.366930     11.646138\\
N        24.178030      12.451600      11.184297\\
N        21.786914      10.408922      11.37922\\
N        22.161827      12.497369      13.55361\\
C        24.588741      12.592328       9.886550\\
C        22.268337      14.860862     13.139982\\
C        22.373332      15.456063      10.839189\\
C        19.523684      11.104514      11.70260\\
C        19.232483      13.433527     11.749181\\
C        25.09639      12.3645052      12.16735\\
C        20.49080      10.0180275      11.58634\\
C        22.742978       9.472848      11.216170\\
C        22.2158688      13.79554      14.077194\\
C        22.061279      11.442149     14.375428\\
C        25.950368      12.624595       9.575258\\
C        22.338266      16.241120      13.468153\\
C        22.466776      16.831687      11.105937\\
H        22.406018     15.093874      9.8124327\\
C        18.144876      10.91960      11.840801\\
C        17.860221      13.301238      11.891505\\
H        19.713271      14.410685     11.714668\\
C        26.457930      12.392714      11.909032\\
H        24.705143      12.270127      13.179991\\
C        20.166184       8.661426     11.677753\\
C        22.470465       8.115723     11.288567\\
H        23.752657       9.844156     11.04111\\
C        22.208568      13.943058      15.49031\\
C        22.037161      11.528552      15.77755\\
H        22.043290      10.468336      13.88811\\
C        26.892079      12.522153      10.58925\\
H        26.275485      12.725973       8.540875\\
C        22.434716      17.195633     12.489741\\
H        22.301097      16.548209     14.51282\\
C        17.306352      12.020455     11.93461\\
H        17.725762       9.914728      11.867497\\
H        17.237266      14.192879      11.971190\\
H        27.1658762      12.317664      12.73528\\
C        21.159532       7.7032378   11.535794\\
H        19.137574       8.3539434      11.862269\\
H        23.277862       7.394318      11.157386\\
C        22.125639     12.851722      16.314877\\
H        22.255252      14.939966      15.926892\\
H        27.957448      12.545087      10.35376\\
H        22.446665      18.248516      12.77230\\
H        16.228990      11.883229      12.04122\\
H        22.082007      13.002376      17.39355\\
H        20.915040       6.642192      11.61009\\
C        22.582306      17.832031      10.043218\\
C        22.074887      17.615307       8.738855\\
C        22.172622      18.590409      7.757468\\
C        22.777658      19.830418      8.029204\\
C        23.303904      20.052908      9.31070\\
C        23.202789      19.079016      10.29585\\
H        21.548491      16.688531      8.504935\\
H        21.760726      18.423360       6.76065\\
H        22.955557      11.251499      18.32673\\
H        23.796731      20.999422       9.53520\\
H        23.646381      19.277364     11.27218\\
C        21.939895      10.343505      16.63031\\
C        21.331941      9.142093      16.18877\\
C        21.229225       8.039524      17.023402\\
C        21.722459       8.080829      18.34002\\
C        22.342985       9.257528      18.78661\\
C        22.443961      10.363797     17.95324\\
H        20.887696       9.089656      15.19341\\
H        20.743777       7.123306      16.68280\\
H        22.750239       9.302974     19.79725\\
H        23.285740      22.650572       6.57082\\
H        21.802202       6.221629      20.96178\\

Rx geometry\\
87\\
E= -21.39227892 a.u.\\
Ru       -3.7326573      16.409606      19.11481\\
N        -5.2422157     15.3154771     18.13735\\
C        -5.4226293      15.553340      16.801928\\
C        -6.3726400      14.911866     16.004975\\
C        -7.1661116      13.971688     16.647345\\
C        -6.9935039      13.710737   18.006280\\
C        -6.0239659      14.388169    18.72910\\
H        -6.4567648     15.1424131     14.943417\\
H        -7.9147591      13.448443     16.050337\\
H        -7.6002998      12.974466     18.53480\\
H        -5.8388466      14.223745     19.790275\\
N        -4.6708282      18.275208     19.18482\\
N        -3.6771459      15.171553   20.848458\\
N        -3.4446699      16.963878     16.950219\\
N        -2.3644212      14.64673      18.582200\\
N        -2.1964895      17.541152     19.921120\\
C        -4.5288009      16.564903     16.230580\\
C        -3.8395012      19.259138    19.689675\\
C        -5.9302280      18.447261     18.75962\\
C        -2.8840863      14.063196      20.84658\\
C        -4.3101057      15.632998      21.94628\\
C        -2.6030600     17.9425048    16.57001\\
C        -2.2621356     13.7154274    19.56809\\
C        -1.8488769      14.372584      17.371953\\
C        -2.5043267      18.884327    20.062064\\
C        -0.9839457      17.096597    20.281525\\
C        -4.7634434      17.116015     14.967546\\
C        -4.3878619      20.551540     19.79400\\
C        -6.4649178      19.735738    18.848490\\
H        -6.4679142      17.599312      18.337562\\
C        -2.7227158      13.322457   22.020236\\
C        -4.1834936      14.911324      23.122378\\
H        -4.8972008     16.5455828     21.832169\\
C        -2.7793560     18.5285337      15.325511\\
H        -1.7844595      18.230956      17.23152\\
C        -1.6171934     12.484037      19.423763\\
C        -1.1819328      13.176601     17.140034\\
H        -1.9817577      15.144413      16.615054\\
C        -1.5218517     19.774399     20.535057\\
C         0.0306727      17.92895      20.76557\\
H        -0.8262616     16.023193      20.19375\\
C        -3.8607408      18.089429      14.559644\\
H        -5.6270804      16.76649     14.40234\\
C        -5.6789862     20.788470      19.37704\\
H        -3.7758251      21.342471      20.225974\\
C        -3.3878979      13.765726      23.154515\\
H        -2.0781822      12.444604      21.998036\\
H        -4.7171678     15.284324      23.997384\\
H        -2.0594328      19.291736      15.027460\\
C        -1.0754805      12.241913      18.16855\\
H        -1.5779946      11.778088      20.252172\\
H        -0.7676446      13.008506      16.14516\\
C        -0.2732598     19.306541     20.879548\\
H        -1.7629267     20.834832      20.60039\\
H        -4.0222494      18.535058     13.576842\\
H        -6.1227471     21.778771     19.488895\\
H        -3.2678781     13.193856     24.076104\\
H         0.5066771     19.996366      21.204576\\
H        -0.5677887     11.288907      18.009664\\
C        -7.8317172      19.92415      18.371964\\
C        -8.8295413      18.93867     18.50841\\
C       -10.1214878      19.13924      18.05350\\
C       -10.4638407      20.348632      17.43513\\
C        -9.4780446     21.331805     17.294965\\
C        -8.1836710     21.13959     17.756174\\
H        -8.5688711      18.02367     19.04258\\
H       -10.9276481      18.412066     18.16054\\
H        -9.7647884     22.26247      16.805955\\
H        -7.4012293      21.88815     17.62292\\
C         1.3305621      17.37779    21.137805\\
C         1.8677926      16.28575     20.42722\\
C         3.0958718      15.768424     20.802167\\
C         3.8076461      16.318090      21.876585\\
C         3.2750333      17.405419      22.57806\\
C         2.0472141     17.93819     22.211418\\
H         1.341993      15.887908     19.55766\\
H         3.542388       14.934722    20.25851\\
H         3.825470     17.82836     23.418071\\
H         1.6183221     18.763004    22.782288\\
C       -11.868647      20.479116     16.95470\\
O       -12.105501      21.72616    16.487097\\
O       -12.707368     19.598622    16.953434\\
H       -13.053970     21.654939     16.20391\\
C         5.11298      15.695157     22.23707\\
O         5.773233      16.39411     23.18825\\
O         5.514531      14.66681   21.726297\\
H        6.611745    15.890057    23.35434\\

MECP geometry\\
87\\
E = -21.41501124 a.u.\\
  Ru   -3.82937049   16.38574319   19.43031042\\
   N   -5.33004161   15.23896091   18.46567411\\
   C   -5.56501426   15.45879348   17.13599675\\
   C   -6.42244167   14.63260579   16.40955120\\
   C   -7.07134664   13.57967851   17.04054491\\
   C   -6.84823972   13.37710576   18.40052204\\
   C   -5.97210970   14.21635290   19.07104999\\
   H   -6.55095828   14.80026308   15.34033172\\
   H   -7.73303409   12.92022191   16.47650891\\
   H   -7.33926730   12.56784584   18.94264916\\
   H   -5.75917083   14.07639601   20.13023303\\
   N   -4.65449440   18.32737149   19.23428743\\
   N   -3.39731277   14.89001041   20.95413412\\
   N   -3.70104544   16.96186222   17.09073427\\
   N   -2.32112456   14.63695656   18.52430444\\
   N   -2.29725405   17.60370677   20.13853688\\
   C   -4.87249975   16.60039906   16.51766522\\
   C   -3.83024377   19.35501923   19.60289859\\
   C   -5.89449566   18.58287616   18.78238194\\
   C   -2.56647119   13.83493658   20.75942398\\
   C   -3.84330744   15.15577534   22.19897451\\
   C   -3.04477633   18.03568483   16.62216717\\
   C   -2.15293086   13.59681368   19.37215155\\
   C   -1.97332373   14.48016478   17.24007477\\
   C   -2.50531436   18.95716533   20.05834068\\
   C   -1.10143437   17.14447125   20.54873838\\
   C   -5.41377436   17.32094353   15.44837669\\
   C   -4.28736146   20.67300777   19.51873887\\
   C   -6.40239179   19.87686373   18.65456739\\
   H   -6.48677081   17.72123610   18.47563016\\
   C   -2.13908611   13.05330156   21.83523558\\
   C   -3.47994121   14.39588738   23.29930699\\
   H   -4.51182314   16.01379314   22.30184329\\
   C   -3.51874743   18.79351927   15.56024213\\
   H   -2.10809174   18.28326753   17.12728951\\
   C   -1.65264004   12.36582645   18.93978324\\
   C   -1.44877038   13.29154010   16.74117155\\
   H   -2.13526586   15.33941723   16.59150137\\
   C   -1.47456465   19.84651172   20.37255089\\
   C   -0.02714506   17.97983643   20.86884124\\
   H   -1.00150466   16.06453612   20.63468944\\
   C   -4.72697792   18.42891794   14.96463822\\
   H   -6.37867406   17.04323027   15.02375149\\
   C   -5.56281387   20.93361015   19.04480262\\
   H   -3.65275403   21.49668997   19.84386468\\
   C   -2.59986127   13.33038987   23.11508842\\
   H   -1.42731857   12.24448403   21.67328747\\
   H   -3.87395505   14.64612336   24.28534502\\
   H   -2.94878880   19.65208822   15.20257050\\
   C   -1.29716568   12.21093071   17.60549866\\
   H   -1.57561278   11.52589064   19.62919646\\
   H   -1.17707618   13.21666278   15.68731037\\
   C   -0.23759808   19.36426214   20.76716932\\
   H   -1.63282920   20.92058614   20.28086017\\
   H   -5.13339753   19.00869644   14.13448663\\
   H   -5.93524738   21.95904781   19.00427706\\
   H   -2.27159411   12.72651670   23.96261790\\
   H    0.58441554   20.05442363   20.96475198\\
   H   -0.91692398   11.25377601   17.24328665\\
   C   -7.74667189   20.10500213   18.11452074\\
   C   -8.82155089   19.28132031   18.48618084\\
   C  -10.08613153   19.50592206   17.96169825\\
   C  -10.29675889   20.54772412   17.05282019\\
   C   -9.23322897   21.37753504   16.68590943\\
   C   -7.96725745   21.15801210   17.21452108\\
   H   -8.67598620   18.49497808   19.22916091\\
   H  -10.93961738   18.89319200   18.25506789\\
   H   -9.40037853   22.19396259   15.98426336\\
   H   -7.13540252   21.79197496   16.90194484\\
   C    1.25805864   17.41132207   21.28827106\\
   C    1.75656748   16.24502706   20.68345770\\
   C    2.97123230   15.71258113   21.08892743\\
   C    3.70753875   16.32985594   22.10555623\\
   C    3.22182773   17.49498549   22.70588198\\
   C    2.00698964   18.03247573   22.29862818\\
   H    1.21143397   15.78140013   19.85880888\\
   H    3.38472027   14.81910968   20.61924488\\
   H    3.79466306   17.97568873   23.49802146\\
   H    1.61886762   18.92196010   22.79774041\\
   C  -11.67702248   20.71828946   16.50718228\\
   O  -11.76808327   21.77129906   15.65894578\\
   O  -12.60131349   19.98459574   16.79175790\\
   H  -12.71429515   21.79325415   15.36107743\\
   C    4.99911203   15.69492629   22.50642385\\
   O    5.65132654   16.40954081   23.45468615\\
   O    5.40535238   14.65362439   22.03143044\\
   H    6.48422888   15.90518874   23.64508430\\

MC geometry\\
87\\
E= -21.42318313 a.u.\\
Ru       -3.51791123      16.49477015      18.81657418\\
N        -5.19451082      15.29322067     18.225139304\\
C        -5.89740833      15.59321733      17.0961250\\
C        -6.9175455      14.75465545     16.6453782\\
C        -7.2403189      13.6134838      17.3684172\\
C        -6.5264891      13.3245807      18.5304765\\
C        -5.5037975      14.1769729     18.9176392\\
H        -7.4343158      14.9850276      15.7136831\\
H        -8.0348026      12.9491968      17.0233798\\
H        -6.7512300      12.4419893      19.1307017\\
H        -4.9044349      13.9905574     19.80981837\\
N        -4.5137577      18.2873315     19.2810831\\
N        -2.9125783      15.3463105     20.8907813\\
N        -4.2675025      17.2817530     16.6117565\\
N        -2.4072563      14.7479952      18.2829391\\
N        -1.9789024      17.7225563      19.5624940\\
C        -5.5390757      16.8500805     16.4208971\\
C        -3.7213572      19.3318788     19.6730172\\
C        -5.7877621      18.5233768      18.931768\\
C        -1.8942275      14.4912906      20.6214971\\
C        -2.9390152      15.9518235      22.0912197\\
C        -3.9211437      18.4836598      16.1189160\\
C        -1.845419      13.96591925      19.2484149\\
C        -2.364087      14.34212973      16.9964416\\
C        -2.360410     18.95693452      20.0142941\\
C        -0.771301      17.2421315      19.8996829\\
C        -6.486413     17.61114666      15.7293049\\
C        -4.226228     20.63468780     19.6791504\\
C        -6.325714     19.8052971      18.818130\\
H        -6.377346      17.652254      18.6580348\\
C        -0.88734     14.22049940      21.5530656\\
C        -1.9661037    15.7421482      23.0601691\\
H        -3.7729357     16.635007      22.2665038\\
C        -4.8112431      19.300250      15.4318266\\
H        -2.889981     18.7971140     16.2958678\\
C        -1.2434636     12.749947     18.9216783\\
C        -1.7552636     13.156904     16.614119\\
H        -2.8378079     15.007145     16.2729397\\
C        -1.4968604     19.711027      20.813106\\
C         0.0818382      17.881252     20.798350\\
H        -0.5177588      16.2649632      19.496866\\
C        -6.1184497     18.85632815      15.2345844\\
H        -7.5144230      17.2622203     15.6296814\\
C        -5.5126291      20.8772595     19.22499473\\
H        -3.5920161      21.4642810     19.9919365\\
C        -0.9199598      14.8611319     22.7855761\\
H        -0.0564394      13.5625342     21.2974302\\
H        -2.0324536      16.2588748    24.0187204\\
H        -4.4793892      20.2674499     15.051379\\
C        -1.1926228      12.3415362     17.5949012\\
H        -0.8444781      12.1150229     19.7128700\\
H        -1.7336586      12.8762753     15.560328\\
C        -0.2927472      19.1663506     21.2289017\\
H        -1.7978148      20.6993506    21.1608968\\
H        -6.8503582      19.4796815     14.7197157\\
H        -5.8979026      21.8975958     19.1952442\\
H        -0.1263288      14.6899344     23.5135867\\
H         0.3641641      19.7339306    21.8897787\\
H        -0.7298137      11.3896772     17.3287469\\
C        -7.6369546      19.9719027     18.1835938\\
C        -8.6473713      19.0106052     18.3518718\\
C        -9.8301969      19.1048893     17.6357398\\
C       -10.0302690      20.1633678     16.7426310\\
C        -9.0447525      21.1436509     16.5941765\\
C        -7.8547705      21.0452477    17.3065913\\
H        -8.5210308      18.20427325     19.0764631\\
H       -10.6309541      18.37385492    17.7590175\\
H        -9.2095296      21.97775126    15.9115997\\
H        -7.0728257      21.78994824     17.148621\\
C         1.2245215      17.1413169    21.344852\\
C         1.918554      16.19754917     20.570149\\
C         2.8793733      15.3855293     21.1516372\\
C         3.171497      15.50340904     22.5152910\\
C         2.516790      16.46934480      23.284601\\
C         1.546575      17.2778593      22.7035899\\
H         1.727298     16.12106400    19.4982972\\
H         3.433459      14.6498933     20.566030\\
H         2.759823      16.57551819    24.3421982\\
H         1.0037284     17.9920791     23.3251981\\
C       -11.3162933     20.1848887     15.981852\\
O       -11.4534573     21.3054625      15.234470\\
O       -12.1309652      19.285597     16.0304221\\
H       -12.3328368      21.223170      14.7822447\\
C         4.1879591      14.568525     23.0841340\\
O         4.4683635      14.840096      24.3807393\\
O         4.6877508     13.6638453      22.4464595\\
H         5.1521660      14.1751787     24.6545700\\

T1-O2 at 2.8\AA \\
89\\
E= -21.78919658 a.u.\\
O        22.24834176       7.12926268      20.42280333\\
O        21.55645657       5.82205553      18.69456216\\
C        21.91864224       6.89660556      19.13563689\\
O        23.59231514      22.10258706       7.30381324\\
O        22.91692229      20.63262518       5.70466657\\
C        23.18782224      20.89418656       6.86191389\\
Ru       22.45970248      12.50798645      11.41916563\\
N        22.58610114      12.63077165       9.34149362\\
C        23.85367619      12.72832287       8.83529395\\
C        24.05779972      12.90376397       7.46432071\\
C        22.96864410      12.96006807       6.60610655\\
C        21.68277305      12.82973939       7.13283245\\
C        21.53275231      12.66664026       8.50126216\\
H        25.06730627      12.99966426       7.06677381\\
H        23.12007462      13.09928779       5.53434774\\
H        20.80082753      12.85450357       6.49140235\\
H        20.54936976      12.57082992       8.96128339\\
N        22.59491214      14.55174903      11.71884178\\
N        20.38817619      12.40978934      11.58207067\\
N        24.51558082      12.49445863      11.12022939\\
N        22.12446434      10.45178121      11.31515768\\
N        22.49937771      12.54022797     13.489545460\\
C        24.92629169      12.63518737       9.82248224\\
C        22.60588792      14.90372143      13.07591464\\
C        22.71088274      15.49892254      10.77512138\\
C        19.86123435      11.14737289      11.63853667\\
C        19.57003419      13.47638561      11.68511370\\
C        25.43394838      12.40736385     12.103288340\\
C        20.82835031      10.06088613     11.52228100\\
C        23.08052825       9.51570733      11.15210214\\
C        22.55341911      13.83839916      14.01312623\\
C        22.39882958     11.48500759      14.31136055\\
C        26.28791856      12.66745442      9.51119001\\
C        22.67581679      16.28397924      13.40408515\\
C        22.80432625      16.87454573     11.04186961\\
H        22.74356904      15.13673279      9.74836462\\
C        18.48242628     10.96246806      11.77673313\\
C        18.19777186     13.34409744      11.8274373\\
H        20.0508214      14.45354374     11.65060076\\
C        26.79548063      12.43557313      11.84496489\\
H        25.04269365      12.31298579      13.11592389\\
C        20.50373504       8.70428487      11.61368519\\
C        22.8080157       8.15858162      11.22449893\\
H        24.0902076       9.88701547      10.97704936\\
C        22.5461184      13.98591728      15.42624363\\
C        22.37471163      11.57141078      15.7134917\\
H        22.38084034      10.51119508      13.82404924\\
C        27.22963014      12.56501193      10.52519049\\
H        26.61303533      12.76883226       8.47680788\\
C        22.77226709     17.23849249      12.42567327\\
H        22.63864814      16.59106769      14.44875449\\
C        17.64390269      12.06331368      11.87054375\\
H        18.06331248      9.95758664      11.80342986\\
H        17.57481633      14.23573796      11.90712286\\
H        27.50342644     12.36052262      12.67122128\\
C        21.49708323       7.7460964      11.47172678\\
H        19.47512429      8.39680198      11.79820096\\
H        23.61541234      7.43717705      11.09331804\\
C        22.46318993      12.89458152      16.25080969\\
H        22.59280267     14.98282522      15.86282392\\
H        28.29499902     12.58794608      10.28969491\\
H        22.78421600      18.29137523      12.70823829\\
H        16.56654061    11.92608825      11.9771617\\
H        22.41955727     13.04523486      17.32948521\\
H        21.25259073      6.68505077      11.54602459\\
C        22.91985639     17.87489033       9.97915082\\
C        22.41243740     17.65816633       8.67478769\\
C        22.51017312     18.63326846       7.69340030\\
C        23.11520890      19.87327690       7.96513599\\
C        23.64145448    20.09576741       9.24663254\\
C        23.54033982     19.12187533      10.23179092\\
H        21.88604179     16.73138993       8.44086765\\
H        22.09827699    18.46621897       6.69658773\\
H        23.29310782     11.29435775      18.26267013\\
H        24.13428219      21.04228120       9.47113957\\
H        23.98393195     19.32022344      11.20811251\\
C        22.27744594     10.38636367      16.56624758\\
C        21.66949146      9.18495164      16.12471091\\
C        21.56677594       8.08238271      16.95933391\\
C        22.06000935       8.12368772      18.27595631\\
C        22.68053537      9.30038750      18.72254711\\
C        22.78151186     10.40665569      17.88917464\\
H        21.22524668      9.13251475      15.12934211\\
H        21.08132822       7.16616534      16.61873339\\
H        23.08778967       9.34583338      19.73318555\\
H        23.62329042      22.69343075       6.50675618\\
H        22.13975312       6.26448832      20.89771543\\
O        19.68340813     13.04835983      15.95220391\\
O        19.75339768      14.01121464      15.22606613\\

\clearpage
\renewcommand{\refname}{Supplementary References}
\putbib[references]
\end{bibunit}

\end{document}